\documentclass[aps,prd,nofootinbib,showkeys,preprint,floatfix]{revtex4-1}

\usepackage{amsmath}
\usepackage{amssymb}
\usepackage{graphicx}
\usepackage[mathscr]{eucal}
\usepackage{color}
\usepackage[T1]{fontenc} 
\usepackage{slashed}
\usepackage{xspace}
\usepackage{hyperref}
\usepackage{xspace}
\usepackage{ulem}
\usepackage{cancel}
\usepackage{enumerate}
\newcommand\SARAH{{\tt SARAH}\xspace}
\newcommand\SPheno{{\tt SPheno}\xspace}
\newcommand\Fortran{{\tt Fortran}\xspace}
\newcommand\Mathematica{{\tt Mathematica}\xspace}

\newcommand{\DRbar}{{\ensuremath{\overline{\mathrm{DR}}}}\xspace}
\newcommand{\MSbar}{{\ensuremath{\overline{\mathrm{MS}}}}\xspace}
\newcommand{\nn}{\nonumber}

\newcommand{\AddrBonn}{%
Bethe Center for Theoretical Physics \& Physikalisches Institut der 
Universit\"at Bonn, \\
 53115 Bonn, Germany
}

\newcommand{\AddrCERN}{%
Theory Division, CERN, 1211 Geneva 23, Switzerland
}

\graphicspath{{./Figs/}}

\begin{document}

\hfill BONN--TH--2015--07, CERN-PH-TH-2015-123 \vspace{0.2cm}

 \title{Precise determination of the Higgs mass \\ in supersymmetric models with vectorlike tops \\ and the impact on naturalness in minimal GMSB}

% \author{Mark D. Goodsell}
% \email{goodsell@lpthe.jussieu.fr}
% \affiliation{\AddrParis}

\author{Kilian Nickel} 
\email{nickel@th.physik.uni--bonn.de}
\affiliation{\AddrBonn}
 
 \author{Florian Staub}
 \email{florian.staub@cern.ch}
 \affiliation{\AddrCERN}

% \pacs{14.60.Pq, 12.60.Jv, 14.80.Cp}

\begin{abstract}
We present a precise analysis of the Higgs mass corrections stemming from vectorlike top partners in supersymmetric models. We reduce the theoretical uncertainty  compared to previous studies in the following aspects:  (i) including the one-loop threshold corrections to SM gauge and Yukawa couplings due to the presence of the new states to obtain  the $\overline{\text{DR}}$ parameters entering all loop calculations, (ii)  including the full momentum dependence at one-loop, and (iii)  including all two-loop corrections but the ones involving $g_1$ and $g_2$. We find that the additional threshold corrections are very important and can give the largest effect on the Higgs mass. However, we identify also parameter regions where the new two-loop effects can be more important than the ones of the MSSM and change the Higgs mass prediction by up to 10 GeV. This is for instance the case in the low $\tan\beta$, small $M_A$ regime. We use these results to calculate the electroweak fine-tuning of an UV complete variant of this model. For this purpose, we add a complete {\bf 10} and {$\overline{\bf  10}$} of $SU(5)$ to the MSSM particle content. We embed this model in minimal Gauge Mediated Supersymmetry Breaking and calculate the electroweak fine-tuning with respect to all important parameters. It turns out that the limit on the gluino mass becomes more important for the fine-tuning than the Higgs mass measurements which is easily to satisfy in this setup. 
\end{abstract}
\maketitle

\section{Introduction}
The discovery of the Higgs boson with a mass of about 125~GeV \cite{Chatrchyan:2012ufa,Aad:2012tfa} has a strong impact on the parameter range of supersymmetric (SUSY) models, especially as its mass value is turning into a precision observable with an uncertainty below 1\%. In particular, in constrained versions of the Minimal Supersymmetric Standard Model (MSSM) large regions of the parameter space are not consistent with this mass range \cite{Arbey:2011ab}. This is in particular the case for models where SUSY breaking is assumed to be transmitted from the hidden to the visible sector via gauge interactions like in minimal Gauge Mediated SUSY Breaking (GMSB). 
Even relaxing the predictive boundary conditions of a constrained model and considering the phenomenological MSSM with many more parameters at the SUSY scale, it is still rather difficult to find regions with the correct Higgs mass. Either, a very large mixing in the stop sector or heavy stop masses are needed to push the Higgs mass to the desired range \cite{Heinemeyer:2011aa,Draper:2011aa,Carena:2011aa,Kadastik:2011aa,Feng:2012jfa,Fowlie:2012im,Arbey:2012dq,Carena:2013qia,Hagiwara:2012mga,Arbey:2012bp,Dumont:2013npa,Djouadi:2013uqa,Djouadi:2013vqa,Djouadi:2013lra,Djouadi:2015jea}. However, the large stop mixing with light stops turns out to be dangerous because of charge and colour breaking minima \cite{Camargo-Molina:2013sta,Blinov:2013fta,Chowdhury:2013dka,Camargo-Molina:2014pwa,Chattopadhyay:2014gfa}. On the other side, very heavy stops introduce again a hierarchy problem which SUSY was supposed to solve. The question about naturalness and fine-tuning is even more pronounced in regions the small $\tan\beta$ region which recently gained some interest because of Higgs fits \cite{Djouadi:2013vqa,Djouadi:2013uqa,Djouadi:2013lra}: in these regions the tree-level Higgs mass is suppressed by a factor $\cos(2\beta)$ and even much bigger loop corrections are needed than for larger values of $\tan\beta$. 

A widely studied ansatz to solve this tension and to reduce the necessary fine-tuning in SUSY models is to enhance the Higgs mass already at tree-level. For this purpose models are considered which give new $F$- \cite{Ellwanger:2009dp,Ellwanger:2006rm,Ross:2011xv,Ross:2012nr,Kaminska:2013mya,Lu:2013cta,Kaminska:2014wia,Ding:2015wma,Benakli:2012cy} or $D$-term contributions to the Higgs mass \cite{Haber:1986gz,Drees:1987tp,Cvetic:1997ky,Ma:2011ea,Zhang:2008jm,Hirsch:2011hg,Krauss:2013jva}. The fine-tuning in these models is often better by a few orders compared to the MSSM. Alternatively, one can also consider models which give new loop-corrections due to the presence of additional large couplings to push the Higgs mass. This happens for instance in inverse-seesaw models \cite{Elsayed:2011de,Chun:2014tfa} or models with vector-like quarks \cite{Moroi:1991mg,Moroi:1992zk,Babu:2004xg,Babu:2008ge,Martin:2009bg,Martin:2010dc,Graham:2009gy,Martin:2010dc,Endo:2011mc,Martin:2012dg,Faroughy:2014oka,Ellis:2014dza,Lalak:2015xea} at the one-loop level, or in models with trilinear $R$-parity violation at the two-loop level \cite{Dreiner:2014lqa}. We are going to concentrate here on models with vectorlike tops partners.  In these models, the effects on the Higgs mass have been so far just studied in the effective potential approach at one-loop. Also a careful analysis of the threshold corrections to the standard model (SM) gauge and Yukawa couplings has been not performed to our knowledge so far. However, it is well known from the MSSM that the SUSY threshold corrections and one-loop momentum dependent effects can alter the Higgs mass by several GeV \cite{Pierce:1996zz}. Of course, also two-loop corrections involving coloured states are crucial in the MSSM and it wouldn't be possible to reach a mass of 125~GeV without them \cite{Heinemeyer:1998jw,Heinemeyer:1999be,Heinemeyer:1998np,Carena:1995wu,Carena:2000dp,Sasaki:1991qu,Carena:1995bx,Carena:2000yi,Carena:2001fw,Brignole:2001jy,Degrassi:2001yf,Brignole:2002bz,Dedes:2002dy,Dedes:2003km}. As soon as the Yukawa-like interactions of the new (s)tops become large, one should expect that effects of a similar size than in the MSSM sector appear. Therefore, we make a careful analysis of all three effects: we calculate the full one-loop threshold corrections to get an accurate prediction of the running gauge and Yukawa couplings at the SUSY scale, we include the entire dependence  of external momenta at the one-loop level, and we add the all two-loop corrections which are independent of electroweak gauge couplings. In this context, all calculations are performed within the \SARAH \cite{Staub:2008uz,Staub:2009bi,Staub:2010jh,Staub:2012pb,Staub:2013tta,Staub:2015kfa} -- \SPheno \cite{Porod:2003um,Porod:2011nf} framework which allows for two-loop calculations in SUSY models beyond the MSSM \cite{Goodsell:2014bna,Goodsell:2015ira}. The obtained precision is comparable to the standard calculations usually employed for the MSSM based on the results of Refs.~\cite{Brignole:2001jy,Degrassi:2001yf,Brignole:2002bz,Dedes:2002dy,Dedes:2003km}. 

Finally, we extend the particle content to have a complete {\bf 10} and {$\overline{\bf 10}$} of $SU(5)$ in addition to the MSSM particle content  to get a model which is consistent with gauge coupling unification. This model has already been studied to some extent after embedding it in minimal supergravity or GMSB \cite{Endo:2011mc,Endo:2011xq,Endo:2012rd}.  We choose here the variant where SUSY breaking is transmitted via gauge mediation and check for the first time for the fine-tuning in regions which are consistent with the Higgs measurements. We show that this gives usually a fine-tuning which can easily compete with other attempts to resurrect natural GMSB by including non-gauge interactions between the messenger particles and MSSM states \cite{Shadmi:2011hs,Evans:2011bea,Jelinski:2011xe,Evans:2012hg,Albaid:2012qk,Abdullah:2012tq,Perez:2012mj,Ding:2013pya,Delgado:2014vha,Basirnia:2015vga,Evans:2015swa}.

This manuscript is organized as follows. We first introduce the minimal SUSY model with vectorlike top partners as well as the UV complete variant embedded in GMSB in sec.~\ref{sec:model}. In sec.~\ref{sec:massspectrum} we summary briefly the main features of the tree-level masses before we explain in large detail the calculation of the one- and two-loop corrections. The numerical results are given in secs.~\ref{sec:results} and \ref{sec:resultsII}. In sec.~\ref{sec:results} we discuss the impact of the different corrections at one- and two-loop on the SM-like Higgs mass using a SUSY scale input, before we analyse in sec.~\ref{sec:resultsII} the fine-tuning of the GMSB embedding. We conclude in sec.~\ref{sec:conclusion}.

\section{The MSSM with vectorlike Tops}
\label{sec:model}
\subsection{The minimal model}
\begin{table}[hbt]
\begin{tabular}{|c|c|c|c|c|c|} 
\hline \hline 
SF & Spin 0 & Spin \(\frac{1}{2}\) & Generations & \((U(1)\otimes\, \text{SU}(2)\otimes\, \text{SU}(3))\) \\ 
\hline 
\(\hat{Q}\) & \(\tilde{q}\) & \(q\) & 3 & \((\frac{1}{6},{\bf 2},{\bf 3}) \) \\ 
\(\hat{L}\) & \(\tilde{l}\) & \(l\) & 3 & \((-\frac{1}{2},{\bf 2},{\bf 1}) \) \\ 
\(\hat{H}_d\) & \(H_d\) & \(\tilde{H}_d\) & 1 & \((-\frac{1}{2},{\bf 2},{\bf 1}) \) \\ 
\(\hat{H}_u\) & \(H_u\) & \(\tilde{H}_u\) & 1 & \((\frac{1}{2},{\bf 2},{\bf 1}) \) \\ 
\(\hat{D}\) & \(\tilde{d}_R^*\) & \(d_R^*\) & 3 & \((\frac{1}{3},{\bf 1},{\bf \overline{3}}) \) \\ 
\(\hat{U}\) & \(\tilde{u}_R^*\) & \(u_R^*\) & 3 & \((-\frac{2}{3},{\bf 1},{\bf \overline{3}}) \) \\ 
\(\hat{E}\) & \(\tilde{e}_R^*\) & \(e_R^*\) & 3 & \((1,{\bf 1},{\bf 1}) \) \\ 
\(\hat T'\) & \(\tilde{t'}^*\) & \({t'}^*\) & 1 & \((-\frac{2}{3},{\bf 1},{\bf \overline{3}}) \) \\ 
\(\hat{\bar{T}}'\) & \(\tilde{\bar{t}}'^*\) & \({\bar{t}}'^*\) & 1 & \((\frac{2}{3},{\bf 1},{\bf 3}) \) \\ 
\hline \hline
\end{tabular} 
\label{tab:particles}
\end{table}

We extend the particle content of the MSSM by a pair of right-handed vectorlike quark superfields $\hat T'$ and $\hat{\bar{T}}'$. The particle content of the model and the naming conventions for all chiral superfields and their spin-0 as well as $\frac12$ components are summarized in Tab.~\ref{tab:particles}. In addition, we have the usual vector superfields $\hat{B}$, $\hat{W}$, $\hat{G}$ which carry the gauge bosons for $U(1)_Y \times SU(2)_L \times SU(3)_C$ as well as the gauginos $\lambda_B$, $\lambda_W$, $\lambda_G$.
The full superpotential for the model reads:
\begin{align}
W = Y^{i j}_{e}\,\hat L_i \hat E_j \hat H_d 
   +Y^{i j}_{d}\, \hat Q_i \hat D_j \hat H_d 
   + 
   Y^{i j}_{u} \hat Q_i \hat U_j \hat H_u +
 \mu\, \hat H_u \hat H_d +  Y^{i}_{t'} \hat Q_i \hat T' \hat H_u + M_{T'} \hat T' \hat{\bar{T}}' + m^i_{t'} \hat U_i \hat{\bar{T}}'
\end{align}
Here, we skipped colour and isospin indices. The Yukawa couplings $Y_e$, $Y_d$ and $Y_u$ are in general complex $3\times 3$ matrices. The new interaction $Y_{t'}$ is a vector, but we concentrate only on cases where the third component $Y_{t'}^3$ has non-vanishing values. To simplify the notation, we define therefore 
\begin{equation}
Y^3_{t'} \equiv Y_{t'} 
\end{equation}
When we speak about the top-Yukawa coupling $Y_t$, we refer to $Y_u^{33}$. \\
The dimensionful parameters in the superpotential are the $\mu$-parameter known from the MSSM, as well as the mass term $M_{T'}$ for the vectorlike top quark superfields, and a bilinear term $m_{t'}$ mixing the new states and the MSSM ones even before electroweak symmetry breaking (EWSB). \\
The soft-SUSY breaking terms for the model are 
\begin{align}
 - \mathscr{L} =  &  \big(T^{ij}_e \tilde{l}_i \tilde{e}_j H_d + T^{ij}_d \tilde{q}_i \tilde{d}_j H_d+T^{ij}_u \tilde{q}_i \tilde{u}_j H_u + B_\mu H_d H_u  + T^i_{t} \tilde{q}_i \tilde{t}' H_u + B_T \tilde{t}' \tilde{\bar{t}}' + B^i_t \tilde{u}_i \tilde{\bar{t}}' + \text{h.c.} \big) \nonumber \\
 &  +  m^2_{ u,{ij}} \tilde{u}^*_i \tilde{u}_j + m^2_{ d,{ij}} \tilde{d}^*_i \tilde{d}_j +m^2_{ q,{ij}} \tilde{q}^*_i \tilde{q}_j + m^2_{ e,{ij}} \tilde{e}^*_i \tilde{e}_j  + m^2_{ l,{ij}} \tilde{l}^*_i \tilde{l}_j  + m_{H_d}^2 |H_d|^2 + m_{H_u}^2 |H_u|^2  \nonumber \\
 & + m^2_{\tilde t'} |\tilde{t}'|^2 + m_{\tilde{\bar{t}}'}^2 |\tilde{\bar{t}}'|^2 + (m^2_{u \tilde{t}'} \tilde u_i^* \tilde{t}' + h.c.) + \left(M_1 \lambda_B \lambda_B + M_2 \lambda_W \lambda_W + M_3 \lambda_G \lambda_G  + h.c. \right) 
\end{align}
In general, the $T$- and $B$- parameters are complex tensors of appropriate dimension, while the mass soft-terms for scalars are hermitian matrices, or vectors or scalars. The gaugino mass terms are complex scalar. However, we are going to neglect CP violation in the soft-sector, i.e. all parameters are taken to be real. For the trilinear soft-term of $Y_{t'}$ we use a similar short-hand notation $T^3_{t'} \equiv T_{t'}$ in the following. 

\subsection{UV completion and fine-tuning}
\subsubsection{Gauge coupling unification}
\begin{figure}[h]
  \centering
  \includegraphics[width=9cm]{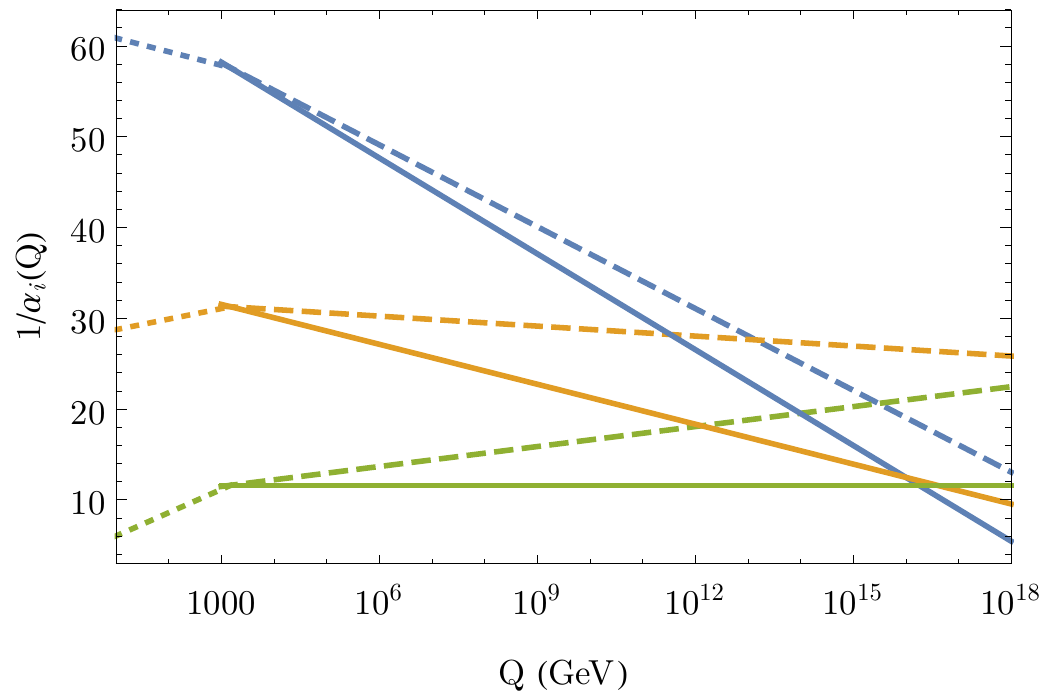}
  \caption{Running of the gauge couplings ($\alpha_i^{-1}(Q),i=1,2,3$, at 1-loop). The dashed lines belong to the minimal vectorlike top model and the full lines to the UV-completed model. The dotted lines represent the SM-only running up to $M_{\text{SUSY}}=1500$~GeV.}
  \label{fig:UVcompleteRunning}
\end{figure}

If we just include the right-handed top superfields, the model is not consistent with gauge coupling unification. To cure this problem, additional fields have to be added. The minimal choice is to add a pair of complete ${\bf 10}$-plets under $SU(5)$ which contain the states we are interested in, but also vectorlike left-handed quarks ($Q'$, $\bar Q'$) and vector-like right-handed leptons ($E'$, $\bar E'$). To generate mass terms for all components of the ${\bf 10}$ and ${\bf \overline {10}}$, the following extension of the superpotential is needed:
\begin{equation}
\Delta W =  M_{Q'} \hat Q' \hat{\bar{Q}}'  + M_{E'} \hat E' \hat{\bar{E}}'.
\end{equation}
Here, the $Q$-fields have quantum numbers \((\frac{1}{6},{\bf 2},{\bf 3}) \), \((-\frac{1}{6},{\bf 2},{\bf \bar 3}) \), while the vector-like leptons $\hat E'$, $\hat{\bar{E}}'$ carry quantum numbers \((\pm 1,{\bf 1},{\bf 1})\) with respect to $U(1)_Y \times SU(2)_L \times SU(3)_c$. We are going to assume that no further interactions between these additional states and the MSSM sector are present, i.e. these particles are only spectators when calculating the SUSY mass corrections. Nevertheless, because of their impact on the SUSY RGEs and also on the threshold corrections to the SM gauge couplings they can play an important role.  We can see this already at the one-loop RGEs of the gauge couplings for the minimal model and the UV complete version:
\begin{align} 
\beta_{g_1}^{(1)} & =  \left(\frac{41}{5} +  \frac{7}{5} \delta_{UV}\right) g_{1}^{3}  \\ 
\beta_{g_2}^{(1)} & =  \left(1 + 3 \delta_{UV} \right) g_{2}^{3} \\ 
\beta_{g_3}^{(1)} & =  \left(-2   +  2 \delta_{UV} \right) g_{3}^{3}, 
\end{align}
where we parametrized the $\beta$ function as 
\begin{equation}
\beta_{g_i} \equiv \frac{1}{16 \pi^2} \beta_{g_i}^{(1)}   + \frac{1}{(16 \pi^2)^2} \beta_{g_i}^{(2)} + \dots
\end{equation}
For $\delta_{UV}=0$ we obtain the minimal model, while $\delta_{UV}=1$ describes the UV complete version. In Fig.~\ref{fig:UVcompleteRunning} the re-established gauge unification can be observed.
 The one-loop $\beta$ functions of the Yukawa couplings are the same in both model variants and read
\begin{align} 
\beta_{Y_d}^{(1)} & =  Y_d \Big(3  Y_{d}^{\dagger}  Y_d  +  Y_{u}^{\dagger}  Y_u+ 3 \mbox{Tr}\Big({Y_d  Y_{d}^{\dagger}}\Big)  -\frac{16}{3} g_{3}^{2} -3 g_{2}^{2} -\frac{7}{15} g_{1}^{2}  + \mbox{Tr}\Big({Y_e  Y_{e}^{\dagger}}\Big)\Big) + Y_{t',{{i_2}}} \Big({Y_d  Y_{t'}^*}\Big)_{{i_1}} \\ 
\beta_{Y_e}^{(1)} & =  
3 {Y_e  Y_{e}^{\dagger}  Y_e}  + Y_e \Big(3 \mbox{Tr}\Big({Y_d  Y_{d}^{\dagger}}\Big) -3 g_{2}^{2} -\frac{9}{5} g_{1}^{2}  + \mbox{Tr}\Big({Y_e  Y_{e}^{\dagger}}\Big)\Big)\\ 
\beta_{Y_{t',{{i_1}}}}^{(1)} & =  
\Big(3 Y_{u}^{T}  Y_u^* + 3 \mbox{Tr}\Big(3 {Y_u  Y_{u}^{\dagger}}\Big) + Y_{d}^{T}  Y_d^*  + 6 \Big({Y_{t'}  Y_{t'}^*}\Big)  -\frac{13}{15} g_{1}^{2} -3 g_{2}^{2}   -\frac{16}{3} g_{3}^{2} \Big)Y_{t',{{i_1}}} \\ 
\beta_{Y_u}^{(1)} & =  
3 Y_{t',{{i_2}}} \Big({Y_u  Y_{t'}^*}\Big)_{{i_1}}   + Y_u \Big(3 Y_{u}^{\dagger}  Y_u + Y_{d}^{\dagger}  Y_d + 3 \mbox{Tr}\Big({Y_u  Y_{u}^{\dagger}}\Big)  + 3 \Big({Y_{t'}  Y_{t'}^*}\Big)  -\frac{13}{15} g_{1}^{2} -3 g_{2}^{2}   -\frac{16}{3} g_{3}^{2} \Big)
\end{align}

\begin{figure}[hbt]
\centering
\includegraphics[width=0.6\linewidth]{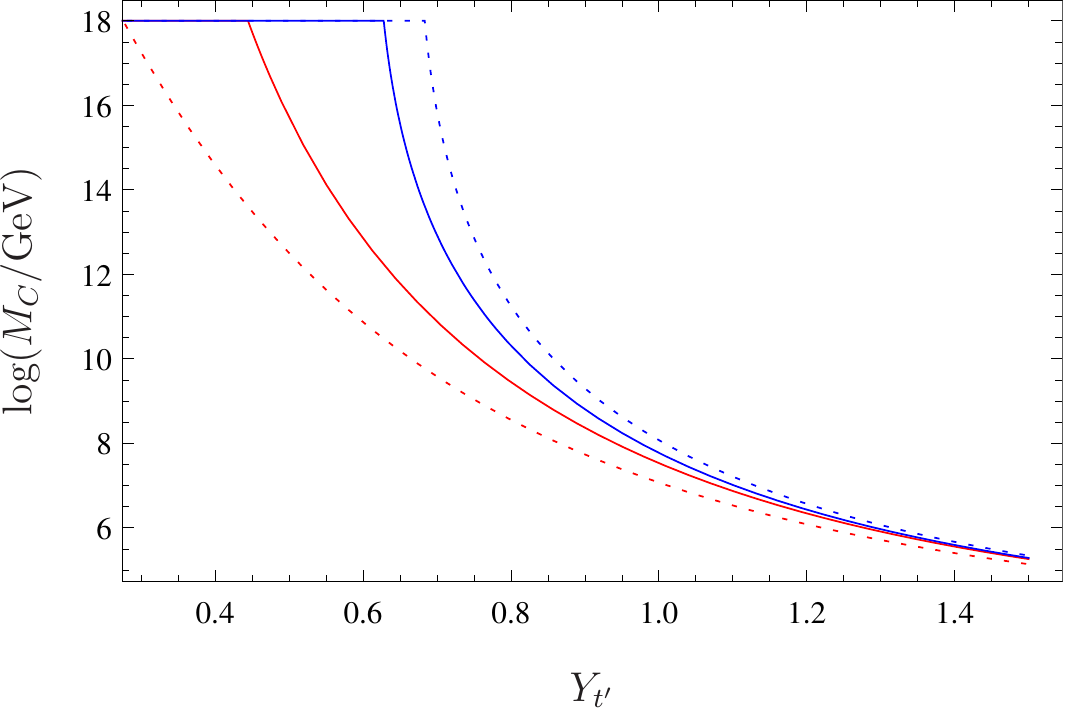}
\caption{This plot shows the scale $M_C$ at which the Landau pole arises as a function of $Y_{t'}$. The red lines are for the minimal model, the blue lines for the UV complete version. For the dotted lines we used $\tan\beta=10$, for the full ones $\tan\beta=60$. }
\label{fig:cutoff}
\end{figure}

We can use these RGEs to make a quick check for the cut off-scale of the theory in the limit of very large $Y_{t'}$. For this purpose, we fix at $M_{SUSY} = 1.5$~TeV the SM gauge couplings as $g_i = (0.47, 0.64, 1.05)$, and consider only third generation Yukawa couplings $Y^{33}_j=\sqrt{2}/246 \cdot (1.8/\cos\beta,2.4/\cos\beta,160/\sin\beta)$ with $j=e,d,u$. Of course, this is a very simplistic setup missing many details like two-loop effects in the running and threshold corrections. These effects will be included in our numerical analysis. Nevertheless, one can already see in Fig.~\ref{fig:cutoff} that the cut-off scale $M_C$ at which the Landau pole arises, given as a function of $Y_{t'}$, is pushed towards higher scales in the UV complete version. \\

The additional soft-terms which appear because of the extended particle content are the following:
\begin{equation}
 - \Delta \mathscr{L} =m^2_{\tilde e'} |\tilde{e}'|^2 + m_{\tilde{\bar{e}}'}^2 |\tilde{\bar{e}}'|^2  + m^2_{\tilde q'} |\tilde{q}'|^2 + m_{\tilde{\bar{q}}'}^2 |\tilde{\bar{q}}'|^2  + (m^2_{e \tilde{e}'} \tilde e_i^* \tilde{e}' + m^2_{q \tilde{q}'} \tilde q_i^* \tilde{q}' +  h.c.)
\end{equation}

We can now embed the UV complete version in a constrained setup to relate the SUSY breaking parameters.  We are going to choose the setup of gauge mediated SUSY breaking (GMSB) which we introduce now briefly.

\subsubsection{Gauge mediated SUSY breaking and boundary conditions}
The mediation of the SUSY breaking from the secluded to the visible sector happens in GMSB by messenger particles charged under SM gauge groups. The minimal model provides a pair of ${\bf 5}$-plets under $SU(5)$ which don't have any interaction with the MSSM sector but due to the gauge couplings. The necessary ingredients to break SUSY are the interaction of the messengers, called $\Phi, \bar \Phi$, and a spurion field $S$ described by 
\begin{equation}
\label{eq:mass_messenger}
W = \lambda S \Phi \bar{\Phi}  \thickspace . 
\end{equation}
$S$ is a gauge singlet and acquires a vacuum expectation value (VEV) along its scalar and auxiliary component due to hidden sector interactions, which we leave here unspecified 
\begin{equation}
\label{eq:vevS}
\langle S \rangle = M + \Theta^2 F \thickspace.
\end{equation}
The coupling $\lambda$ of eq.~(\ref{eq:mass_messenger}) can be absorbed into the redefinitions of $M \equiv \lambda M$ and $F\equiv \lambda F$. With these conventions, we find that the fermionic components of the messengers have a mass $M$, while the scalars get masses 
\begin{equation}
\label{eq:ScalarMessengerMass}
\tilde{\phi}_{+,-} = \frac{1}{\sqrt{2}} \left(\tilde{\phi}_M \pm \tilde{\bar{\phi}}_M\right) \thickspace, \hspace{1cm} m_{+,-} = \sqrt{M^2 \pm F} \thickspace.
\end{equation}
This gives the condition $M^2 > F$. The soft breaking masses of the MSSM fields are generated via loop diagrams involving the messenger particles. The gauginos receive masses $M_{\tilde{\lambda}}$ at one-loop level while the scalar masses $m^2_{\tilde{f}}$ are generated at the two-loop.
The leading approximations for the soft breaking masses are
\begin{equation}
\label{eq:SoftBreakingGMSB}
M_{\tilde{\lambda}_i}(t) =  \frac{\alpha_i(t)}{4 \pi} \Lambda_G \thickspace , \hspace{1cm}
m_{\tilde{f}_i}^2(t) = 2 \sum_{r=1}^3 C_r(\tilde{f}) \frac{\alpha(t)_r^2}{16 \pi^2}  \Lambda_S^2 
\end{equation}
\(\alpha_i(t) = g_i^2/(4\pi)\) are the running coupling constants at the scale $t$ and $C_r$ is the Casimir of the representation $r$. The SUSY soft breaking scales \(\Lambda_G\) and \(\Lambda_S\) depend  on  $F$ and $M$ as follows:
\begin{equation}
 \Lambda_G = \frac{F}{M} g\left(\frac{F}{M}^2\right) \thickspace , \hspace{1cm} \Lambda_S^2 = \frac{F^2}{M^2} f\left(\frac{F}{M^2}\right)
\end{equation}
with
\begin{equation}
 g(x) \simeq 1 + \frac{x^2}{6} + \frac{x^4}{15} + \frac{x^6}{28} + O(x^8) \thickspace, \hspace{0.5cm} f(x)\simeq 1 + \frac{x^2}{36} - \frac{11 x^4}{450} - \frac{319 x^6}{11760} + O(x^8) \thickspace. 
\end{equation}
It is convenient to define
\begin{equation}
 \Lambda \equiv \frac{F}{M}
\end{equation}
For $F \ll M^2$ this leads to $\Lambda_G = \Lambda_S = \Lambda$. 
Applying the general results to our (UV complete) model, we have the following boundary conditions at the messenger scale $M$ for the scalar soft masses
\begin{align}
m_{l,jj}^2 = m_{H_u}^2 = m_{H_d}^2 =& \left(\frac{3}{10} g_1^4 + \frac{3}{2} g_2^4\right) \Lambda_S^2 \\
m_{q,jj}^2 = m_{\tilde{q}'}^2 = m_{\tilde{\bar{q}}'} =  & \left(\frac{1}{30} g_1^4 + \frac{3}{2} g_2^4 + g_3^4 \right) \Lambda_S^2 \\
m_{u,jj}^2 = m_{\tilde{t}'}^2 = m_{\tilde{\bar{t}}'} =  & \left(\frac{8}{15} g_1^4 + \frac83 g_3^4 \right) \Lambda_S^2 \\
m_{e,jj}^2 = m_{\tilde{e}'}^2 = m_{\tilde{\bar{e}}'} =  & \frac{6}{5} g_1^4  \Lambda_S^2 \\
m_{d,jj}^2 = & \left(\frac{2}{15} g_1^4 + \frac83 g_3^4 \right) \Lambda_S^2
\end{align}
with $j=1,2,3$. All off-diagonal entries  are staying zero at the messenger scale. For the gaugino mass terms, we have the MSSM results 
\begin{align}
M_i = g_i^2 \Lambda_G 
\end{align}
while all other soft-terms vanish up to two-loop 
\begin{align}
T_x =&  0 \hspace{1cm} x = d,u,e,t'\\
B_X =&  0 \hspace{1cm} X = Q', E', T' \\
m^2_{u \tilde{t}'} = m^2_{q \tilde{q}'} = m^2_{e \tilde{e}'} = B_t =& 0 \\
\end{align}
Furthermore, we assume that the bilinear mass terms for the vector states unify at the messenger scale 
\begin{equation}
M_{T'} = M_{Q'} = M_{E'} \equiv M_{V'} 
\end{equation}
We make no attempt to explain the size of $\mu$ or $B_\mu$ in this setup. There are several proposals how these parameters receive numerical values needed for phenomenological reasons \cite{Dvali:1996cu,Dimopoulos:1997je,Giudice:1998bp}. We take it as given that one of these ideas is working and calculate the $\mu$ and $B_\mu$ from the vacuum conditions. Similarly, we are also agnostic concerning the cosmological gravitino problem usually introduced in GMSB by the Gravitino LSP and possible solutions for it \cite{Dimopoulos:1996gy,Han:1997wn,Baltz:2001rq,Fujii:2002fv,Fujii:2002yx,Jedamzik:2005ir,Staub:2009ww}.\\

Thus, our full set of input parameters in this setup is
\begin{eqnarray}
&M \,, \quad \Lambda \,, \quad \tan\beta \,, \quad M_{V'} \,,\quad Y_{t'} &
\end{eqnarray}

\subsubsection{Fine-tuning}
Fine-tuning addresses the question how to quantify if a model and a particular parameter point is natural or not. For this purpose, different measures are proposed to calculate the fine-tuning (FT). We are using the measure for the electroweak fine-tuning introduced in Refs.~\cite{Ellis:1986yg, Barbieri:1987fn} 
\begin{equation} 
\label{eq:measure}
\Delta_{FT} \equiv \max {\text{Abs}}\big[\Delta _{\alpha}\big],\qquad \Delta _{\alpha}\equiv \frac{\partial \ln
  M_Z^{2}}{\partial \ln \alpha} = \frac{\alpha}{M_Z^2}\frac{\partial M_Z^2}{\partial \alpha} \;.
\end{equation}
In this setup, the sensitivity of the $Z$ mass on the fundamental parameters at the UV scale is calculated. $\alpha$ is a set of independent parameters at this scale and $\Delta_\alpha^{-1}$ gives an estimate of the accuracy to which the parameter $\alpha$ must be tuned to get the correct electroweak breaking scale \cite{Ghilencea:2012qk}. The smaller $\Delta_{FT}$, the more natural is the model under consideration. We use the messenger scale $M$ in GMSB as a reference scale and calculate the FT with respect to 
\begin{equation}
\alpha = \{\Lambda,\ M_{V'},\ Y_{t'},\ Y_t,\ g_3,\ \mu,\ B_\mu \}.
\end{equation}
The practical calculation of the FT in our numerical calculation works as follows: we vary these parameters at the messenger scale $M$ and run the two-loop RGEs down to the SUSY scale. At the SUSY scale, the electroweak VEVs are calculated numerically using the minimization conditions of the potential and the resulting variation in the $Z$ mass is derived.

\section{The mass spectrum of the minimal model}
\label{sec:massspectrum}
To get a good estimate of the fine-tuning by including the Higgs constraint, it is necessary to reduce the theoretical uncertainty of the Higgs mass prediction. Our aim is to get the same uncertainty as for the MSSM, namely to consider the Higgs mass in the range 
\begin{equation}
m_h = ( 125 \pm 3 )~\text{GeV}
\end{equation}
This precision can only be reached if a full one-loop calculation is done, and the dominant two-loop corrections are included. Since this has not been done before in literature for the considered model, we discuss our calculation of the mass spectrum, in particular of the threshold corrections and two-loop Higgs corrections, in detail.  

\subsection{Tree-level properties}
When electroweak symmetry gets broken, the neutral Higgs states receive VEVs $v_d$ and $v_u$ and split in their CP even and odd components:
\begin{equation}
H_d^0 \to \frac{1}{\sqrt{2}}\left(\phi_d + i \sigma_d + v_d\right) \,,\quad  H_u^0 \to \frac{1}{\sqrt{2}}\left(\phi_u + i \sigma_u + v_u\right).
\end{equation}
We have $\tan\beta = \frac{v_u}{v_d}$ and $v=\sqrt{v_d^2+ v_u^2} \simeq 246$~GeV. Using these conventions, the tree-level mass matrix squared for the scalar Higgs particles is the same as in the MSSM. It reads in the basis $(\phi_{d}, \phi_{u})$
\begin{equation} 
m^{2,(T)}_{h} = \left( 
\begin{array}{cc}
\frac{1}{8} \Big(g_{1}^{2} + g_{2}^{2}\Big)\Big(3 v_{d}^{2}  - v_{u}^{2} \Big) + m_{H_d}^2 + |\mu|^2 &-\frac{1}{4} \Big(g_{1}^{2} + g_{2}^{2}\Big)v_d v_u  - {\Re\Big(B_{\mu}\Big)} \\ 
-\frac{1}{4} \Big(g_{1}^{2} + g_{2}^{2}\Big)v_d v_u  - {\Re\Big(B_{\mu}\Big)}  &-\frac{1}{8} \Big(g_{1}^{2} + g_{2}^{2}\Big)\Big(-3 v_{u}^{2}  + v_{d}^{2}\Big) + m_{H_u}^2 + |\mu|^2\end{array} 
\right) 
 \end{equation} 
This matrix is diagonalized by \(Z^H\): 
\begin{equation} 
Z^H m^2_{h} Z^{H,\dagger} = m^{dia}_{2,h} 
\end{equation} 
Two of the parameters in this matrix can be eliminated by the tadpole conditions for EWSB:
\begin{align} 
\label{eq:tad1}
T_d \equiv \frac{\partial V}{\partial \phi_{d}} &= -\frac{1}{2} v_u \Big(B_{\mu} + B_{\mu}^*\Big) + \frac{1}{8} \Big(g_{1}^{2} + g_{2}^{2}\Big)v_d \Big(- v^2_u  + v^2_d\Big) + v_d \Big(m_{H_d}^2 + |\mu|^2\Big)=0\\
\label{eq:tad2}
T_u \equiv  \frac{\partial V}{\partial \phi_{u}} &= \frac{1}{8} \Big(g_{1}^{2} + g_{2}^{2}\Big)v_u \Big(- v_{d}^{2}  + v_{u}^{2}\Big) - v_d {\Re\Big(B_{\mu}\Big)}  + v_u \Big(m_{H_u}^2 + |\mu|^2\Big)=0
\end{align} 
We are going to solve these equations for the squared soft-masses $m_{H_d}^2$ and $m_{H_u}^2$ when we consider a SUSY scale input. That leaves three free parameters in the Higgs sector at tree-level: $\tan\beta$, $\mu$ and $B_\mu$. The last one is related to the tree-level mass squared $M^2_A$ of the physical pseudo-scalar via
\begin{equation}
B_\mu = \frac{1}{\tan\beta + 1/\tan\beta} M_A^2
\end{equation}
However, when we consider the UV completion, $m_{H_d}^2$ and $m_{H_u}^2$ are fixed at the SUSY scale and we are going to solve the above equations (\ref{eq:tad1}) and (\ref{eq:tad2}) for $\mu$ and $B_\mu$. \\

Also, the mass matrices for the CP-odd and charged Higgs bosons, for down (s)quarks, charged and neutral (s)leptons, as well as for neutralino and charginos are identical to the MSSM. Only in the up (s)quark sector things change because of the additional top-like states. The scalar mass matrix that links the left- and right-handed MSSM up-squarks and the new vector-like states is given in the basis of $\left(\tilde{u}_{L,i}, \tilde{u}_{R,i}, \tilde{t'}, \tilde{\bar{t}}'^*\right)$ by
\begin{equation} 
m^2_{\tilde{u}} = \left( 
\begin{array}{cccc}
m_{\tilde{u}_L\tilde{u}_L^*} & \cdot & \cdot & \cdot \\ 
\frac{1}{\sqrt{2}} \Big( v_u T_u - v_d Y_u \mu^* \Big) &m_{\tilde{u}_R\tilde{u}_R^*} & \cdot & \cdot \\ 
 \frac{1}{\sqrt{2}}  \Big( v_u T_{t'}- v_d \mu^* Y_{t'} \Big) & \frac{1}{2} \Big(2 \Big(M_{T'} m^*_{{t'}}  + m_{\tilde u \tilde t'}^{2}\Big) + v_{u}^{2} Y^*_{u} Y_{t'}  \Big) &m_{\tilde{t'}\tilde{t'}^*} & \cdot  \\ 
 \frac{1}{\sqrt{2}} v_u \Big(M_{T'}^* Y_{t'}  + Y^T_{u} m^*_{{t'}}  \Big) &B^*_{{t'}}   &B_{T'}^*   &m_{\tilde{\bar{t}}'^*\tilde{\bar{t}}'}\end{array} 
\right).
 \end{equation} 
with the diagonal entries
\begin{align} 
m_{\tilde{u}_L\tilde{u}_L^*} &= -\frac{1}{24} \Big(-3 g_{2}^{2}  + g_{1}^{2}\Big){\bf 1} \Big(- v_{u}^{2}  + v_{d}^{2}\Big)  + \frac{1}{2}  \Big(2 m_q^2  + v_{u}^{2} \Big(Y^*_{{t'}} Y_{t'}  + {Y_{u}^{\dagger}  Y_u}\Big)\Big)\\ 
% m_{\tilde{u}_L\tilde{u}_R^*} &= \frac{1}{\sqrt{2}} \Big(- v_d Y_u \mu^*  + v_u T_u \Big)\\ 
m_{\tilde{u}_R\tilde{u}_R^*} &= \frac{1}{2} \Big(2 \Big(m^*_{{t'}} m_{t'}  + m_u^2\Big) + v_{u}^{2} {Y_u  Y_{u}^{\dagger}} \Big) + \frac{1}{6} g_{1}^{2} {\bf 1} \Big(- v_{u}^{2}  + v_{d}^{2}\Big)\\ 
% m_{\tilde{u}_L\tilde{t'}^*} &= \frac{1}{\sqrt{2}}  \Big(- v_d \mu^* Y_{t'}  + v_u T_{t'} \Big)\\ 
% m_{\tilde{u}_R\tilde{t'}^*} &= \frac{1}{2} \Big(2 \Big(M_{T'} m^*_{{t'}}  + m_{\tilde u \tilde t'}^{2}\Big) + v_{u}^{2} Y^*_{u} Y_{t'}  \Big)\\ 
m_{\tilde{t'}\tilde{t'}^*} &= \frac{1}{2}  \Big(2 \Big(m_{\tilde{t}'}^2 + |M_{T'}|^2\Big) + v_{u}^{2} |Y_{t'}|^2 \Big) + \frac{1}{6} g_{1}^{2} \Big(- v_{u}^{2}  + v_{d}^{2}\Big)\\ 
% m_{\tilde{u}_L\tilde{\bar{t}}'} &= \frac{1}{\sqrt{2}} v_u \Big(M_{T'}^* Y_{t'}  + Y^T_{u} m^*_{{t'}}  \Big)\\ 
m_{\tilde{\bar{t}}'^*\tilde{\bar{t}}'} &=  \Big(m_{\tilde{\bar{t}}'}^2 + |M_{T'}|^2 + |m_{t'}|^2\Big) + \frac{1}{6} g_{1}^{2} \Big(- v_{d}^{2}  + v_{u}^{2}\Big)
\end{align} 
This matrix is diagonalized by \(Z^U\): 
\begin{equation} 
Z^U m^2_{\tilde{u}} Z^{U,\dagger} = m^{dia}_{2,\tilde{u}} 
\end{equation}
and we have eight mass eigenstates called $\tilde{u}_i$ in the following. Similarly, in the fermionic counterpart  we choose the basis $\left(u_{L,i}, {\bar{t}}'^*\right)/\left(u^*_{R,i}, {t'}^*_{{{\beta_2}}}\right) $. The mass matrix in this basis reads
\begin{equation} 
m_{u} = \left( 
\begin{array}{cc}
\frac{1}{\sqrt{2}} v_u  Y_{u}^{T}  &\frac{1}{\sqrt{2}} v_u  Y_{t'} \\ 
 m_{t'}  &M_{T'}  \end{array} 
\right).
 \end{equation} 
Here, we need two rotation matrices \(U^u_L\) and \(U^u_R\) to diagonalize this matrix,
\begin{equation} 
U^{u,*}_L m_{u} U_{R}^{u,\dagger} = m^{\text{dia}}_{u} .
\end{equation} 
The four generations of mass eigenstates are called $u_i$ where the first three generations correspond to the up, charm and top quark.

\subsection{Calculation of the Higgs masses at one- and two-loop}
In this section we give details about the calculation of the Higgs masses at the one- and two-loop level. We have performed all calculations with the combination of the software packages \SARAH and \SPheno which automatize all relevant steps. There are three changes compared to the calculation of the Higgs masses in the MSSM:
\begin{enumerate}
 \item The new vectorlike states change the threshold corrections at $M_Z$ to derive the gauge and Yukawa couplings in \DRbar scheme from the measured SM couplings and fermion masses. \SARAH and \SPheno applies and generalizes the procedure of Ref.~\cite{Pierce:1996zz} to make this matching. 
 We give more details about the main differences compared to the MSSM in sec.~\ref{sec:thresholds}.
 \item At the one-loop level new contributions of $O(\alpha_{t'})$ arise. These corrections are widely discussed in literature and are known to be able to give a push of many GeV to the Higgs mass. While these corrections so far have just been calculated in the effective potential approach, \SARAH and \SPheno perform the full one-loop corrections in a diagrammatic way including the dependence of the external momenta. This calculation is again a generalization of the renormalization procedure presented in Ref.~\cite{Pierce:1996zz}. We explain this calculation and the difference to the MSSM more detailed in sec.~\ref{sec:oneloop}.
 \item At the two-loop level, new corrections $O(\alpha_{t'} (\alpha_S + \alpha_t + \alpha_b + \alpha_{t'}))$ arise. The importance of these corrections was unknown up to now. However, with the generic results of Ref.~\cite{Martin:2001vx} for the two-loop effective potential implemented into \SARAH \cite{Goodsell:2014bna}, a numerical derivation in analogy to Ref.~\cite{Martin:2002wn} allows to obtain the two-loop self-energies at vanishing external momentum for the scalars which get a VEV. Moreover, since Ref.~\cite{Goodsell:2015ira}, a fully equivalent and diagrammatic calculation in the limit $p^2 =0$ can also be performed by \SARAH and \SPheno. Both approaches are used to cross-check the two-loop results. We give more details about this calculation in sec.~\ref{sec:twoloop}.
\end{enumerate}

\subsubsection{Threshold corrections}
\label{sec:thresholds}
The presence of additional vectorlike states change the relations between the running \DRbar parameters and the measured SM parameters. In the gauge sector, the relation between the SM couplings ($\overline{\text{MS}}$ scheme with five flavours) and the \DRbar ones are
\begin{align}
  \alpha^{\DRbar}(M_Z) &= \frac{\alpha^{(5),\overline{\text{MS}}}(M_Z)}{1 - \Delta\alpha(M_Z)} ,\\
  \alpha_S^{\DRbar}(M_Z) &= \frac{\alpha_S^{(5),\overline{\text{MS}}}(M_Z)}{1 - \Delta\alpha_S(M_Z)} 
\end{align}
Here, $\alpha_S^{(5),\overline{\text{MS}}}$ and $\alpha^{(5),\overline{\text{MS}}}$ are taken as input and receive corrections from the top loops as well as from new physics. For the minimal model, the thresholds read
\begin{eqnarray}
\Delta\alpha(\mu) =& \frac{\alpha}{2\pi} \left(\frac{1}{3}- \frac{16}{9} \sum_{i=3}^4  \log{\frac{m_{u_i}}{\mu}} - \frac{4}{9}  \sum_{i=1}^8 \log{\frac{m_{\tilde u_i}}{\mu}} + \Delta\alpha^{\text{MSSM}}(\mu) \right) \\
\Delta\alpha_S(\mu) =& \frac{\alpha_S}{2\pi} \left( - \frac{2}{3} \sum_{i=3}^4  \log{\frac{m_{u_i}}{\mu}} - \frac{1}{6} \sum_{i=1}^8  \log{\frac{m_{\tilde u_i}}{\mu}} + \Delta\alpha_S^{\text{MSSM}}(\mu)  \right)
\end{eqnarray}
We absorbed all corrections which don't change with respect to the MSSM in $\Delta\alpha_S^{\text{MSSM}}(\mu)$ and $\Delta\alpha^{\text{MSSM}}(\mu)$. Note, this does not include the up-squark sector, now consisting of 8 squarks, to prevent double counting. In the case of the UV complete model, additional terms of the same form show up. \\
To relate $\alpha$ to the running couplings $g_1$ and $g_2$, the running Weinberg angle $\sin\Theta$ and the electroweak VEV in \DRbar scheme are needed. Also here the vector-like tops enter because of the new loop corrections to the mass shifts $\delta M_Z^2$ and $\delta M_W^2$ of the gauge bosons. The corrections from the extended (s)top sector to the transversal self-energies are 
\begin{align} 
\Delta \Pi^{T,Z}(p^2) &= 
 +3 \sum_{a=1}^{8}{A_0\Big(m^2_{\tilde{u}_{{a}}}\Big)} {\Gamma_{Z,Z,\tilde{u}^*_{{a}},\tilde{u}_{{a}}}}  -12 \sum_{a=1}^{8}\sum_{b=1}^{8}|{\Gamma_{Z,\tilde{u}^*_{{a}},\tilde{u}_{{b}}}}|^2 {B_{00}\Big(p^{2},m^2_{\tilde{u}_{{a}}},m^2_{\tilde{u}_{{b}}}\Big)}  \nonumber \\
 & +3 \sum_{a=1}^{4}\sum_{b=1}^{4} \Big[\Big(|{\Gamma^L_{Z,\bar{u}_{{a}},u_{{b}}}}|^2 + |{\Gamma^R_{Z,\bar{u}_{{a}},u_{{b}}}}|^2\Big){H_0\Big(p^{2},m^2_{u_{{a}}},m^2_{u_{{b}}}\Big)} \nonumber \\ & +4 {B_0\Big(p^{2},m^2_{u_{{a}}},m^2_{u_{{b}}}\Big)} m_{u_{{a}}} m_{u_{{b}}} {\Re\Big({\Gamma^{L*}_{Z,\bar{u}_{{a}},u_{{b}}}} {\Gamma^R_{Z,\bar{u}_{{a}},u_{{b}}}} \Big)} \Big] \\
\Delta  \Pi^{W,T}(p^2) &= -12 \sum_{a=1}^{8}\sum_{b=1}^{6}|{\Gamma_{W^+,\tilde{u}^*_{{a}},\tilde{d}_{{b}}}}|^2 {B_{00}\Big(p^{2},m^2_{\tilde{d}_{{b}}},m^2_{\tilde{u}_{{a}}}\Big)}    + 3 \sum_{a=1}^{8}{A_0\Big(m^2_{\tilde{u}_{{a}}}\Big)} {\Gamma_{W^-,W^+,\tilde{u}^*_{{a}},\tilde{u}_{{a}}}}  \nonumber \\
& + 3 \sum_{a=1}^{4}\sum_{b=1}^{3} \Big[\Big(|{\Gamma^L_{W^+,\bar{u}_{{a}},d_{{b}}}}|^2 + |{\Gamma^R_{W^+,\bar{u}_{{a}},d_{{b}}}}|^2\Big){H_0\Big(p^{2},m^2_{u_{{a}}},m^2_{d_{{b}}}\Big)} \nonumber \\ & +4 {B_0\Big(p^{2},m^2_{u_{{a}}},m^2_{d_{{b}}}\Big)} m_{d_{{b}}} m_{u_{{a}}} {\Re\Big({\Gamma^{L*}_{W^+,\bar{u}_{{a}},d_{{b}}}} {\Gamma^R_{W^+,\bar{u}_{{a}},d_{{b}}}} \Big)} \Big]
\end{align} 
with 
\begin{align}
H_0 (p,m_1,m_2) =& 4B_{22}(p,m_1,m_2) + G_0(p,m_1,m_2)\ ,\\
G_0(p,m_1,m_2) =&
(p^2-m_1^2-m_2^2)B_0(p,m_1,m_2)-A_0(m_1)-A_0(m_2)\ ,\\
B_{22}(p, m_1,m_2) =& \frac{1}{6}\ \Bigg\{\,
\frac{1}{2}\biggl(A_0(m_1)+A_0(m_2)\biggr)
+\left(m_1^2+m_2^2-\frac{1}{2}p^2\right)B_0(p,m_1,m_2)\nonumber \\ &+
\frac{m_2^2-m_1^2}{2p^2}\ \biggl[\,A_0(m_2)-A_0(m_1)-(m_2^2-m_1^2)
B_0(p,m_1,m_2)\,\biggr] \nonumber \\
& +  m_1^2 + m_2^2
-\frac{1}{3}p^2\,\Bigg\}.
\end{align}

The appearing vertices are given in appendix~\ref{app:VBvertices}. All other contributions are identical to the MSSM and given for instance in Ref.~\cite{Pierce:1996zz}.  With that information, $v$ and $\sin^2\Theta^{\DRbar}_W$ are calculated by
\begin{align}
v^2 =& (M_Z^2 + \delta M_Z^2) \frac{(1- \sin^2\Theta^{\DRbar}_W)\sin^2\Theta^\DRbar_W}{\pi \alpha^{\DRbar}} \\
\sin^2\Theta^{\DRbar}_W =&  \frac{1}{2} - \sqrt{\frac{1}{4} - \frac{\pi \alpha^{\DRbar}}{\sqrt{2} M_Z^2 G_F (1-\delta_r)}}
\end{align}
Here, $G_F$ is the Fermi constant and $\delta_r$ doesn't receive new corrections compared to the MSSM (Expressions for $\delta_r$ can be found in \cite{Chankowski:1993eu}). Also here the spectator fields in the UV complete version will show up in a similar way because their contributions don't vanish even in the limit that all superpotential and soft-breaking interactions of those are assumed to vanish. 

The running Yukawa couplings are also calculated in an iterative way. We concentrate on the quark sector, because the leptons don't get new contributions from the new vector-like quarks at one-loop. This is also true for the UV complete model because these contributions are proportional to the superpotential interactions which we assume to vanish for the $E'$ and $Q'$ states.  The starting point are the running fermion masses in \DRbar\ obtained from the pole masses given as input:
\begin{align}
\label{eq:drbardown} m_{d,s,b}^{\DRbar,\text{SM}} =& m_{d,s,b} \left(1- \frac{\alpha^\DRbar_S}{3\pi}-\frac{23 \alpha_S^{\DRbar,2}}{72 \pi^2} + \frac{3}{128 \pi^2} g^{\DRbar,2}_2 - \frac{13}{1152 \pi^2} g^{\DRbar,2}_1\right)\\
\label{eq:drbarup} m_{u,c}^{\DRbar,\text{SM}} =& m_{u,c} \left(1- \frac{\alpha^\DRbar_S}{3\pi}-\frac{23 \alpha^{\DRbar,2}_S}{72 \pi^2} + \frac{3}{128 \pi^2} g^{\DRbar,2}_2 - \frac{7}{1152 \pi^2} g^{\DRbar,2}_1\right)\\
\label{eq:drbartop} m^{\DRbar,\text{SM}}_t =& m_t \left[1 + \frac{1}{16\pi^2} \left(\Delta m_t^{(1),qcd} +\Delta m_t^{(2),qcd} + \Delta m_t^{(1),ew}\right)\right]
\end{align}
with 
\begin{align}
 \Delta m_t^{(1),qcd} &= -\frac{16 \pi \alpha_S^\DRbar }{3} \left(5 + 3 \log\frac{M_Z^2}{m_t^2} \right) \\
 \Delta m_t^{(2),qcd} &=  -\frac{64 \pi^2 \alpha_S^{\DRbar,2} }{3} \left(\frac{1}{24}+\frac{2011}{384\pi^2}+\frac{\ln2}{12}-\frac{\zeta(3)}{8\pi^2}+\frac{123}{32\pi^2} \log\frac{M_Z^2}{m_t^2}+\frac{33}{32\pi^2} \left(\log\frac{M_Z^2}{m_t^2}\right)^2 \right)\\
 \Delta m_t^{(1),ew} &= - \frac{4}{9} g_2^{\DRbar,2} \sin^2 \Theta^\DRbar_W \left(5 + 3 \log\frac{M_Z^2}{m_t^2}\right)
\end{align}
The two-loop parts are taken from Ref.~\cite{Avdeev:1997sz,Bednyakov:2002sf}. 
The $\DRbar$ masses are matched to the eigenvalues of the loop-corrected fermion mass matrices calculated as
\begin{equation}
\label{eq:oneloopMF}
m_f^{(1L)}(p^2_i) =  m_f^{(T)} - \tilde{\Sigma}_S(p^2_i)
 - \tilde{\Sigma}_R(p^2_i) m_f^{(T)} - m_f^{(T)} \tilde{\Sigma}_L(p^2_i) 
\end{equation}
{\allowdisplaybreaks
Here, the pure QCD and QED corrections are dropped in the self-energies $\tilde \Sigma$ because they are already absorbed in the running \DRbar\ masses. The self-energy contributions from the extended (s)top sector to down-quarks are
\begin{align}
\Sigma^{d,S}_{i,j}(p^2) &= \sum_{a=1}^{2}\sum_{b=1}^{4}{B_0\Big(p^{2},m^2_{u_{{b}}},m^2_{H^-_{{a}}}\Big)} {\Gamma^{L*}_{\bar{\check{d_j}},H^-_{{a}},u_{{b}}}} m_{u_{{b}}} {\Gamma^R_{\bar{\check{d_{{i}}}},H^-_{{a}},u_{{b}}}} \nn \\
&+\sum_{a=1}^{8}\sum_{b=1}^{2}{B_0\Big(p^{2},m^2_{\tilde{\chi}^-_{{b}}},m^2_{\tilde{u}_{{a}}}\Big)} {\Gamma^{L*}_{\bar{\check{d_{{j}}}},\tilde{u}_{{a}},\tilde{\chi}^-_{{b}}}} m_{\tilde{\chi}^-_{{b}}} {\Gamma^R_{\bar{\check{d_{{i}}}},\tilde{u}_{{a}},\tilde{\chi}^-_{{b}}}} \nonumber \\ 
 &-4 \sum_{b=1}^{4}\Big(  + {B_0\Big(p^{2},m^2_{u_{{b}}},m^2_{W^-}\Big)}\Big){\Gamma^{R*}_{\bar{\check{d_{{j}}}},W^-,u_{{b}}}} m_{u_{{b}}} {\Gamma^L_{\bar{\check{d_{{i}}}},W^-,u_{{b}}}}  \\ 
\Sigma^{d,R}_{i,j}(p^2) &= -\frac{1}{2} \sum_{a=1}^{2}\sum_{b=1}^{4}{B_1\Big(p^{2},m^2_{u_{{b}}},m^2_{H^-_{{a}}}\Big)} {\Gamma^{R*}_{\bar{\check{d_{{j}}}},H^-_{{a}},u_{{b}}}} {\Gamma^R_{\bar{\check{d_{{i}}}},H^-_{{a}},u_{{b}}}}  \nn \\
&-\frac{1}{2} \sum_{a=1}^{8}\sum_{b=1}^{2}{B_1\Big(p^{2},m^2_{\tilde{\chi}^-_{{b}}},m^2_{\tilde{u}_{{a}}}\Big)} {\Gamma^{R*}_{\bar{\check{d_{{j}}}},\tilde{u}_{{a}},\tilde{\chi}^-_{{b}}}} {\Gamma^R_{\bar{\check{d_{{i}}}},\tilde{u}_{{a}},\tilde{\chi}^-_{{b}}}}  \nonumber \\ 
&- \sum_{b=1}^{4}{B_1\Big(p^{2},m^2_{u_{{b}}},m^2_{W^-}\Big)} {\Gamma^{L*}_{\bar{\check{d_{{j}}}},W^-,u_{{b}}}} {\Gamma^L_{\bar{\check{d_{{i}}}},W^-,u_{{b}}}}  \\ 
\Sigma^{d,L}_{i,j}(p^2) &= \Sigma^{d,R}_{i,j}(p^2)\Big|_{(L \leftrightarrow R)}  
\end{align}
The full self-energies in the up-quark sector read now
\begin{align} 
\label{eq:sigmauS}\Sigma^{u,S}_{i,j}(p^2) &={B_0\Big(p^{2},m^2_{d_{{b}}},m^2_{H^-_{{a}}}\Big)} {\Gamma^{L*}_{\bar{\check{u_{{j}}}},H^+_{{a}},d_{{b}}}} m_{d_{{b}}} {\Gamma^R_{\bar{\check{u_{{i}}}},H^+_{{a}},d_{{b}}}}+{B_0\Big(p^{2},m^2_{u_{{b}}},m^2_{h_{{a}}}\Big)} {\Gamma^{L*}_{\bar{\check{u_{{j}}}},h_{{a}},u_{{b}}}} m_{u_{{b}}} {\Gamma^R_{\bar{\check{u_{{i}}}},h_{{a}},u_{{b}}}} \nonumber \\ 
 &+ m_{\tilde{\chi}^-_{{a}}} {B_0\Big(p^{2},m^2_{\tilde{\chi}^-_{{a}}},m^2_{\tilde{d}_{{b}}}\Big)} {\Gamma^{L*}_{\bar{\check{u}}_{{j}},\bar{\tilde{\chi}}^-_{{a}},\tilde{d}_{{b}}}} {\Gamma^R_{\bar{\check{u}}_{{i}},\bar{\tilde{\chi}}^-_{{a}},\tilde{d}_{{b}}}}+m_{u_{{a}}} {B_0\Big(p^{2},m^2_{u_{{a}}},m^2_{A^0_{{b}}}\Big)} {\Gamma^{L*}_{\bar{\check{u}}_{{j}},u_{{a}},A^0_{{b}}}} {\Gamma^R_{\bar{\check{u}}_{{i}},u_{{a}},A^0_{{b}}}}  \nonumber \\ 
 &+{B_0\Big(p^{2},m^2_{\tilde{\chi}^0_{{b}}},m^2_{\tilde{u}_{{a}}}\Big)} {\Gamma^{L*}_{\bar{\check{u_{{j}}}},\tilde{u}_{{a}},\tilde{\chi}^0_{{b}}}} m_{\tilde{\chi}^0_{{b}}} {\Gamma^R_{\bar{\check{u_{{i}}}},\tilde{u}_{{a}},\tilde{\chi}^0_{{b}}}}+\frac{4}{3} m_{\tilde{g}}{B_0\Big(p^{2},m^2_{\tilde{g}},m^2_{\tilde{u}_{{a}}}\Big)} {\Gamma^{L*}_{\bar{\check{u_{{j}}}},\tilde{u}_{{a}},\tilde{g}_{{1}}}} {\Gamma^R_{\bar{\check{u_{{i}}}},\tilde{u}_{{a}},\tilde{g}_{{1}}}}  \nonumber \\ 
 &-4 \Big( {B_0\Big(p^{2},m^2_{d_{{b}}},m^2_{W^-}\Big)}\Big){\Gamma^{R*}_{\bar{\check{u_{{j}}}},W^+,d_{{b}}}} m_{d_{{b}}} {\Gamma^L_{\bar{\check{u_{{i}}}},W^+,d_{{b}}}}-\frac{16}{3}  {B_0\Big(p^{2},m^2_{u_{{b}}},0\Big)}{\Gamma^{R*}_{\bar{\check{u_{{j}}}},g,u_{{b}}}} m_{u_{{b}}} {\Gamma^L_{\bar{\check{u_{{i}}}},g,u_{{b}}}}    \nonumber \\ 
 & -4  {B_0\Big(p^{2},m^2_{u_{{b}}},0\Big)}{\Gamma^{R*}_{\bar{\check{u_{{j}}}},\gamma,u_{{b}}}} m_{u_{{b}}} {\Gamma^L_{\bar{\check{u_{{i}}}},\gamma,u_{{b}}}}-4 {B_0\Big(p^{2},m^2_{u_{{b}}},m^2_{Z}\Big)}{\Gamma^{R*}_{\bar{\check{u_{{j}}}},Z,u_{{b}}}} m_{u_{{b}}} {\Gamma^L_{\bar{\check{u_{{i}}}},Z,u_{{b}}}} \\
\label{eq:sigmauR}\Sigma^{u,R}_{i,j}(p^2) &= -\frac{1}{2} {B_1\Big(p^{2},m^2_{d_{{b}}},m^2_{H^-_{{a}}}\Big)} {\Gamma^{R*}_{\bar{\check{u_{{j}}}},H^+_{{a}},d_{{b}}}} {\Gamma^R_{\bar{\check{u_{{i}}}},H^+_{{a}},d_{{b}}}} 
-\frac{1}{2} {B_1\Big(p^{2},m^2_{u_{{b}}},m^2_{h_{{a}}}\Big)} {\Gamma^{R*}_{\bar{\check{u_{{j}}}},h_{{a}},u_{{b}}}} {\Gamma^R_{\bar{\check{u_{{i}}}},h_{{a}},u_{{b}}}}  \nonumber \\ 
 &-\frac{1}{2} {B_1\Big(p^{2},m^2_{\tilde{\chi}^-_{{a}}},m^2_{\tilde{d}_{{b}}}\Big)} {\Gamma^{R*}_{\bar{\check{u_{{j}}}},\bar{\tilde{\chi}^-}_{{a}},\tilde{d}_{{b}}}} {\Gamma^R_{\bar{\check{u_{{i}}}},\bar{\tilde{\chi}^-}_{{a}},\tilde{d}_{{b}}}} 
-\frac{1}{2} {B_1\Big(p^{2},m^2_{u_{{a}}},m^2_{A^0_{{b}}}\Big)} {\Gamma^{R*}_{\bar{\check{u_{{j}}}},u_{{a}},A^0_{{b}}}} {\Gamma^R_{\bar{\check{u_{{i}}}},u_{{a}},A^0_{{b}}}}  \nonumber \\ 
 &-\frac{1}{2} {B_1\Big(p^{2},m^2_{\tilde{\chi}^0_{{b}}},m^2_{\tilde{u}_{{a}}}\Big)} {\Gamma^{R*}_{\bar{\check{u_{{j}}}},\tilde{u}_{{a}},\tilde{\chi}^0_{{b}}}} {\Gamma^R_{\bar{\check{u_{{i}}}},\tilde{u}_{{a}},\tilde{\chi}^0_{{b}}}}  -\frac{2}{3} {B_1\Big(p^{2},m^2_{\tilde{g}},m^2_{\tilde{u}_{{a}}}\Big)} {\Gamma^{R*}_{\bar{\check{u_{{j}}}},\tilde{u}_{{a}},\tilde{g}_{{1}}}} {\Gamma^R_{\bar{\check{u_{{i}}}},\tilde{u}_{{a}},\tilde{g}_{{1}}}}  \nonumber \\ 
 & - {B_1\Big(p^{2},m^2_{d_{{b}}},m^2_{W^-}\Big)} {\Gamma^{L*}_{\bar{\check{u_{{j}}}},W^+,d_{{b}}}} {\Gamma^L_{\bar{\check{u_{{i}}}},W^+,d_{{b}}}}  -\frac{4}{3} {B_1\Big(p^{2},m^2_{u_{{b}}},0\Big)} {\Gamma^{L*}_{\bar{\check{u_{{j}}}},g,u_{{b}}}} {\Gamma^L_{\bar{\check{u_{{i}}}},g,u_{{b}}}}  \nonumber \\ 
 &- {B_1\Big(p^{2},m^2_{u_{{b}}},0\Big)} {\Gamma^{L*}_{\bar{\check{u_{{j}}}},\gamma,u_{{b}}}} {\Gamma^L_{\bar{\check{u_{{i}}}},\gamma,u_{{b}}}}  - {B_1\Big(p^{2},m^2_{u_{{b}}},m^2_{Z}\Big)} {\Gamma^{L*}_{\bar{\check{u_{{j}}}},Z,u_{{b}}}} {\Gamma^L_{\bar{\check{u_i}},Z,u_{{b}}}}  \\ 
\label{eq:sigmauL}\Sigma^{u,L}_{i,j}(p^2) &= \Sigma^{u,R}_{i,j}(p^2)\Big|_{(L \leftrightarrow R)}
\end{align} 
Because of the length of the expressions eqs.~(\ref{eq:sigmauS}-\ref{eq:sigmauL}), the sums over internal generation indices $a$ and $b$ are understood. 
}
All necessary vertices are listed in Appendix~\ref{app:QuarkVertices}\footnote{the rotation matrices of the external states (marked as $\check{x}$ in the expressions for $\Sigma$) have to replaced by the identity matrix since the corrections to the mass matrices are calculated}. The eigenvalues of $m_f^{(1L)}(p^2_i)$ must fulfill
\begin{align}
 \text{Eig}\left[m_d^{(1L)}(p^2 = m^2_{d_i})\right] = (m_{d}^{\DRbar,\text{SM}},m_{s}^{\DRbar,\text{SM}},m_{b}^{\DRbar,\text{SM}} ) \\ 
  \text{Eig}\left[m_u^{(1L)}(p^2 = m^2_{u_i})\right] = (m_{u}^{\DRbar,\text{SM}},m_{c}^{\DRbar,\text{SM}},m_{t}^{\DRbar,\text{SM}}, m_{t'}^{\DRbar} ) 
\end{align}
with the $\DRbar$--masses taken from eqs.~(\ref{eq:drbardown}-\ref{eq:drbartop}).
In addition, the rotation matrices diagonalizing $m_d^{(1L)}$ and $m_u^{(1L)}$ are constrained by the measurement of the CKM matrix. 
One can use these conditions and invert eq.~(\ref{eq:oneloopMF}) to get expressions for the tree-level mass matrices, which are then used to calculated  $Y_d^{\DRbar}$ and $Y_u^{\DRbar}$. Since the self-energies depend on the Yukawa matrices, the entire calculation has to be numerically iterated until a stable point is reached.  \\

After the calculation of the gauge and Yukawa couplings at $M_Z$ is finished, the two-loop RGEs shown in Appendix~\ref{app:RGEs} are used to run the couplings up to $M_{\text{SUSY}}$. Since in all calculations the masses of the SUSY states at $M_Z$ are needed, also a two-loop running of all parameters from $M_{\text{SUSY}}$ to $M_Z$ is done to get the running tree-level masses at $M_Z$. 

\begin{figure}[hbt]
\includegraphics[width=0.45\linewidth]{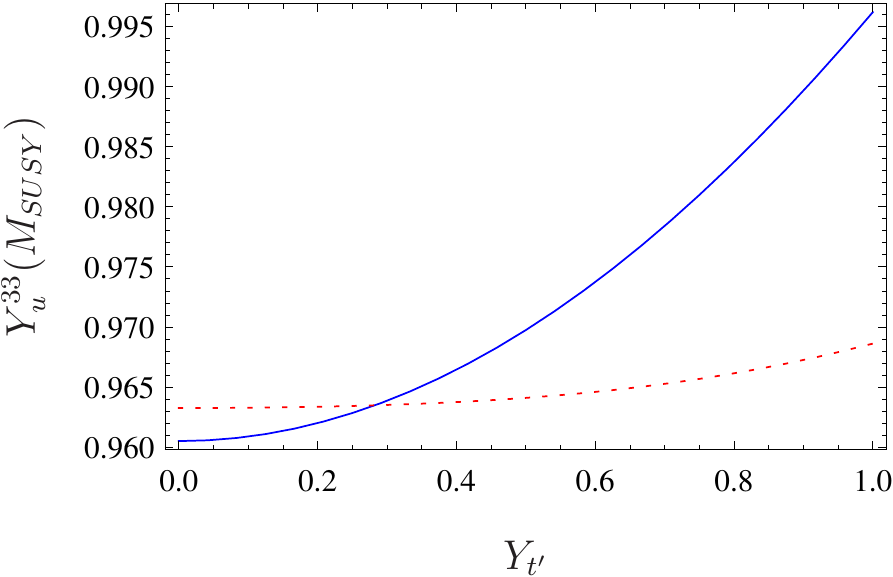}
\caption{Running top Yukawa coupling ($Y_u^{33}$) at the SUSY scale as function of $Y_{t'}$ for two different values of $M_{T'}$: 1.0~TeV (blue) and 3.0~TeV (dotted red).}
\label{fig:YtYu}
\end{figure}

The effect of the threshold corrections on the running value of the top Yukawa coupling ($Y_u^{33}$) at the SUSY scale as a function of $Y_{t'}$ is shown in Fig.~\ref{fig:YtYu}. We have used two different values of $M_{T'}$: 1 and 3~TeV. In addition, we fixed $\tan\beta=3$ and all soft-masses to 1.5~TeV. In total, this effect can be as large as a  few percent and is larger for smaller $M_{T'}$ because the $t-t'$ mixing becomes larger. This already gives an important change in the MSSM-like corrections to the Higgs states which turn out to be of order of a few GeV, as we will see. One might wonder why the values for the top Yukawa don't agree for $Y_{t'}=0$. 
%The reason are the threshold corrections to $g_3$ which are always there independent of the couplings of the vectorlike states. 
The reason is that the threshold corrections to $g_3$ are always present and they depend on $M_{T'}$, even if other couplings of the vectorlike states are absent.
This changes the prediction for $g_3$ which enters (i) the SM and MSSM part of the thresholds corrections, and (ii) the RGEs when running from $M_Z$ to $M_{\text{SUSY}}$.

\begin{figure}[hbt]
\includegraphics[width=0.45\linewidth]{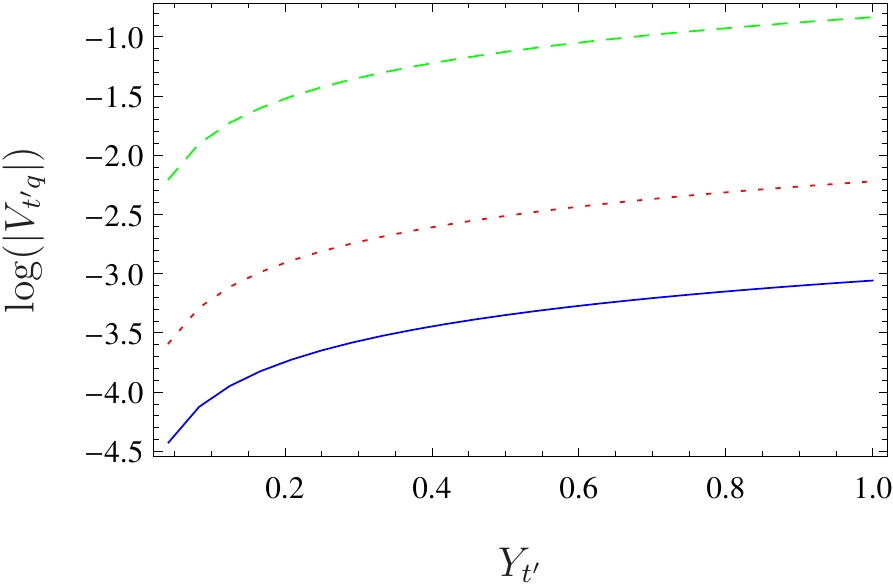}
\caption{Absolute size  $|V_{t'q}|$ of the CKM entries between the vectorlike top states and the SM down quarks $q=u,s,b$. The colour code is $|V_{t'd}|$ (full blue), $|V_{t's}|$ (dotted red), and $|V_{t'b}|$ (dashed green). We fixed here $M_{T'} = 1$~TeV.}
\label{fig:YtCKM}
\end{figure}

While a study of flavour physics in this model is beyond the scope of this paper, we want to briefly comment on the expected effects. The CKM matrix in this model is a $4\times 3$ matrix and we adjust the Yukawa  couplings  $Y_d$ and $Y_u$ in our study in a way that the $3\times 3$ sub-matrix assigning the couplings between SM-quarks is in agreement with measurements. The last column  of the CKM matrix carries the elements $V_{t'q}$ which define the size of the flavour changing charged currents between the vectorlike top and the SM down-quarks. The size of $|V_{t'q}|$ is constrained by the measurements of flavour violating processes which are known to a high precision and which are in agreement with SM predictions. In Ref.~\cite{Alok:2015iha} the following limits were derived at $3\sigma$:
\begin{eqnarray}
& |V_{t'd}| < 0.01 \,,\quad |V_{t's}| < 0.01 \,,\quad |V_{t'b}| < 0.27 & 
\end{eqnarray}
We show the prediction of these elements as a function of $Y_{t'}$ in Fig.~\ref{fig:YtCKM} for $M_{T'} = 1$~TeV. One can see that the obtained values are well below the current bounds. The main reason for this is that we assume $Y_{t'}^1$ and $Y_{t'}^2$ to vanish.  

\subsubsection{One-loop corrections}
\label{sec:oneloop}
A generic one-loop calculation with \SARAH and \SPheno was introduced in Ref.~\cite{Staub:2010ty}. The procedure for this is as follows. First, all running tree-level parameters are calculated at the SUSY scale. The $g_i$ ($i=1,2,3$) and $Y_i$ ($i=e,d,u$) are obtained by running up the \DRbar values calculated at $M_Z$, the Higgs soft-masses $m_{H_d}^2$ and $m_{H_u}^2$ are derived from the tadpole equations eqs.~(\ref{eq:tad1})--(\ref{eq:tad2}). Using these values all tree-level masses are obtained and  $\delta M^2_Z$ is calculated. This quantity is needed to get the correct electroweak VEVs at the SUSY scale from the $Z$-boson pole mass $M^{2,\text{pole}}_Z$ and $\tan\beta$ via
\begin{equation}
v^{\text{SUSY}}=\sqrt{\frac{g_1^2+g_2^2}{4}(M_Z^{2,\text{pole}} - \delta M_Z^2)}\,,\quad v_d = v^{\text{SUSY}} \cos\beta \,,\quad v_u = v^{\text{SUSY}} \sin\beta
\end{equation}
With these values the tree-level masses are re-calculated and the calculation of the one-loop corrections is started. Here, first the one-loop corrections $\delta t_i^{(1)}$ to the tadpole equations $T_i$ are needed. The changes compared to the MSSM stemming from vectorlike tops are: 
\begin{align} 
\delta^{u,\tilde{u}} t^{(1)}_i = & +6 \sum_{a=1}^{4}{A_0\Big(m^2_{u_{{a}}}\Big)} m_{u_{{a}}} \Big({\Gamma^L_{\phi_i,\bar{u_{{a}}},u_{{a}}}} + {\Gamma^R_{\phi_i,\bar{u_{{a}}},u_{{a}}}}\Big)   -3 \sum_{a=1}^{8}{A_0\Big(m^2_{\tilde{u}_{{a}}}\Big)} {\Gamma_{\phi_i,\tilde{u}^*_{{a}},\tilde{u}_{{a}}}}   
\end{align}
with $i=u,d$.
All other corrections are identical to the results of Ref.~\cite{Pierce:1996zz}. Afterwards, we need the one-loop corrections to the scalar Higgs mass matrix. Here, the vectorlike top quarks contribute to the scalar self-energy $\Pi(p^2)$
\begin{align} 
\Pi^{u,\tilde{u}}_{ij}(p^2) &= -6 \sum_{a=1}^{4}m_{u_{{a}}} \sum_{b=1}^{4}{B_0\Big(p^{2},m^2_{u_{{a}}},m^2_{u_{{b}}}\Big)} m_{u_{{b}}} \Big({\Gamma^{L*}_{\phi_i,\bar{u}_{{a}},u_{{b}}}} {\Gamma^R_{\phi_j,\bar{u}_{{a}},u_{{b}}}}  + {\Gamma^{R*}_{\phi_i,\bar{u}_{{a}},u_{{b}}}} {\Gamma^L_{\phi_j,\bar{u}_{{a}},u_{{b}}}} \Big) \nonumber \\ 
 &-3 \sum_{a=1}^{8}{A_0\Big(m^2_{\tilde{u}_{{a}}}\Big)} {\Gamma_{\phi_i,\phi_j,\tilde{u}^*_{{a}},\tilde{u}_{{a}}}}  +3 \sum_{a=1}^{8}\sum_{b=1}^{8}{B_0\Big(p^{2},m^2_{\tilde{u}_{{a}}},m^2_{\tilde{u}_{{b}}}\Big)} {\Gamma^*_{\phi_i,\tilde{u}^*_{{a}},\tilde{u}_{{b}}}} {\Gamma_{\phi_j,\tilde{u}^*_{{a}},\tilde{u}_{{b}}}}  
\end{align} 
The necessary vertices to calculate $\delta^{t,t'} t^{(1)}$ and $\Pi^{t,t'}(p^2)$ are given in Appendix~\ref{app:HiggsVertices}. We can now express the one-loop corrected mass matrix of the scalar Higgs by
\begin{align}
m^{2,(1L)}_h(p^2) =&  m^{2,(T)}_h + \Pi^{u,\tilde{u}}(p^2) + \left(\begin{array}{cc} \frac{1}{v_d} \delta^{u,\tilde{u}} t^{(1)}_d & 0 \\ 0 & \frac{1}{v_u} \delta^{u,\tilde{u}} t^{(1)}_u  \end{array}\right) \nonumber \\
 & \hspace{2cm}+ \Pi^{\text{MSSM}}_{\cancel{u},\cancel{\tilde{u}}}(p^2) + \left(\begin{array}{cc} \frac{1}{v_d} \delta^{\text{MSSM}}_{\cancel{u},\cancel{\tilde{u}}} t^{(1)}_d & 0 \\ 0 & \frac{1}{v_u} \delta^{\text{MSSM}}_{\cancel{u},\cancel{\tilde{u}}} t^{(1)}_u\end{array}\right)
\end{align}
Here, $\Pi^{\text{MSSM}}_{\cancel{u},\cancel{\tilde{u}}}$ and $\delta^{\text{MSSM}}_{\cancel{u},\cancel{\tilde{u}}} t^{(1)}_{d,u}$ are the MSSM results without any contributions from up (s)quarks. The eigenvalues $m^2_{h_i}$  of $m^{2,(1L)}_h$ correspond to the loop corrected Higgs masses. Since, $m^{2,(1L)}_h(p^2)$ is a function of the external momentum, this calculation is usually iterated until a stable solution $m^{2,(1L)}_h(m^2_{h_i})$ for each eigenvalue is found. \\
Previously, the one-loop corrections in this model have been calculated in the effective potential approach \cite{Martin:2009bg}. This calculation is equivalent to ours in the limit $p^2 \to 0$. Thus, by checking this limit we can easily estimate the error introduced in these calculations by that approximation.  Since the additional fermions and the scalars are usually heavier than the desired Higgs mass of 125~GeV, one can expect that the momentum effects are rather moderate. However, before we discuss this in detail, we go one step further to the two-loop corrections. 

\subsubsection{Two-loop corrections}
\label{sec:twoloop}
It is very well known that two-loop corrections in the MSSM are crucial: they can give a large push to the Higgs mass and are the only chance to get agreement between the Higgs mass in the MSSM for moderate SUSY masses ($< 2$~TeV) and the measurement of about 125~GeV. This mass is out of reach only using one-loop corrections. This is not necessarily the case for models with vectorlike quarks: if the new couplings to the SM-like Higgs are large enough, even one-loop corrections might be sufficient to find a sufficiently large Higgs mass. Nevertheless, there are good reasons to consider also the two-loop corrections: to be able to make any meaningful statement in the considered model if a point is excluded, the difference to the measurement must be larger than the theoretical uncertainty. At one-loop the theoretical uncertainty in the Higgs mass prediction can easily be 10~GeV or more, i.e. it is not possible at all to restrict many regions of the parameter space by a one-loop calculation.  Of course, also the opposite might happen: points which are in good agreement at one-loop can be ruled out by a two-loop calculation. 

For this reason, we are going to give details about a two-loop calculation including the dominant corrections. 'Dominant' in this context means all contributions excluding those of the electroweak gauge couplings $g_1$ and $g_2$. That's the same precision which is also usually considered for the MSSM. The remaining electroweak corrections, together with the missing momentum dependence and the unknown higher-order corrections are estimated to a remaining uncertainty of about 3~GeV. 
\begin{figure}[hbt]
\includegraphics[width=0.7\linewidth]{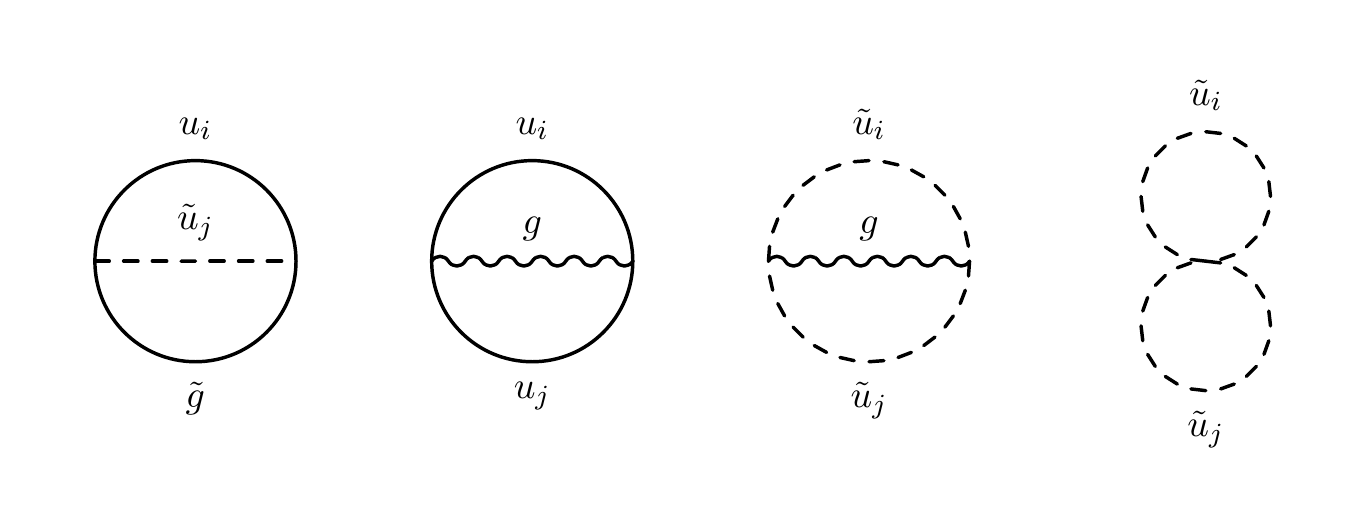}
\caption{Two-loop diagrams giving contributions to the effective potential $O(\alpha_t \alpha_s)$ and $O(\alpha_{t'} \alpha_s)$. Here, the indices of up-quark generations ($u_i$) run from 1 to 4, and those of up-squark generations ($\tilde u_i$) from 1 to 8.}
\label{fig:DiaAlphaS}
\end{figure}
In the MSSM the most dominant two-loop corrections are those involving the strong coupling constant $g_3$ because of large colour factors. The diagrams which contribute in the MSSM are depicted in Fig.~\ref{fig:DiaAlphaS}. In the model at hand with vectorlike tops, the diagrams are actually the same but with a sum over a larger number of (s)quark generations. The obtained corrections from these diagrams are $O(\alpha_t \alpha_s)$ and $O(\alpha_{t'} \alpha_s)$ with $\alpha_t = (Y_u^{33})^2/4\pi$, $\alpha_{t'} = (Y_{t'}^3)^2/4\pi$.  

\begin{figure}[hbt]
\includegraphics[width=0.95\linewidth]{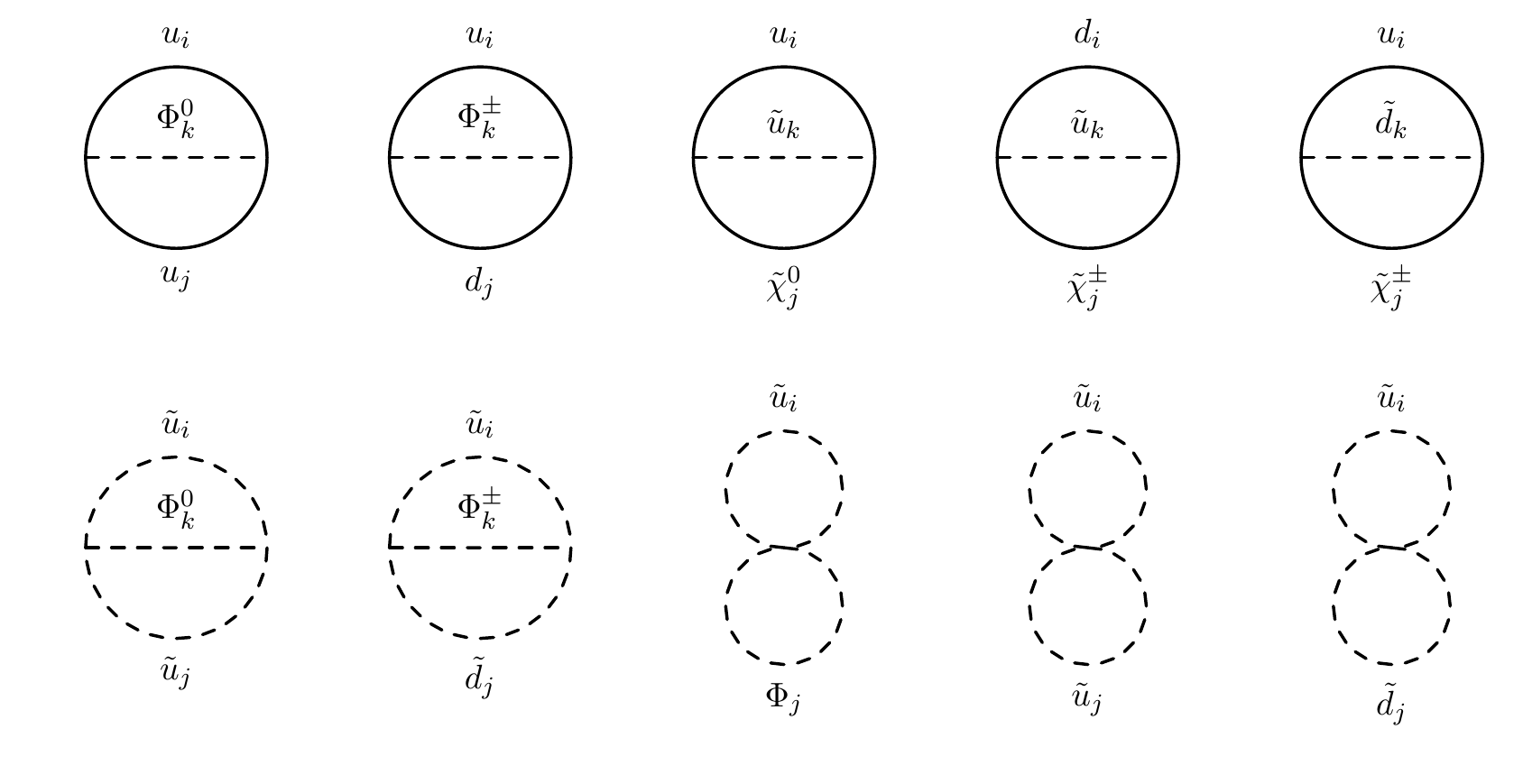}
\caption{Two-loop diagrams giving contributions to the effective potential $O(\alpha^2_t)$, $O(\alpha^2_{t'})$, and $O(\alpha_t \alpha_{t'})$. Here, $\Phi^0 = \{h, H, G^0, A^0\}$, $\Phi^\pm = \{H^\pm, G^\pm\}$, $\Phi= \{\Phi^0, \Phi^\pm\}$. The index ranges are: $\Phi (1,2)$; $\tilde{\chi^0} (1-4)$;  $\tilde{\chi^\pm} (1,2)$; $u (1-4)$; $d (1-3)$; $\tilde{u} (1-8)$; $\tilde{d} (1-6)$. }
\label{fig:DiaAlphaT}
\end{figure}
The next important contributions from the MSSM are those of $O(\alpha_t^2)$. These come from diagrams involving (s)tops and Higgs states respectively Higgsinos. 
Also here, the diagrams shown in Fig.~\ref{fig:DiaAlphaT} are the same as in the MSSM, but the sums over (s)fermion generations are extended. These diagrams give contributions of the order $O(\alpha_t^2)$, $O(\alpha_{t'} \alpha_t)$ and $O(\alpha_{t'}^2)$. 
Also the corrections $O(\alpha_t(\alpha_b+\alpha_\tau))$ with $\alpha_b=(Y_d^{33})^2/(4\pi)$, $\alpha_\tau = (Y_\tau^{33})^2/(4\pi)$ are known in the MSSM. Especially for moderate values of $\tan\beta$ these corrections are less important. Nevertheless, in our calculations also these corrections together with the counterparts $O(\alpha_{t'}(\alpha_b+\alpha_\tau))$ are included. 

\SARAH and \SPheno offer two possibilities to calculate the two-loop corrections to scalar Higgs masses: either a purely effective potential calculation can be done. In that case, the diagrams as shown in Figs.~\ref{fig:DiaAlphaS} and \ref{fig:DiaAlphaT} are calculated to get $V^{\text{eff},(2L)}$, and the derivatives of the results with respect to the Higgs VEVs are taken to get the two-loop corrections to the tadpoles and self-energies
\begin{equation}
 \delta t^{(2L)}_i = \frac{\partial V^{\text{eff},(2L)}}{\partial v_i} \hspace{1cm} \Pi^{(2L)}_{ij} =  \frac{\partial^2 V^{\text{eff},(2L)}}{\partial v_i \partial v_j}
\end{equation}
However, this involves a numerical derivation which sometimes suffers from numerical problems and rather large uncertainties. Thus, the second method implemented in \SARAH and \SPheno is often the preferred one: this method employs a diagrammatic calculation where the external Higgs legs explicitly show up. Even if this leads to a much bigger set of two-loop diagrams, the calculation is not necessarily slower. All diagrams are evaluated in the limit $p^2\to 0$, i.e. the results give  equivalent results for $\delta t^{(2L)}_i$ and $\Pi^{(2L}_{ij}$ as the first method does. \\

Given the two-loop corrections, the loop-corrected Higgs mass can be expressed by
\begin{equation}
m^{2,(2L)}_h(p^2) = m^{2,(T)}_h + \Pi^{(1L)}(p^2) + \Pi^{(2L)}(0)- \left(\begin{array}{cc} \frac{1}{v_d} (\delta t^{(1L)}_d+\delta t^{(2L)}_d) & 0 \\ 0 & \frac{1}{v_u} (\delta t^{(1L)}_u + \delta t^{(2L)}_u) \end{array}\right) 
\end{equation}
Here, we have no longer distinguished between corrections involving vectorlike tops or not, but used $\Pi^{(XL)}$ and $\delta t^{(XL)}$ for the sum of all contributions. The eigenvalues $m_{h_i}^2$ fulfilling  $\text{Eig}(m^{2,(2L)}_h(m_{h_i}^2)) = m_{h_i}^2$ are associated with the scalar pole masses. In the following, the smaller value $m_{h_1}^2$ corresponds to the SM-like Higgs boson and we are going to use the short notation $m_h \equiv \sqrt{m_{h_1}^2}$ for it. 

\begin{figure}[hbt]
\centering
\includegraphics[width=0.4\linewidth]{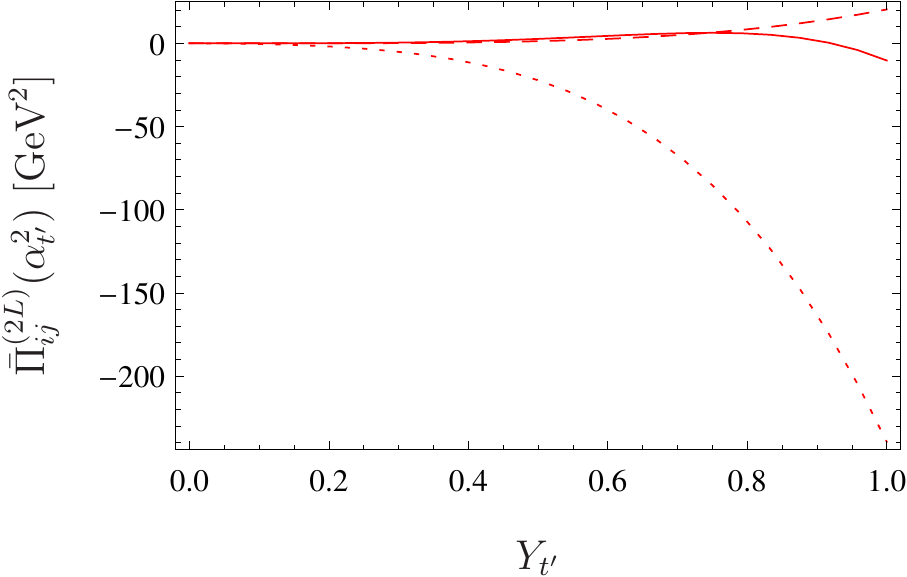}\hspace{1cm}\includegraphics[width=0.4\linewidth]{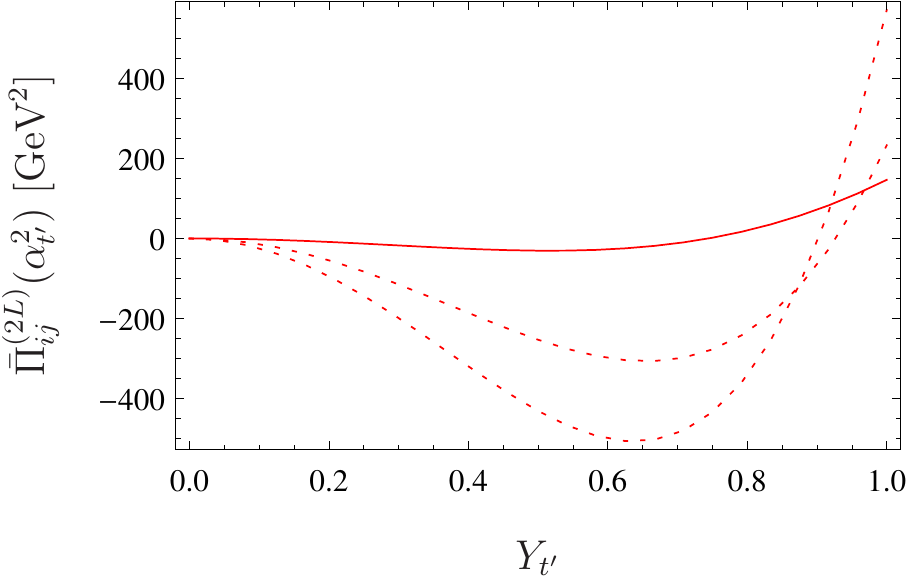}\\
\includegraphics[width=0.4\linewidth]{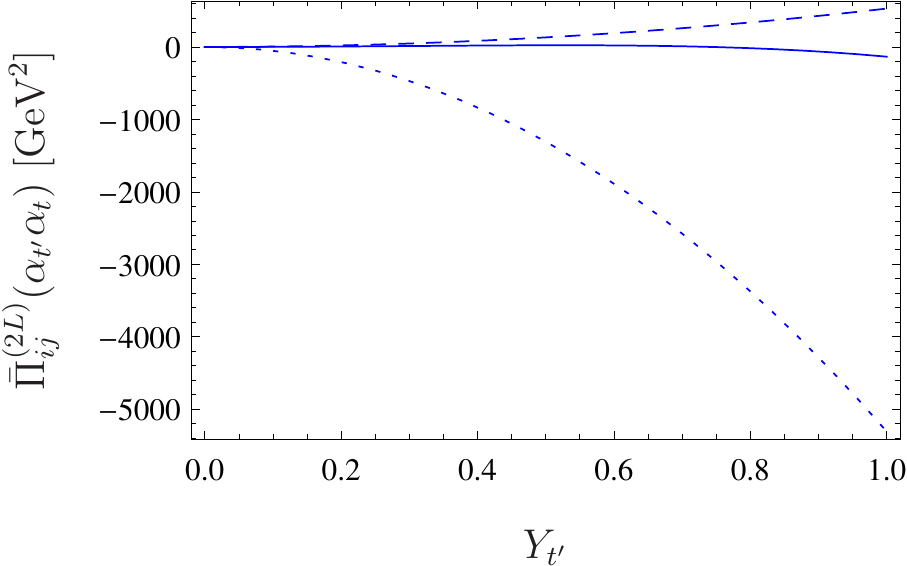}\hspace{1cm}\includegraphics[width=0.4\linewidth]{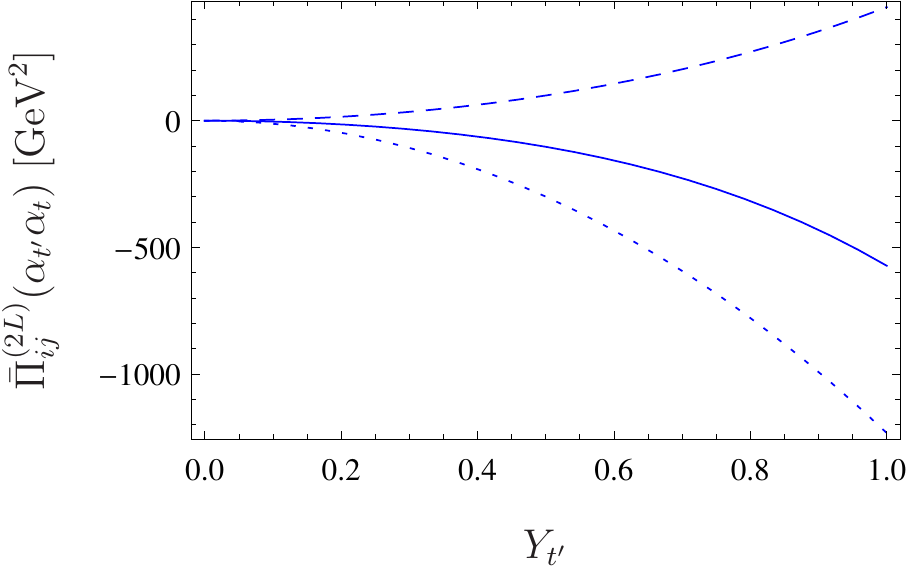}\\
\includegraphics[width=0.4\linewidth]{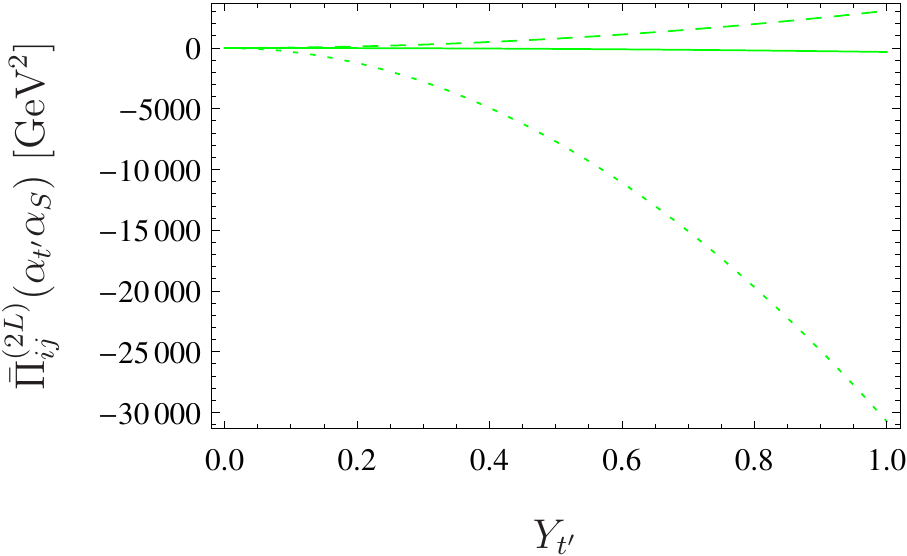}\hspace{1cm}\includegraphics[width=0.4\linewidth]{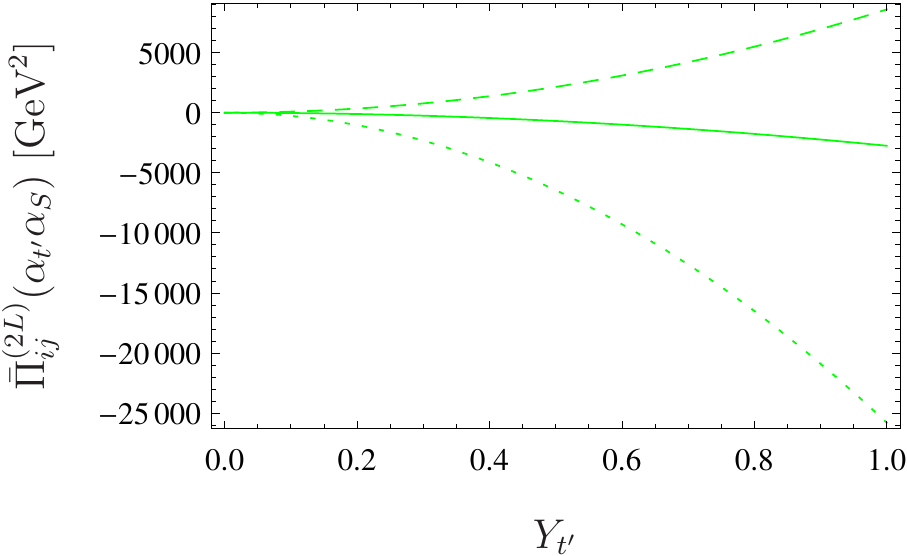}\\
\includegraphics[width=0.4\linewidth]{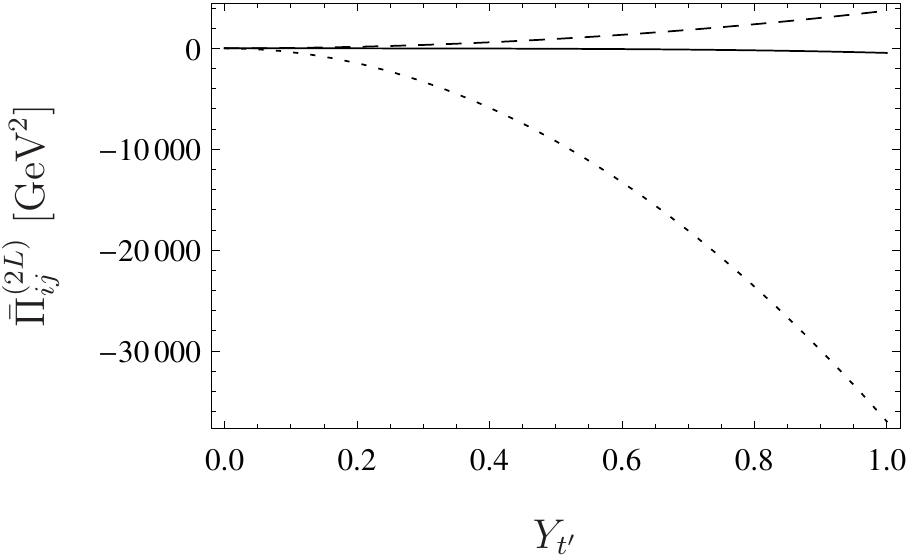}\hspace{1cm}\includegraphics[width=0.4\linewidth]{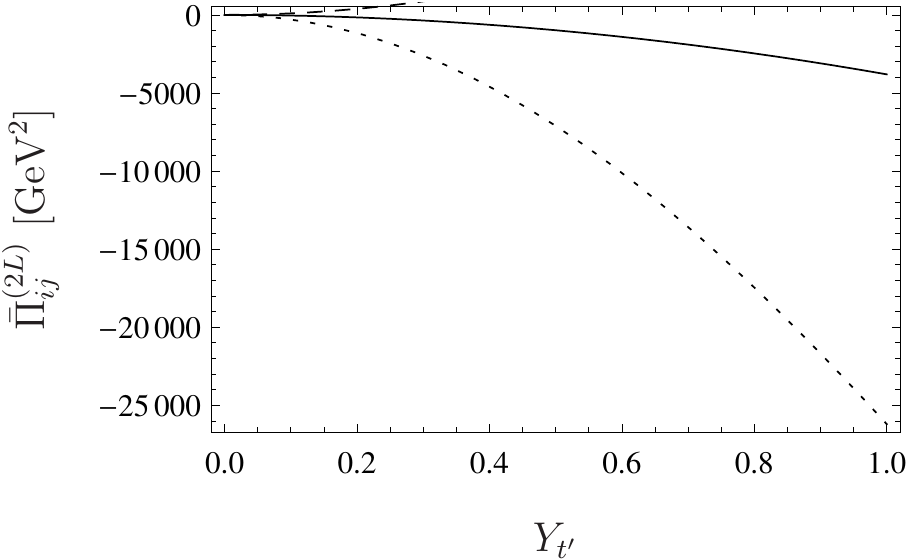}%\\
\caption{Two-loop contributions the Higgs mass matrix involving vectorlike (s)quark. We used here $M_{T'}=1.0$~TeV and put all soft-mass terms to 1.5~TeV. On the left, we set $T_{t'}=0$, on the right $T_{t'}=2.0~\text{TeV}\cdot Y_{t'}$. Dashed lines are for the (1,1) element, full lines for the (2,2) one, and dotted lines for the off-diagonal contribution. In the first three rows we plot the individual contributions $O(\alpha_{t'}^2)$, $O(\alpha_{t'}\alpha_t)$, $O(\alpha_{t'} \alpha_S)$, while the last row shows the sum of all contributions. }
\label{fig:Pi}
\end{figure}

Before we turn to the full calculation, we want to discuss briefly the importance of the different contributions at two-loop.  For this purpose we depict in Fig.~\ref{fig:Pi} the different two-loop contributions to the Higgs mass matrix:
\begin{equation}
\bar\Pi_{ij} \equiv \Pi_{ij}^{(2L)} - \delta_{ij} \frac{1}{v_i} \delta t_i^{(2L)} \hspace{1cm} i=d,u  
\end{equation}
It turns out that the corrections $O(\alpha_{t'} \alpha_b)$ are negligible. The corrections $O(\alpha_{t'} \alpha_\tau)$ are even much smaller and therefore not shown in Fig.~\ref{fig:Pi}.  We consider here two different cases: vanishing $T_{t'}$ and $T_{t'} = 2000~\text{GeV} \cdot Y_{t'}$. In both cases we find that the most dominant contributions are those involving the strong interaction what's similar to the MSSM.  The next important ones are those $O(\alpha_{t'} \alpha_t)$, while the $O(\alpha_{t'}^2)$ contributions are moderately small. Here, the difference compared to the MSSM corrections $O(\alpha_s \alpha_t)$ and $O(\alpha_t^2)$ which often cancel to some extent, is that here the contributions come with the same sign. 
We also see that for most contributions the impact on the (1,1) element is the largest one, i.e. the dominant part of these contributions come from $F$-terms $\simeq \mu Y_{t'}$. Thus, the new two-loop corrections are expected to be more important for parameter regions where the light Higgs has a larger $H_d$ fraction. The main differences between the cases of vanishing and non-vanishing $T_{t'}$ is that the corrections involving the strong interaction to (1,1) become smaller, while those to the (2,2) increase. Also the $O(\alpha_{t'} \alpha_t)$ contributions to the (2,2) are enhanced.  Thus, another region where the new two-loop corrections are expected to become important are those with large trilinear soft-terms $T_{t'}$.

\section{Results -- Part I: the Higgs mass}
\label{sec:results}
Before we turn to our main results, namely the discussion of the fine-tuning in the UV complete model, we want to discuss the importance of the different Higgs mass corrections we have included. For this reason we consider first the minimal model with the MSSM extended by vector-like tops only. To deal with the large number of free parameters at the SUSY scale when not considering an UV embedding, we make the following assumptions about the MSSM soft masses:
\begin{eqnarray*}
&m_{\tilde u}^2 = m_{\tilde q}^2 = m_{\tilde d}^2 = m_{\tilde e}^2 = m_{\tilde l}^2 = {\bf 1} \cdot (1.5 \text{TeV})^2 &\\
&M_1 = 0.5~\text{TeV} \,, \quad M_2 = 1.0~\text{TeV}\,,\quad M_3= 2.0~\text{TeV}&\\
& T_u=T_d=T_e=0&
\end{eqnarray*}
Moreover, we fix usually the MSSM parameters
\begin{eqnarray*}
&\mu = 1.0~\text{TeV}\,,\quad M^2_A = (1~\text{TeV})^2 & 
\end{eqnarray*}
and for the new sector we assume if not stated otherwise  
\begin{eqnarray*}
&T_{t'} = m_t = B_t = 0& \\
&m^2_{\tilde t'} = m^2_{\tilde {\bar{t}}'} = (1.5~\text{TeV})^2&
\end{eqnarray*}
In addition, the most important SM parameters were chosen as
\begin{eqnarray*}
\alpha_S^{\MSbar}(M_Z) = 0.1180 \,,\quad m_b^\MSbar(m_b) = 4.2~\text{GeV}\,,\quad m^{\text{pole}}_t = 173.2~\text{GeV} 
\end{eqnarray*}

As already mentioned we employ the combination of the computer tools \SPheno and \SARAH for all numerical calculations: we have implemented the minimal model with vectorlike tops as well as the UV complete variant in \SARAH version {\tt 4.5.3} and the model files will become public with the next release of \SARAH. \SARAH was used to generate \Fortran code for \SPheno. The obtained \Fortran routines include automatically all new features from vectorlike stops  discussed in the last sections which are necessary for the precise Higgs mass calculation. Also routines for the calculation of  flavour observables and decays widths are generated by \SARAH. However, we will not go into details in these aspects of this model here. We are just using the {\tt FlavorKit} results \cite{Porod:2014xia} to double check that all points are in agreement with current bounds from flavour observables. This is, of course, expected as we already discussed in sec.~\ref{sec:thresholds}. The \Fortran code written by \SARAH was compiled together with \SPheno version {\tt 3.3.6}. For all parameter scans in the following we have used the \Mathematica package {\tt SSP} \cite{Staub:2011dp}.

\subsection{The difference between one-loop effective potential, full one-loop and two-loop}
\begin{figure}[hbt]
 \includegraphics[width=0.45\linewidth]{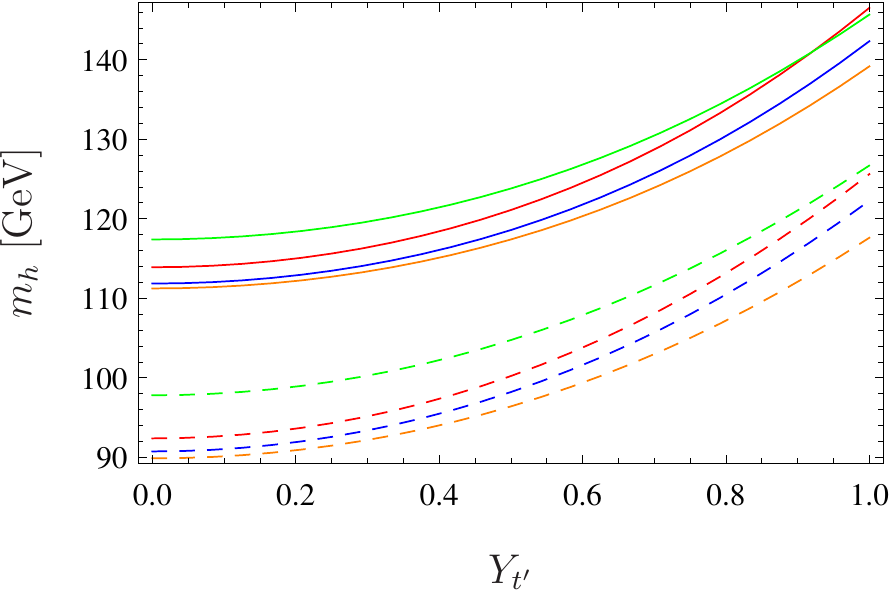}\hfill
 \includegraphics[width=0.45\linewidth]{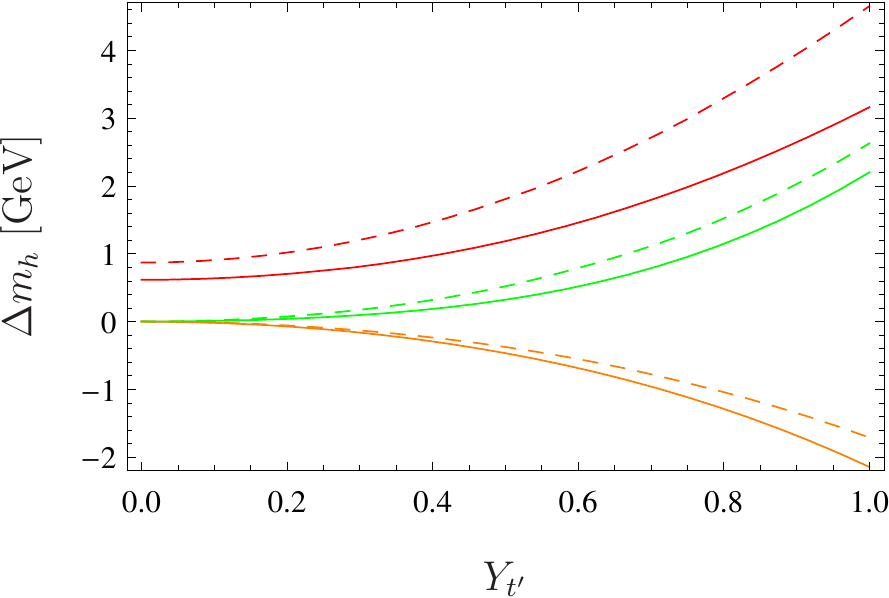} %\\
  \includegraphics[width=0.45\linewidth]{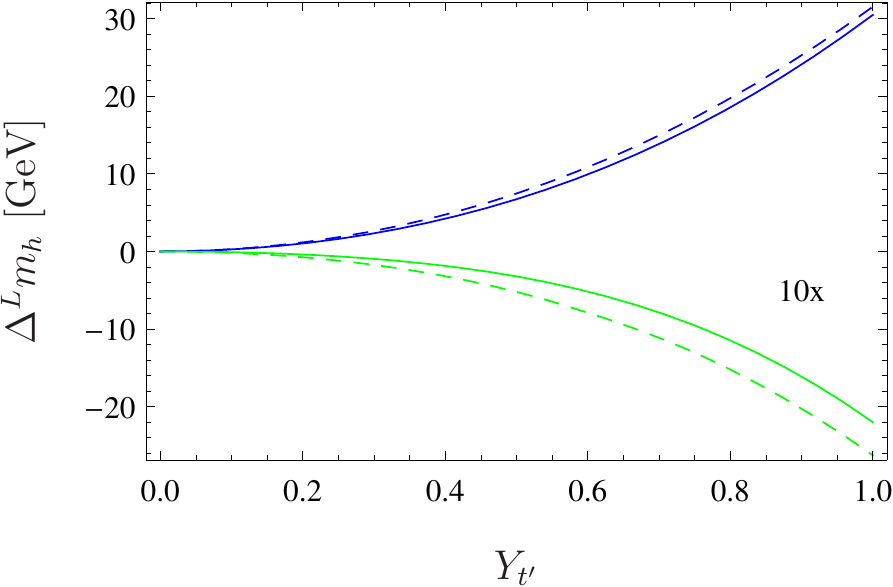}\hfill
  \includegraphics[width=0.45\linewidth]{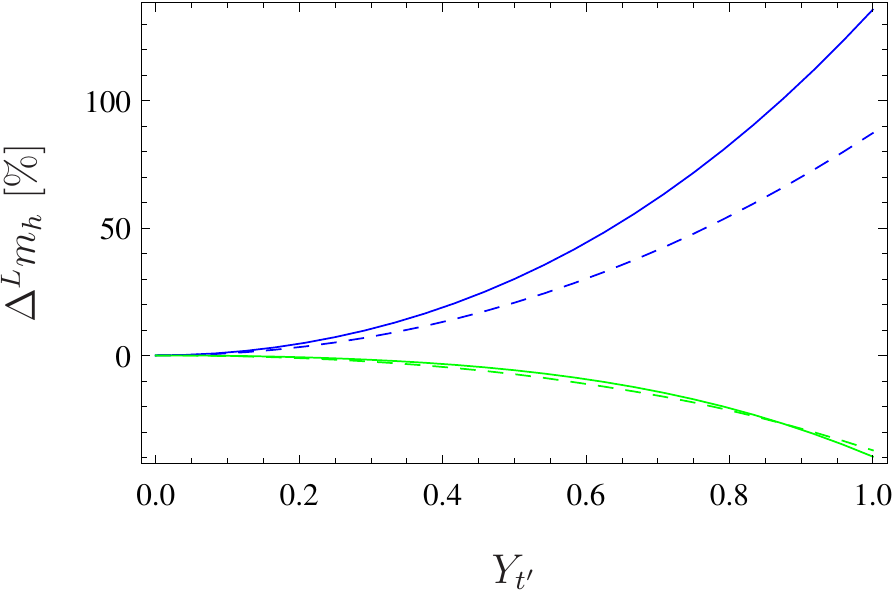} 
 \caption{
Top left: light Higgs mass as function of $Y_{t'}$. The red line corresponds to the effective potential calculation at one-loop, orange is the one-loop corrections with external momenta but neglecting the new threshold correction stemming from vectorlike states,  blue is the full one-loop calculation including the momentum dependence and all thresholds, and green includes the dominant two-loop corrections together with the full one-loop correction. Top right: impact of the threshold corrections (red), the momentum dependence at one-loop (orange) and the two-loop corrections (green), given as the difference $\Delta m_h=m_h - m_h({\text{1L},p^2\neq 0,\text{all thresholds}})$. Bottom left: absolute size of the one- (blue) and two-loop (green) corrections stemming from the vectorlike states. Note, for better readability we re-scaled the two-loop corrections by a factor of 10. 
Bottom right: relative importance of the one- (blue) and two-loop (green) corrections normalized to the size of the purely MSSM-like corrections. 
The full lines are for $\tan\beta=10$ and the dotted one are for $\tan\beta=2$.  We used here  $M_{T'}=1.0$~TeV, $B_{T'} = 0$. }
 \label{fig:YtTB}
\end{figure}

\begin{figure}[hbt]
 \includegraphics[width=0.45\linewidth]{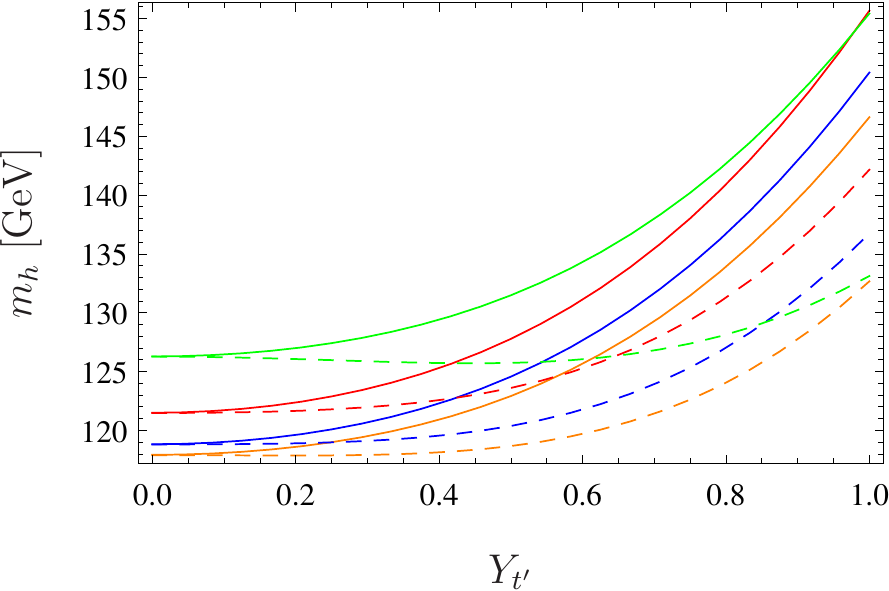}\hfill
 \includegraphics[width=0.45\linewidth]{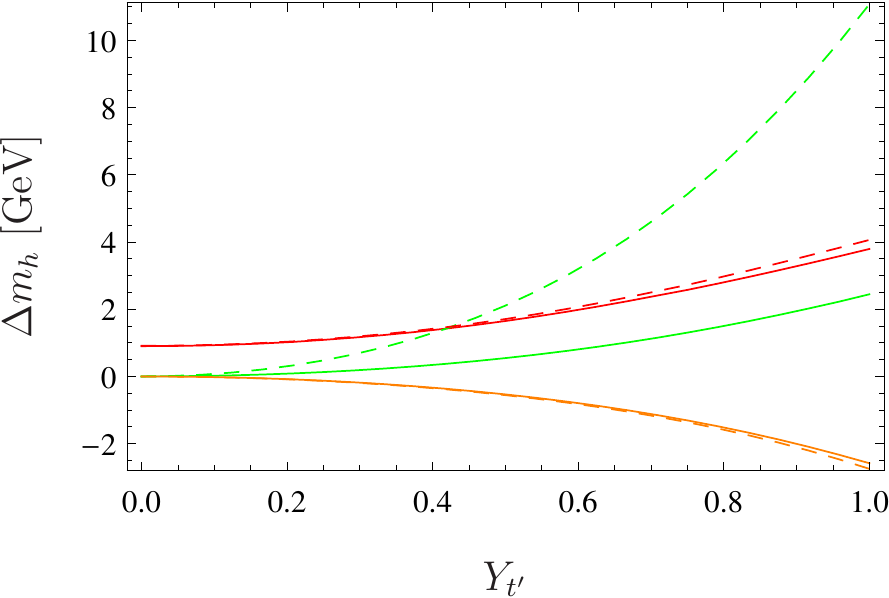} %\\
  \includegraphics[width=0.45\linewidth]{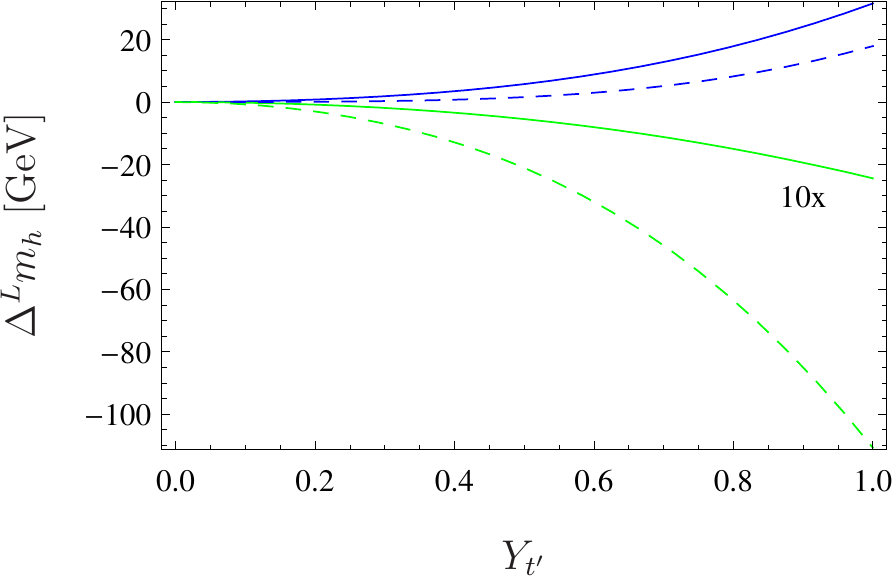}\hfill
  \includegraphics[width=0.45\linewidth]{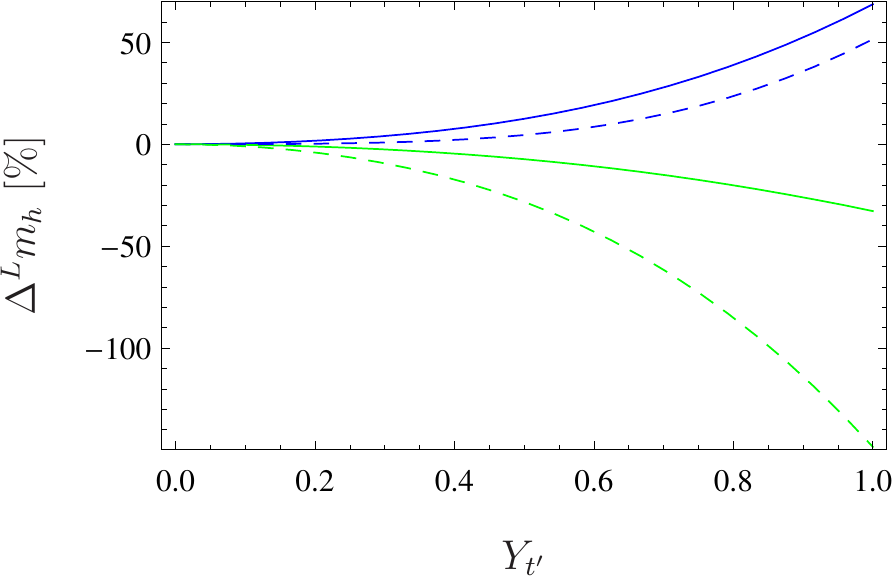} 
 \caption{ The plots show the same results as in Fig.~\ref{fig:YtTB} when including non-vanishing $T_{t'}$. We used $T_{t'}=2.0~\text{TeV} \cdot Y_{t'}$, $\tan\beta=5$ and $T_{u,33} = -2500~\text{GeV}$. The full lines are for $B_{T'} = 0$, while the dotted ones correspond to $B_{T'} = (1.5~\text{TeV})^2$. 
}
%We used here $\tan\beta = 5$ and $T_{u,33} = -2500~\text{GeV}$.}
\label{fig:TtBt}
\end{figure}

\begin{figure}[hbt]
 \includegraphics[width=0.45\linewidth]{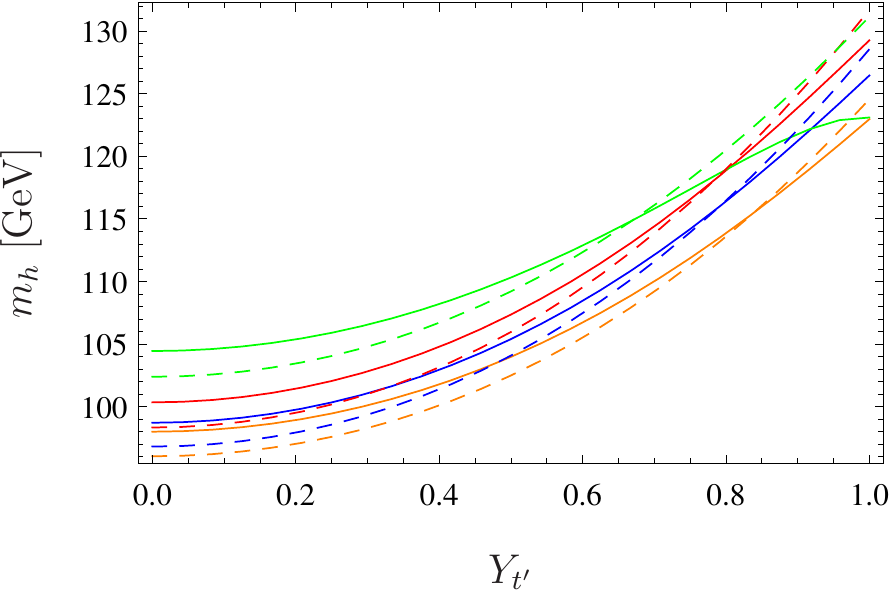}\hfill
 \includegraphics[width=0.45\linewidth]{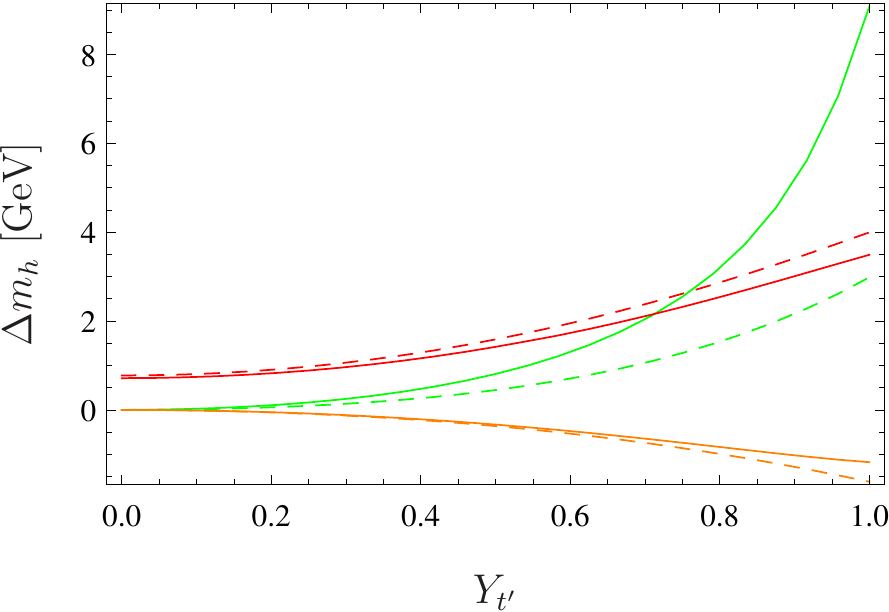} %\\
  \includegraphics[width=0.45\linewidth]{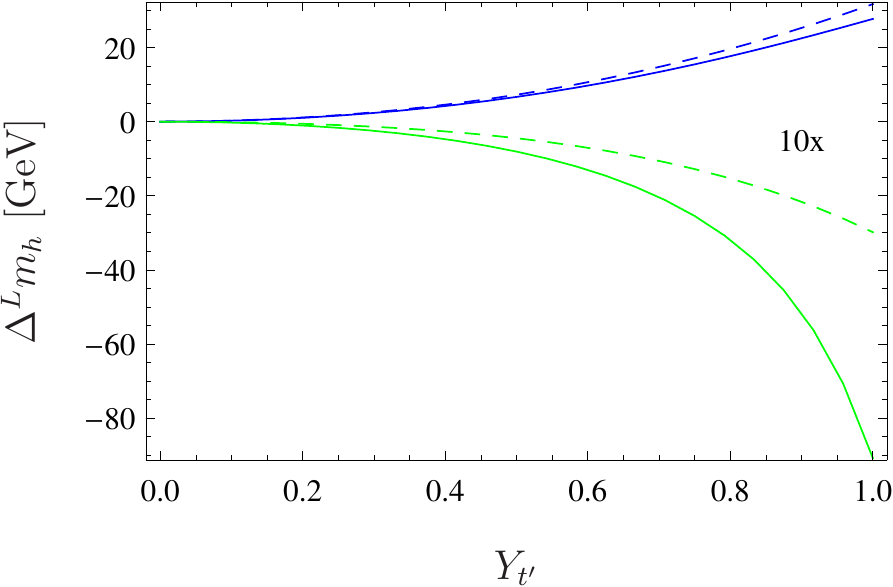}\hfill
  \includegraphics[width=0.45\linewidth]{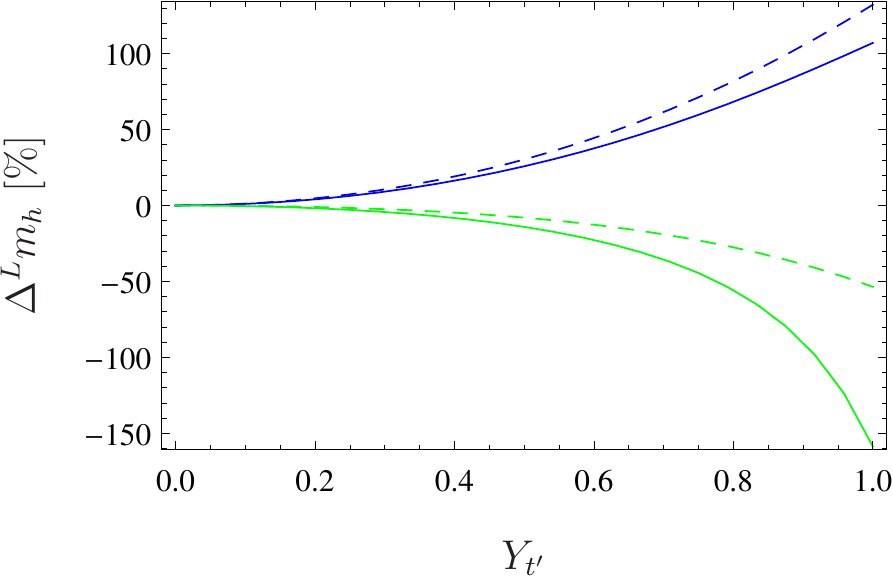} 
 \caption{
 The plots show the same results as in Fig.~\ref{fig:YtTB} for smaller $M_A^2 = 10^5~\text{GeV}^2$. We put $T_{t'}=T_u=0$, and $\tan\beta=3$. The dashed lines are for $B_{T'} = 0$, while the full ones correspond to $B_{T'} = (1.5~\text{TeV})^2$. }
 \label{fig:YtSmallMA}
\end{figure}

We check the importance of the corrections calculated here for the first time. For this purpose we compare in Figs.~\ref{fig:YtTB} -- \ref{fig:YtSmallMA}  the prediction for the Higgs mass calculated 
\begin{enumerate}[(i)]
\item at one-loop with vanishing external momenta but including thresholds,
\item at one-loop with full momentum dependence  but neglecting the threshold corrections to SM gauge and Yukawa couplings,
\item at full one-loop including the full momentum dependence and all threshold corrections,
\item at full one-loop with dominant two-loop corrections.
\end{enumerate}
The one-loop calculation without external momenta is equivalent to the calculation performed in the effective potential approach. For all three Figures we have put $M_{T'} = 1$~TeV. \\ 
In Fig.~\ref{fig:YtTB} we compare the results for two different values of $\tan\beta$: 2 and 10. While there is a large difference already at tree-level, the impact of the loop corrections is similar for both values of $\tan\beta$. Thus, we find that $m_h \simeq 125$~GeV is found for $Y_{t'} \sim 0.9$ (0.6) for $\tan\beta=2$ (10). 
Including the momentum dependence in the one-loop calculation of the vectorlike states can account for changes up to 2 GeV for large $Y_{t'}$  and are negative. In contrast, for the considered scenario the two-loop corrections are of a similar size, but positive. However, the biggest difference are caused by the threshold corrections. Since these can have a large impact on the top Yukawa couplings, we find that the prediction of the SM-like Higgs mass can deviate by up to 5~GeV. This effect is more pronounced for smaller $\tan\beta$. Note, even in the limit $Y_{t'} \to 0$, we find a shift by about 1~GeV compared to the calculation using only MSSM results. The reason is that the threshold corrections to $g_3$ don't vanish even in this limit. Therefore, the running value of the top Yukawa coupling entering the loop calculations changes slightly, which has still a visible effect on the Higgs mass. The absolute size of the one-loop corrections can grow up to 30~GeV for both values of $\tan\beta$, while the two-loop corrections are smaller by about a factor of 10. When we compare these numbers with the purely MSSM corrections, we see that the one-loop corrections can become as important as the MSSM ones, while the two-loop corrections can reach about half the size of the MSSM two-loop corrections. \\

We have identified in sec.~\ref{sec:twoloop} two regions where the new two-loop corrections are expected to be even more important. The first region is the one with non-vanishing $T_{t'}$. This is studied in Fig.~\ref{fig:TtBt} where we set $T_{t'} = 2000~\text{GeV} \cdot Y_{t'}$. In addition, we check also the effect of $B_{T'}$. For $B_{T'} = 0$ the differences to the results with $T_{t'} = 0$ are not very large: the corrections from the momentum dependence and the two-loop terms are of the same size and come with different signs. The largest effect is again from the threshold corrections. However, if $B_{T'}$ becomes large and causes a mass splitting for the vectorlike stops, the picture changes. Now, the most important effect comes from the two-loop corrections which can become as important as the MSSM ones. For $Y_{t'}$ values of $O(1)$ this can reduce the Higgs mass prediction by more than 10~GeV and easily over-compensate the two-loop corrections from the MSSM sector.

The other region we identified where the two-loop corrections can be important is the one where the SM-like Higgs has a larger down-type fraction. This happens if $M_A^2$ becomes small. We discuss this case in Fig.~\ref{fig:YtSmallMA} for zero and non-zero $B_{T'}$ again. In particular for the  large $B_{T'}$ the two-loop contributions can clearly make the biggest effect compared to the incomplete calculations used so far. These are again negative and can reduce the SM-like Higgs mass by up to 8 GeV. Thus, while it seems that one can reach the preferred mass of 125~GeV at one-loop with $Y_{t'} < 1$, with the two-loop corrections this is not possible for the considered combination of parameters. Although if $B_{T'}$ is taken to be zero, the effect can still be large and the overall size of the new two-loop corrections is still in the ballpark of the MSSM corrections.

\subsection{Dependence on the vectorlike masses, stop masses, and the  gaugino mass}
\begin{figure}[hbt]
 \includegraphics[width=0.33\linewidth]{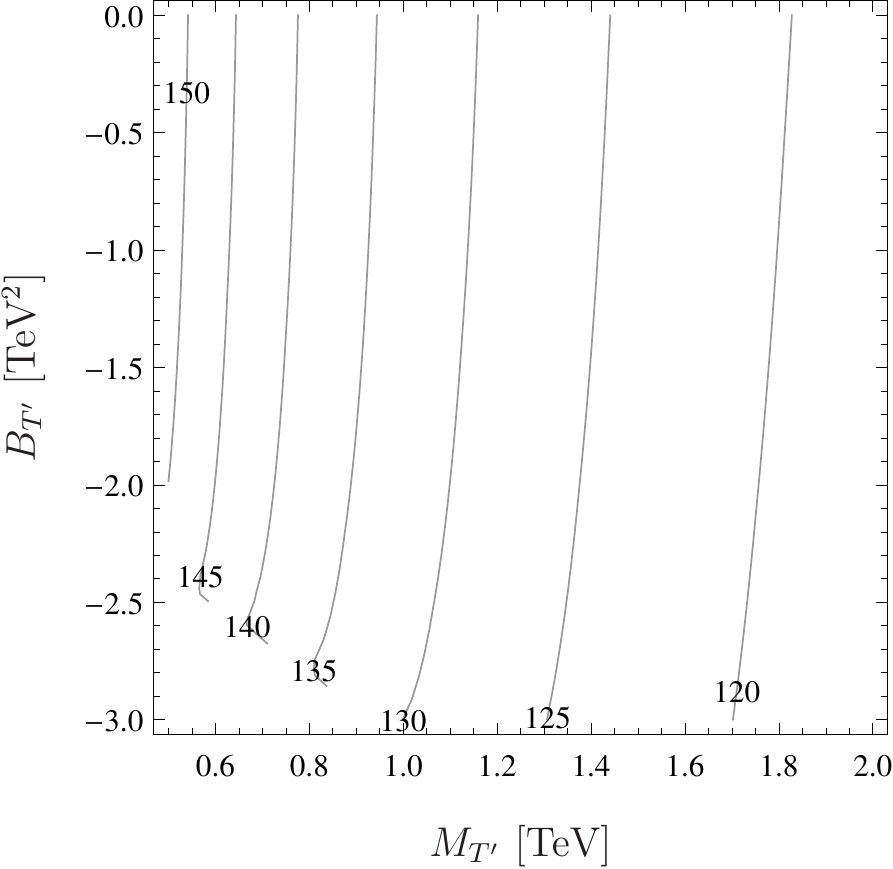}\hfill
 \includegraphics[width=0.33\linewidth]{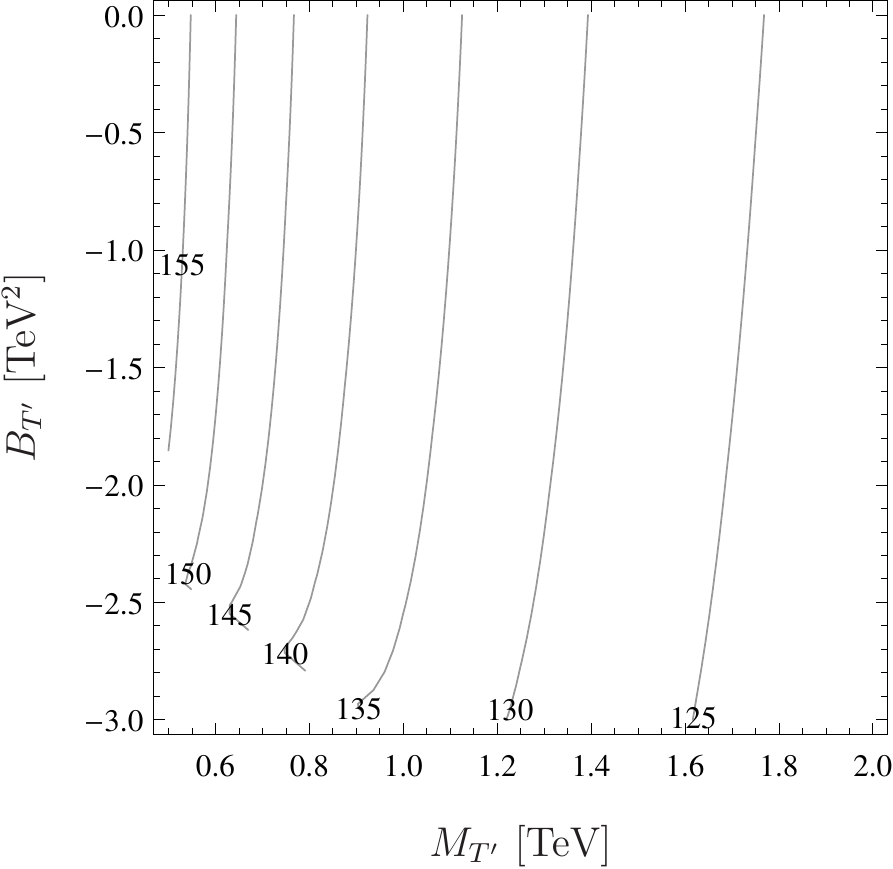}\hfill
 \includegraphics[width=0.33\linewidth]{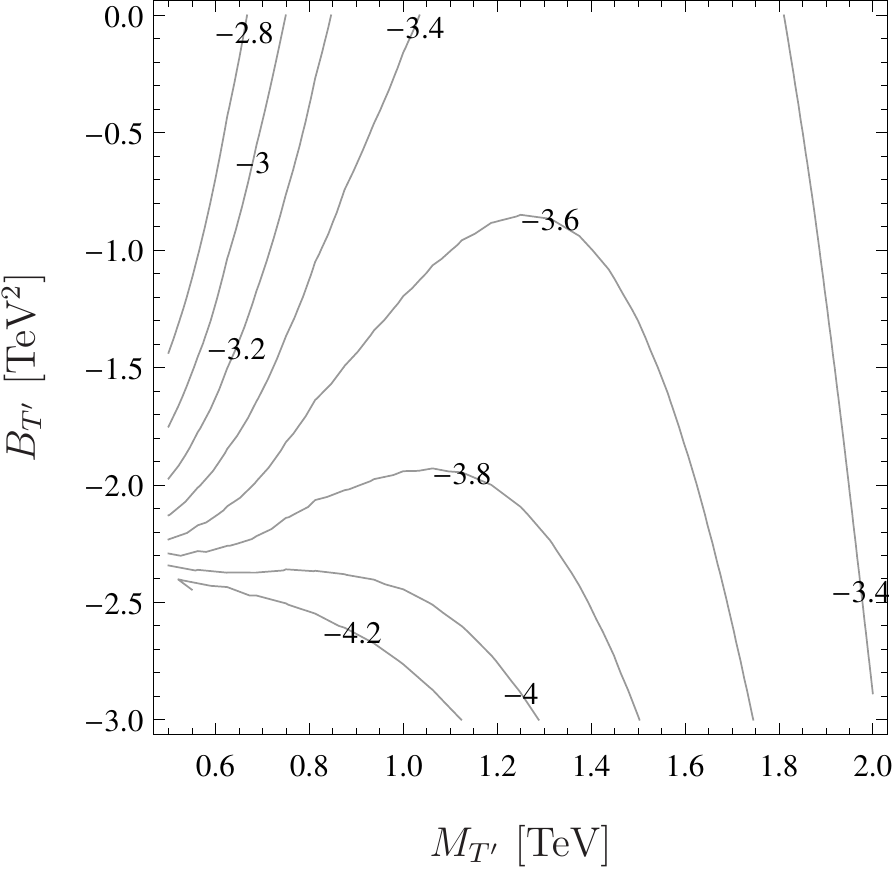}\\
 \includegraphics[width=0.33\linewidth]{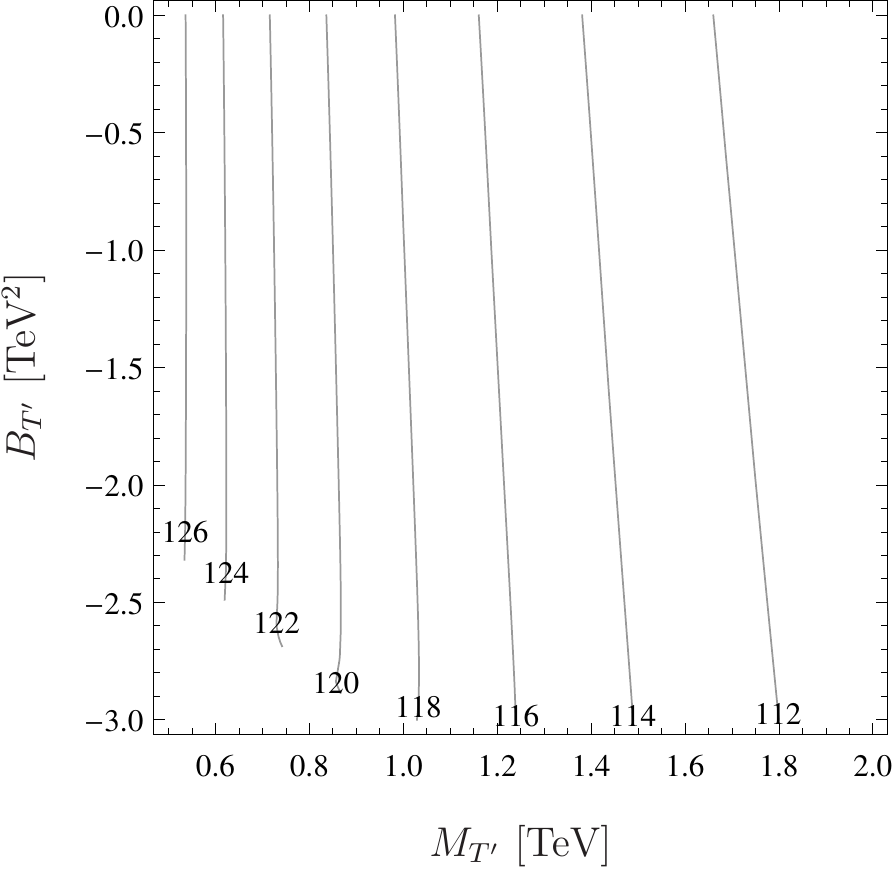}\hfill
 \includegraphics[width=0.33\linewidth]{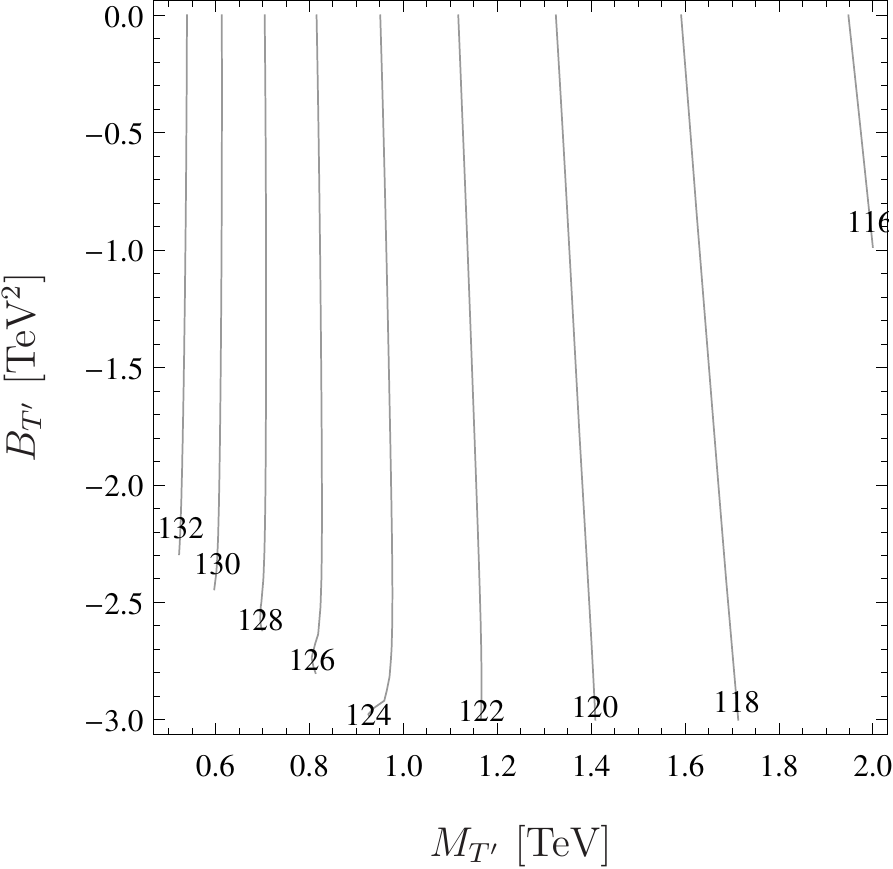}\hfill
 \includegraphics[width=0.33\linewidth]{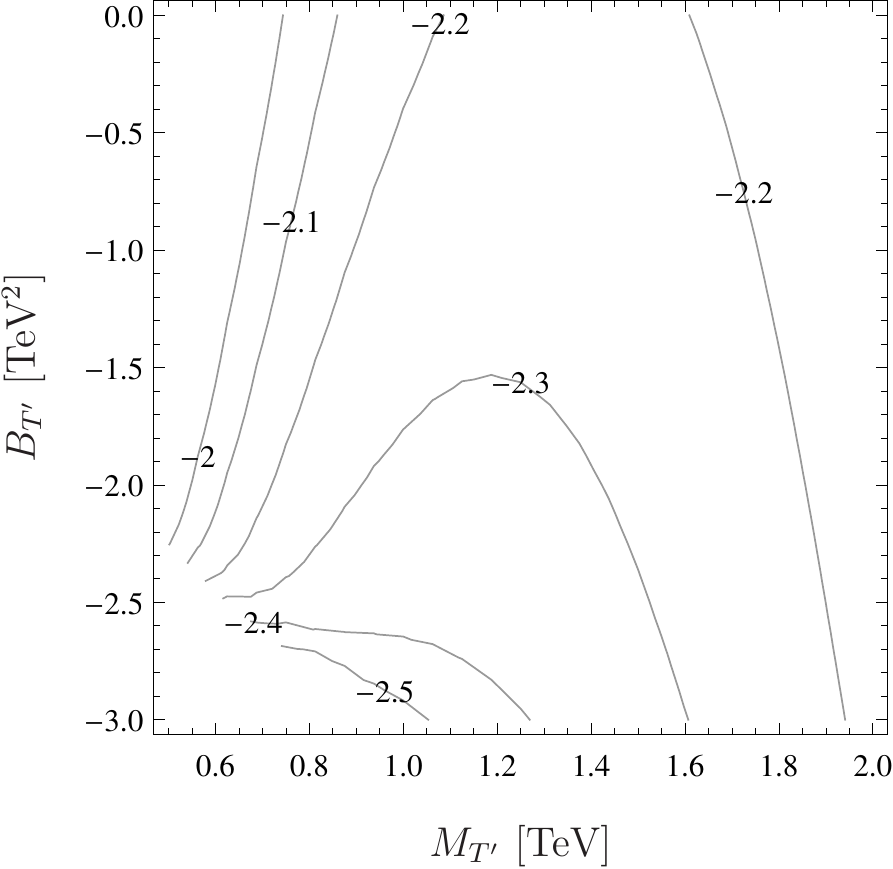}
 \caption{Contour lines of constant $m_h$ at one- (left)  and two-loop (middle) in the $(M_{T'},m_{\tilde{t}'})$ plane. The plots on the right column show the size of the two-loop corrections involving vectorlike states.  The plots in the first row are for $Y_{t'}=1.0$ with $T_{t'}=0$ and in the second for $Y_{t'}=0.7$ with $T_{t'}=1400~\text{GeV}$ .}
 \label{fig:MSMF}
\end{figure}
As a next step we want to understand the dependence of the loop corrections on the involved masses a bit more. We start with the dependence on the vectorlike mass parameter $M_{T'}$ and $B_{T'}$ and show in Fig.~\ref{fig:MSMF} the Higgs mass at the one- and two-loop level. At one-loop we have the well-known picture that the corrections quickly decrease with increasing mass of the vectorlike states, while the dependence on $B_{T'}$ is small and just shows up for smallish $M_{T'}$ of 1~TeV and smaller and large $|B_{T'}| > 2.0~\text{TeV}^2$ for $Y_{t'} = 1.0$ and $T_{t'}=0$. This general picture does, of course, not change at two-loop but we find a shift by several GeV usually dominated by the MSSM-like corrections. The two-loop corrections from the vectorlike states are singled out in the right column of Fig.~\ref{fig:MSMF}. They don't show this strong $M_{T'}$ dependence as the one-loop corrections do, and actually slightly increase with larger $M_{T'}$. Also the dependence on $B_{T'}$ is more pronounced at two-loop. If we go for smaller $Y_{t'}$ and turn on $T_{t'}$ the one-loop corrections in total become smaller and are less dependent on $B_{T'}$. However, the sensitivity at two-loop and $M_{T'}$ and $B_{T'}$ is nearly the same, but just the total size of the corrections decreases.

\begin{figure}[hbt]
 \includegraphics[width=0.45\linewidth]{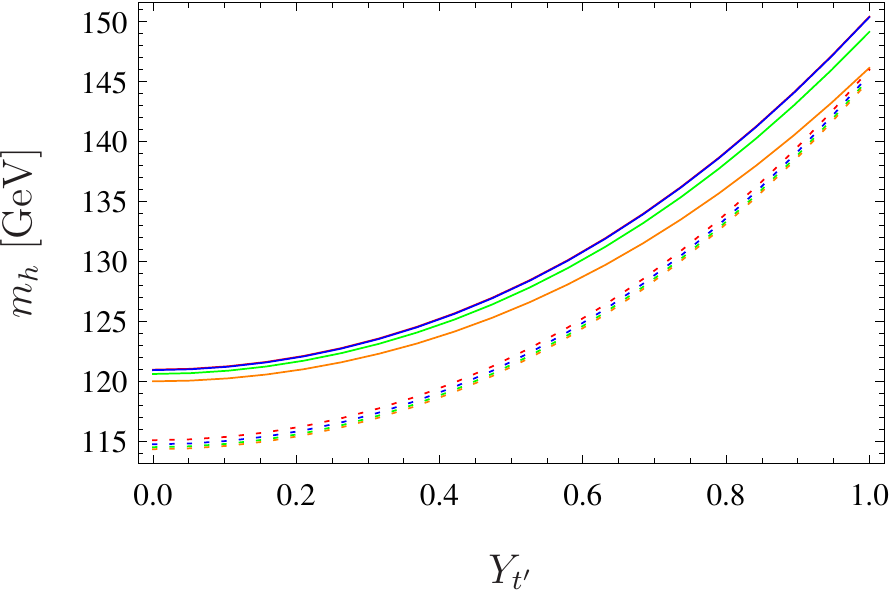}\hfill
  \includegraphics[width=0.45\linewidth]{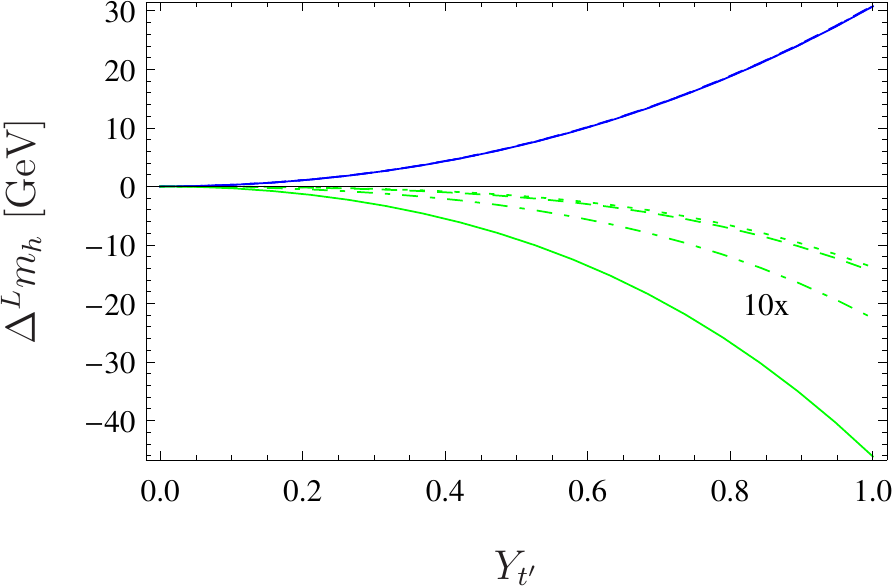}
 \caption{On the left: the light Higgs mass $m_h$ as function of $Y_{t'}$. Here, we used different values for $M_3$: 1~TeV (red), 2~TeV (blue), 3~TeV (green), 4~TeV (orange). The full lines are the two-loop results, the dotted ones the one-loop. On the right the absolute size of the one- (blue) and two-loop (green) corrections involving vectorlike states. The line coding is dashed, dotted, dot-dashed, full for increasing $M_3$.}
 \label{fig:M3}
\end{figure}

We have so far just concentrated on the dependence  of the Higgs mass corrections on the new parameters absent in the MSSM. We want to finalize our discussion of the loop corrections by also briefly commenting on the impact of at least two MSSM parameters: the gluino mass parameter $M_3$ and the soft-mass for the left-handed stop, $m_{q,33}$. We start with the dependence on the gluino mass shown in Fig.~\ref{fig:M3}. Here, we vary $Y_{t'}$ and use gluino masses between 1 and 4~TeV. At the one-loop level there is of course just a tiny impact on the Higgs mass. The small difference 
comes from SUSY threshold corrections. For $M_{T'}=1.5$~TeV and 3.0~TeV we find that with increasing $M_3$ the  two-loop corrections $O(\alpha_S \alpha_{t'})$ become larger. Since they are negative, the prediction for $m_h$ becomes smaller. However, for large $M_{T'}$ the dominance of the corrections $O(\alpha^2_{t'})$ is so large that this effect nearly doesn't play any role. 

\begin{figure}[hbt]
 \includegraphics[width=0.45\linewidth]{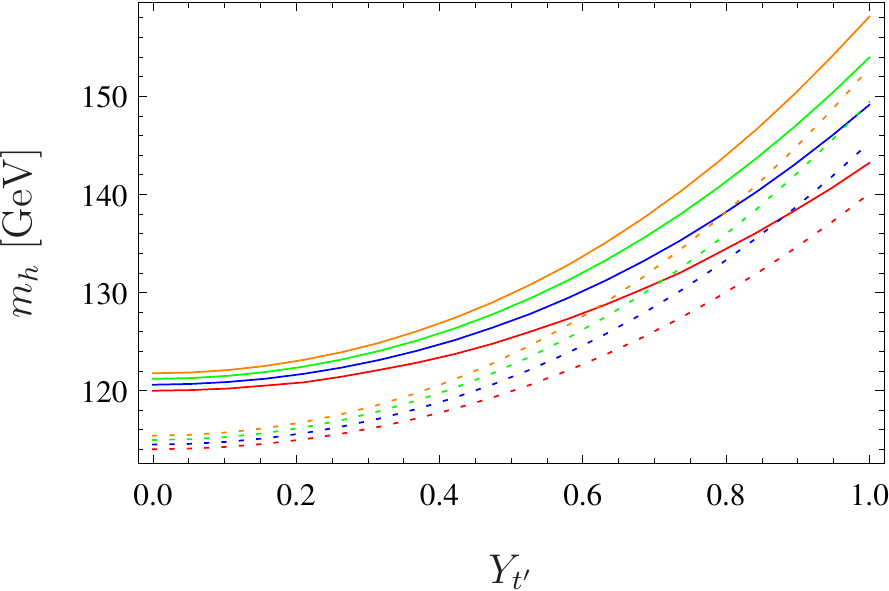}\hfill
 \includegraphics[width=0.45\linewidth]{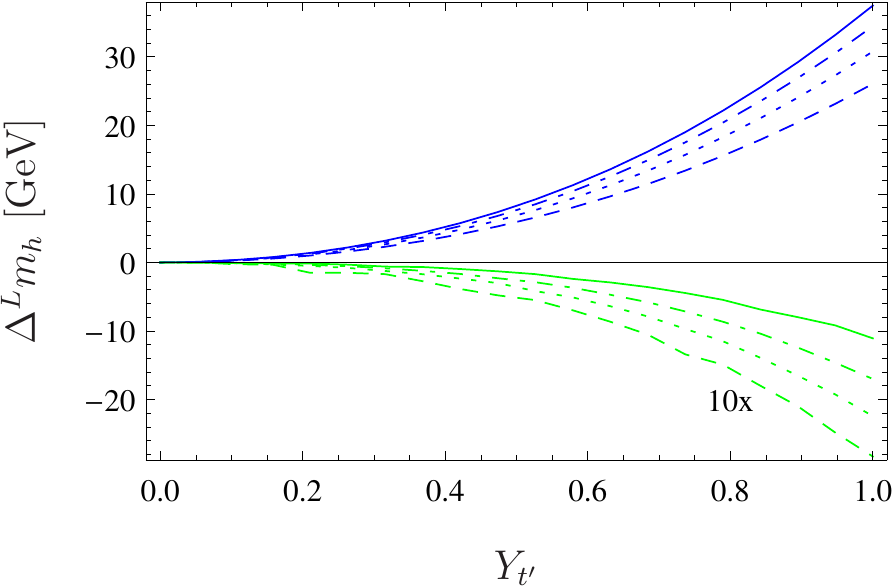}%\\
 \caption{On the left: the light Higgs mass $m_h$ as function of $Y_{t'}$. Here, we used different values for $m_{q,33}$: 1~TeV (red), 2~TeV (blue), 3~TeV (green), 4~TeV (orange). The full lines are the two-loop results, the dotted ones the one-loop. On the right the absolute size of the one- (blue) and two-loop (green) corrections involving vectorlike states is shown. The line coding is dashed, dotted, dot-dashed, full for increasing $m_{q,33}$.}
 \label{fig:mq33}
\end{figure}

Finally, we check the impact of the soft-masses for the left-handed stops. The one- and two-loop corrections as function of $Y_{t'}$ and $m_{\tilde q,33}=1,2,3,4$~TeV are summarized in Fig.~\ref{fig:mq33}. We see that this parameter plays an important role at one-and two-loop: at one-loop, the  corrections increase by a factor 2 when going from 1 to 4~TeV. At two-loop this effect is even more important and the corrections change by nearly a factor of 3. Interestingly, the one-loop corrections are larger for larger squark soft-terms, while the two-loop corrections increase with decreasing squark masses.

\section{Results -- Part II: The fine-tuning in Gauge mediated SUSY breaking}
\label{sec:resultsII}

\begin{figure}[hbt]
 \includegraphics[width=0.4\linewidth]{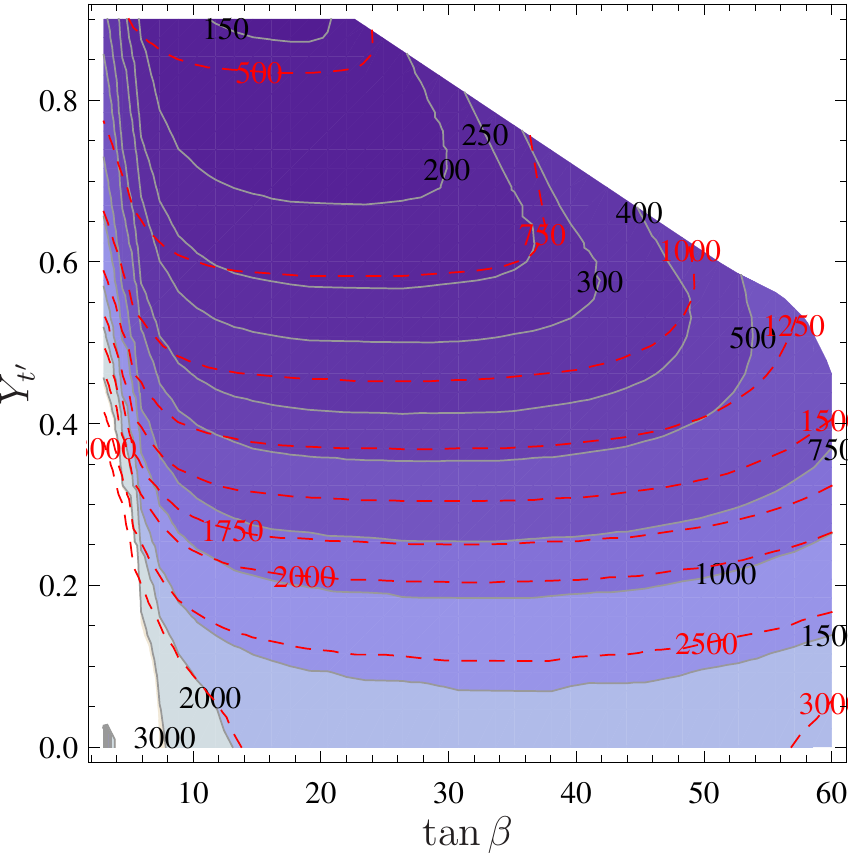}
\hfill
  \includegraphics[width=0.4\linewidth]{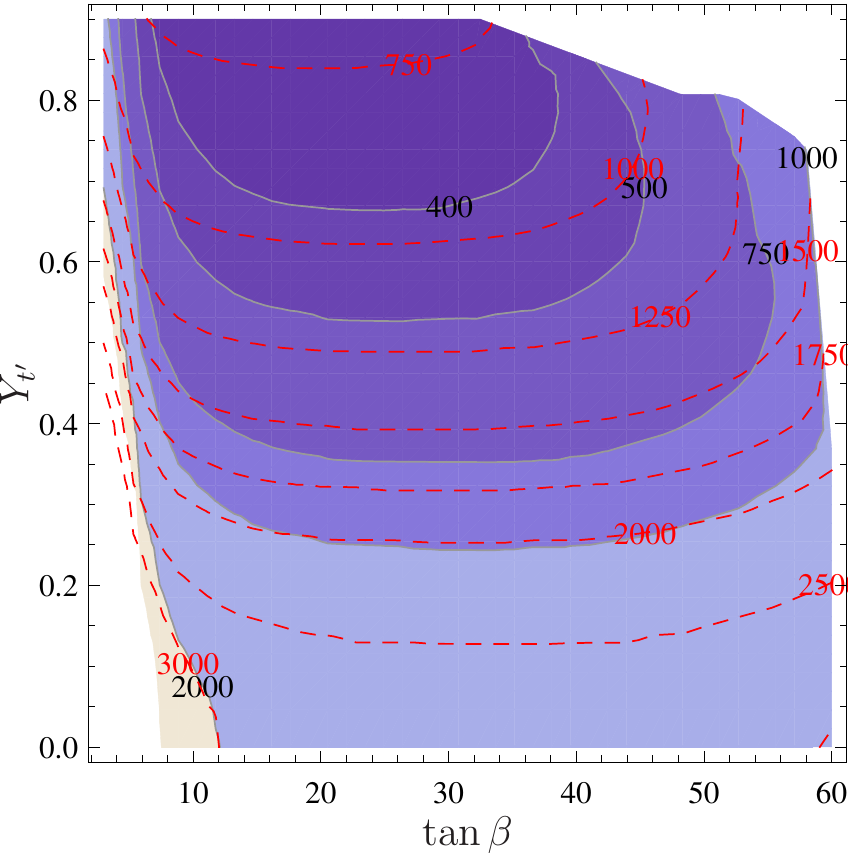} \\
 \includegraphics[width=0.4\linewidth]{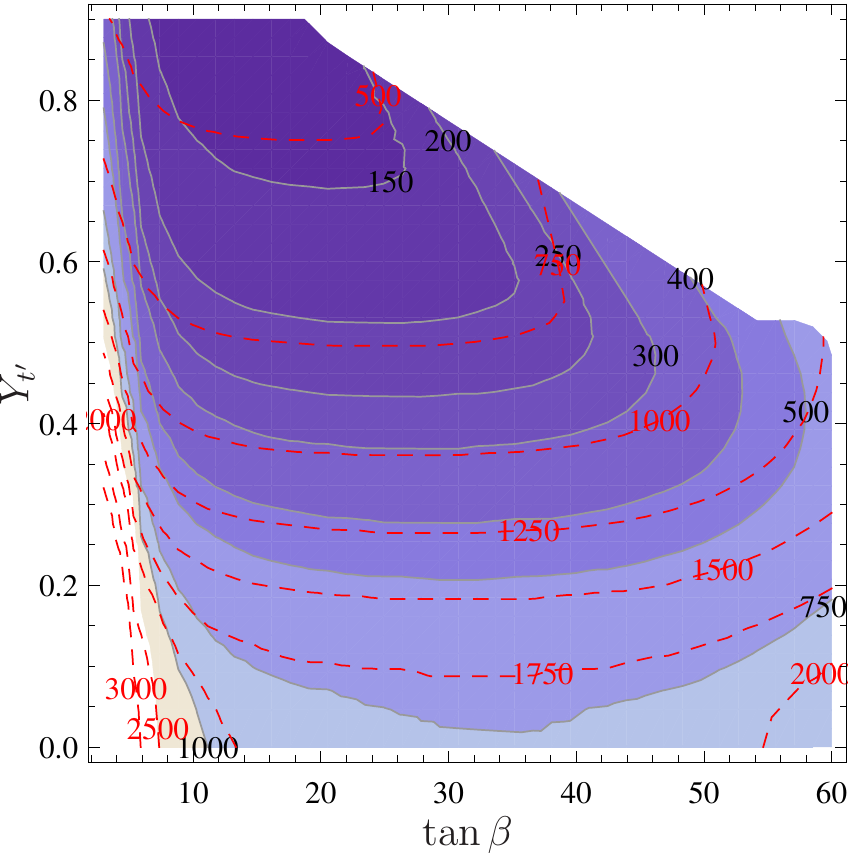} \hfill
  \includegraphics[width=0.4\linewidth]{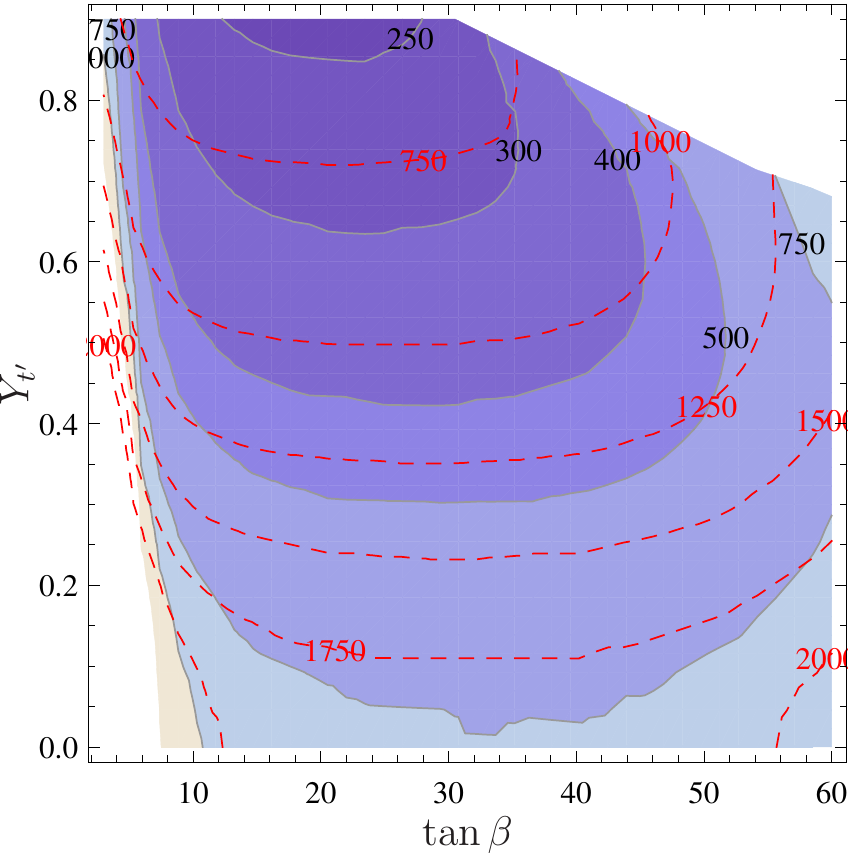} \\
   \includegraphics[width=0.4\linewidth]{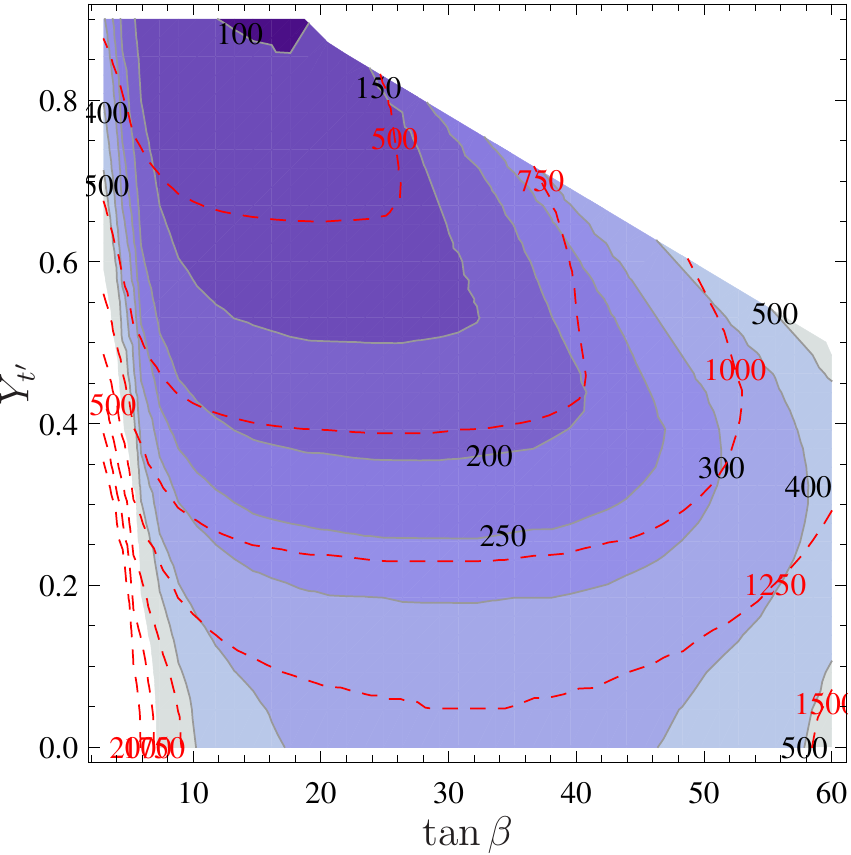} \hfill
  \includegraphics[width=0.4\linewidth]{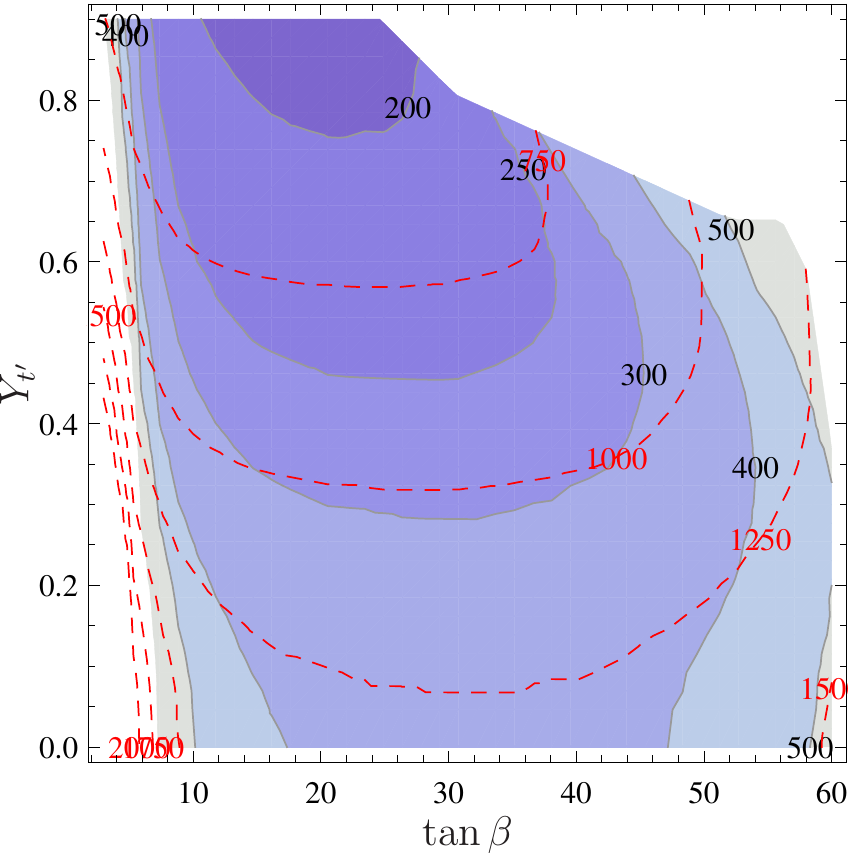} 
  \caption{Contours of overall fine-tuning $\Delta$ in the $(\tan\beta, Y_{t'})$-plane demanding a Higgs mass $m_h = 128$~GeV (top), $m_h = 125$~GeV (middle), and $m_h = 122$~GeV (bottom) for the UV complete variant of the model. We fixed here $M=10^7$~GeV and $M_{V'} = 0.5$~TeV (left column), respectively, $M_{V'} = 1.0$~TeV (right column). The red dashed lines indicate the gluino mass in GeV.}  
\label{fig:FTUVM7} 
\end{figure}

We now turn to the consequence of the loop corrections for the fine-tuning in minimal GMSB. The intrinsic problem of minimal GMSB in the MSSM is that it predicts very small trilinear couplings. Thus, the only chance to enhance the Higgs mass via loop corrections is to go to very large values of $\Lambda$ and $M$ to get sufficiently heavy stops. When calculating the fine-tuning for this setup and demanding $m_h \simeq 125$~GeV, one finds that the fine-tuning $\Delta$ is well above 1000. Of course, in the presence of large  loop corrections due to vectorlike states, the need of superheavy stops is relaxed and the fine-tuning is expected to improve. We show in Fig.~\ref{fig:FTUVM7} the fine-tuning in the $(\tan\beta, Y_{t'})$ plane for different constraints for the Higgs mass within the theoretical uncertainty: (i) $m_h = 122$~GeV, (ii) $m_h = 125$~GeV, (iii) $m_h = 128$~GeV. For the vectorlike states, masses of 500 and 1000~GeV were used  at the messenger scale. 

One finds that the fine-tuning quickly drops with increasing $Y_{t'}$ because lighter SUSY states are sufficient to push the Higgs mass to the desired level. For very large $Y_{t'}$ of $O(1)$ and the looser constraint of $m_h > 122$~GeV, even a fine-tuning of about 100 seems possible. 
There is also another, very interesting observation: even for $Y_{t'} = 0$ the fine-tuning in this model is not as bad as one expects it from the MSSM. The reason is that the strong interaction at the messenger scale is larger compared to MSSM expectations because of the different running. Therefore, for the same value of $\Lambda$, the squarks are already significantly heavier and lead to larger Higgs mass corrections. \\

\begin{figure}[hbt]
\centering
 \includegraphics[width=0.6\linewidth]{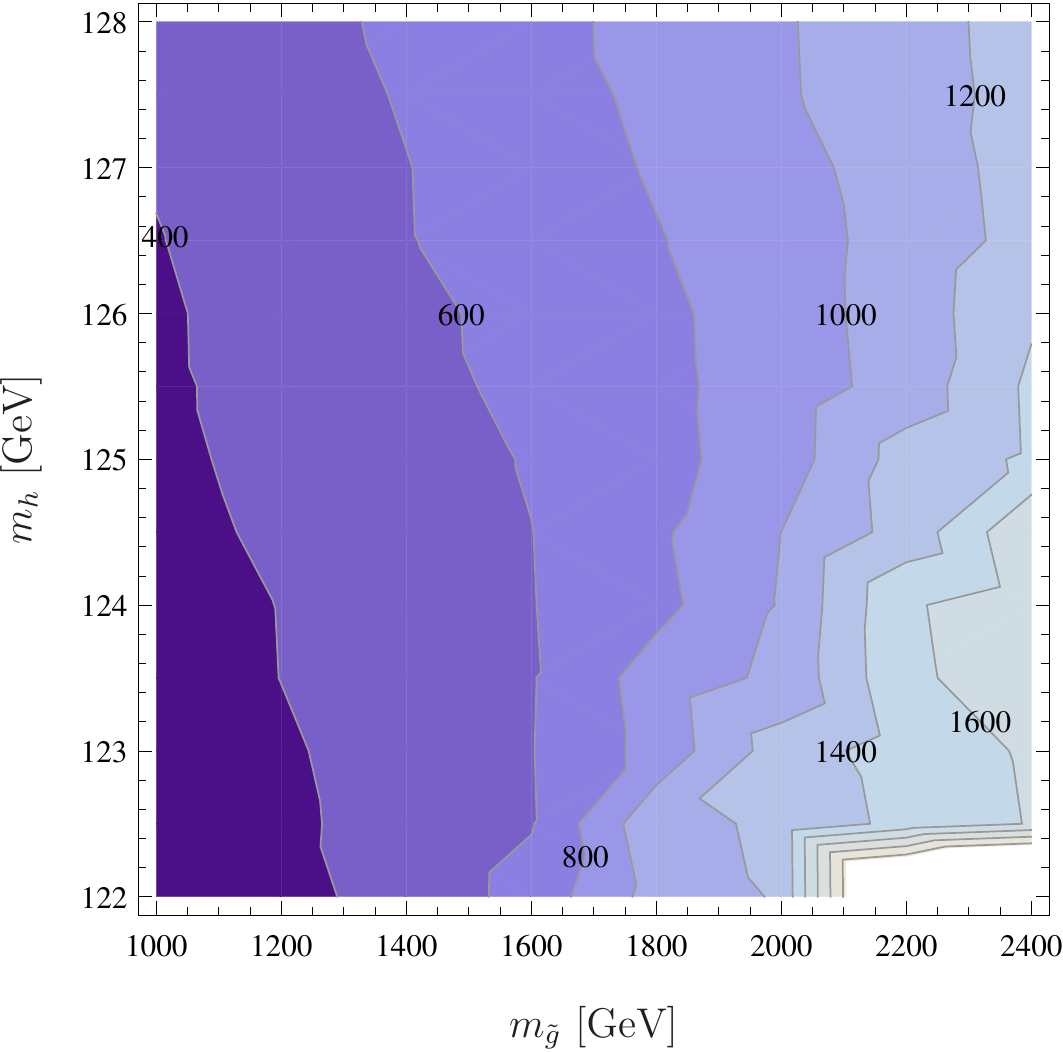} 
\caption{Minimal fine-tuning for given Higgs mass $m_h$ and gluino mass $m_{\tilde g}$. We fixed here $M=10^7$~GeV and $M_{T'}=1$~TeV and scanned over $\tan\beta$, $Y_{t'}$ and $\Lambda$.}  
\label{fig:mhmglu} 
\end{figure}

However, including the bounds from direct SUSY searches has a large impact: the points with a small fine-tuning are excluded because of the light gluino mass. That's completely different to the GMSB variant of the MSSM where the Higgs mass pushes the fine-tuning of the model to higher values. In this model, the vanishing trilinear couplings at the messenger scale just play a subdominant role concerning the fine-tuning, but the gluino mass demands a larger SUSY scale $\Lambda$, which increases the fine-tuning.  The situation wouldn't change if we go to larger Messenger masses to increase the running because the one-loop $\beta$-function of $M_3$ vanishes in this model and the mass is actually slightly decreasing with increasing $M$. Moreover, it's a general feature of GMSB that the gaugino masses are not very sensitive to the messenger scale because the leading dependence in the RGE running always drops out. The running gaugino mass at the SUSY scale is related to the one at the messenger scale by the ratio of the corresponding gauge coupling at both scales:
\begin{equation}
M_i(Q) = M_i(M) \frac{g^2_i(Q)}{g^2_i(M)} = g^2_i(Q) \Lambda_G 
\end{equation}
% We find that for both constraints on the Higgs mass the fine-tuning depends as follows on the gluino mass, when setting $M_{V'} = 1$~TeV.:
% \begin{align}
%  m_{\tilde g} > 1000~\text{GeV} \quad & \to \quad \Delta \simeq (275, 360, 430) \\
%  m_{\tilde g} > 1200~\text{GeV}\quad  & \to \quad \Delta \simeq (350, 440, 525) \\
%  m_{\tilde g} > 1400~\text{GeV} \quad & \to \quad \Delta \simeq (460, 520, 650) \\
%  m_{\tilde g} > 1600~\text{GeV} \quad & \to\quad  \Delta \simeq (670, 610, 725)
% \end{align}
% The first number in brackets refers to $m_h = 122$~GeV, the second to $m_h = 125$~GeV and the third one to $m_h = 128$~GeV. 
We show the minimal fine-tuning in the $(m_{\tilde g}, m_h)$ plane in Fig.~\ref{fig:mhmglu}.
It is interesting that the fine-tuning for $m_h = 125$~GeV can be smaller than for $m_h = 122$~GeV  and $m_h = 128$~GeV when the gluino mass is sufficiently large.

\begin{figure}[hbt]
 \includegraphics[width=0.45\linewidth]{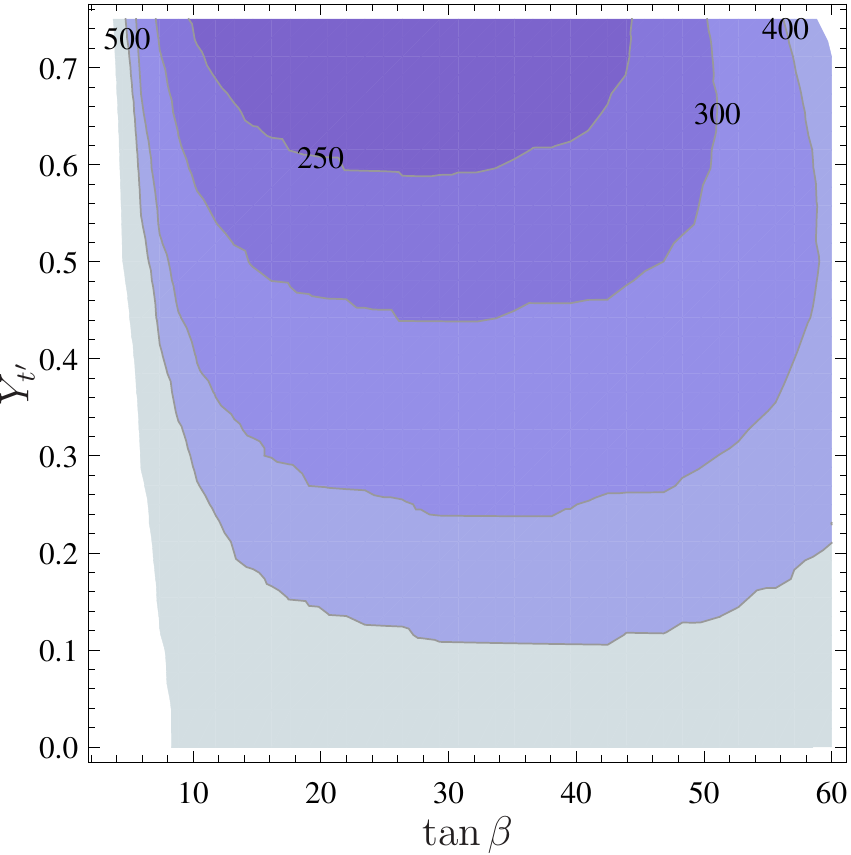} \hfill
  \includegraphics[width=0.45\linewidth]{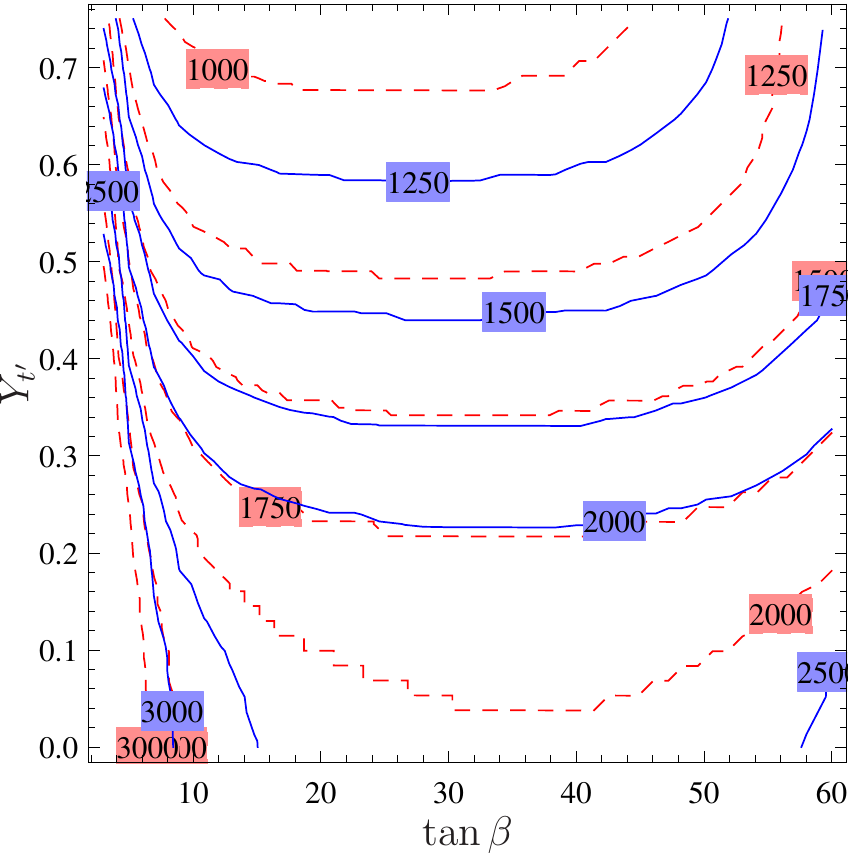}
\caption{Contours of the overall fine-tuning $\Delta$ (left) and the mass of the lightest up-squark (right, full blue lines) and gluino (right, dashed red lines) in the $(\tan\beta, Y_{t'})$-plane demanding a Higgs mass $m_h > 122$~GeV for the variant of the model without spectator fields. We fixed here $M=10^7$~GeV.}  
\label{fig:FTnoUVM7} 
\end{figure}

\begin{figure}[hbt]
 \includegraphics[width=0.45\linewidth]{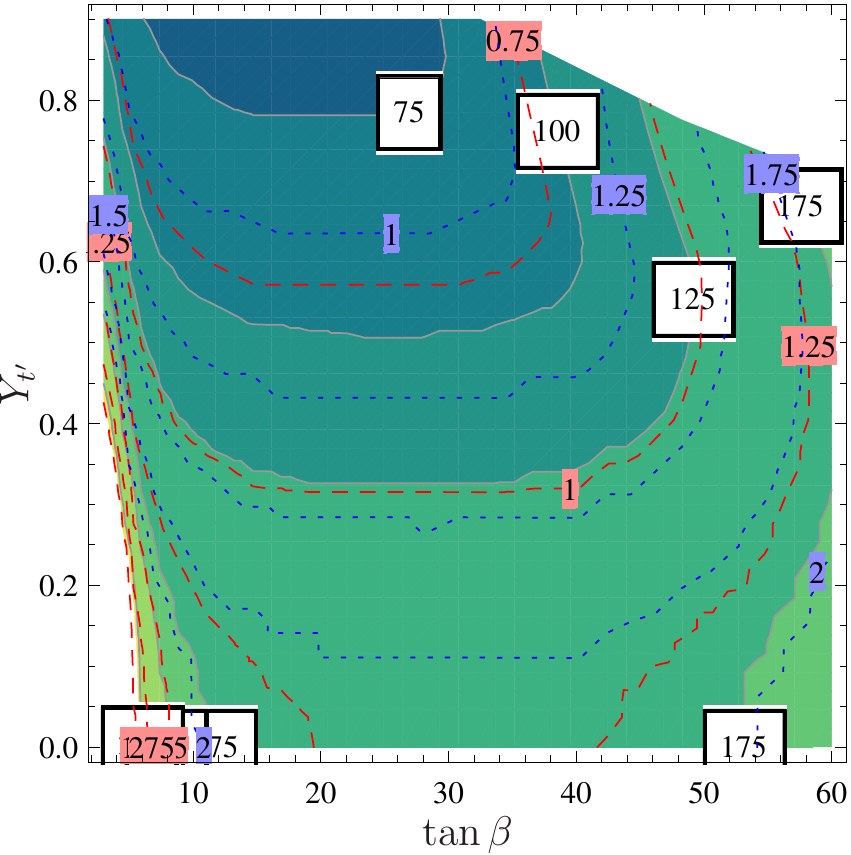} \hfill
  \includegraphics[width=0.45\linewidth]{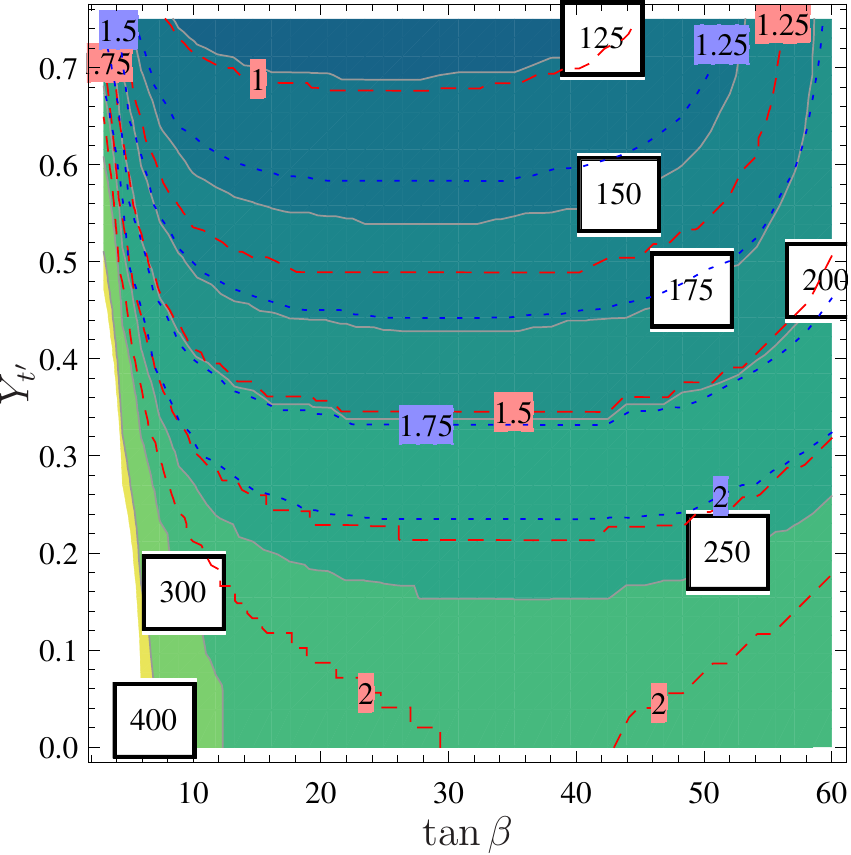}
\caption{Contours of constant $\Lambda$ (black),  the lightest top-squark mass (right, full blue lines) and gluino mass (right, dashed red lines) in the $(\tan\beta, Y_{t'})$-plane demanding a Higgs mass $m_h > 122$~GeV. All contours are given in units of TeV. On the left for the UV complete model, on the right for the model with only vectorlike tops.  We fixed here $M=10^7$~GeV.}  
\label{fig:FTLam} 
\end{figure}

For very large $Y_{t'}$ where the FT becomes the best, the theory is not perturbative up to the GUT scale. Since there is a cut-off anyway in the theory, there is no real need to maintain gauge coupling unification by adding the spectator fields at the SUSY scale. Therefore, one might wonder what the FT of the minimal model is. This is depicted in Fig.~\ref{fig:FTnoUVM7}. In this setup, the squarks are lighter for the same values of $M$ and $\Lambda$ because of the smaller strong coupling at the messenger scale. Thus, in general larger $\Lambda$ is needed to increase the Higgs mass. This leads also to larger gluino masses. This is shown in Fig.~\ref{fig:FTLam} where we compare the minimal value of $\Lambda$ to get a Higgs mass larger than 122~GeV in the $(\tan\beta,Y_{t'})$ plane for a messenger scale of again $10^7$~GeV, and the resulting stop and gluino masses triggered by this $\Lambda$. We find for the minimal model the following fine-tuning 
\begin{equation}
\Delta \simeq (230, 275, 320, 380)
\end{equation}
for $m_{\tilde g} > (1000,1200,1400,1600)$~GeV and $m_h > 122$~GeV. 
\section{Conclusion}
\label{sec:conclusion}
We discussed the loop corrections to the light Higgs mass in the MSSM extended by a pair of vectorlike top quarks. We have improved previous calculations in literature in three respects: (i) we included the additional threshold corrections from the vectorlike states to SM gauge and Yukawa couplings, (ii) we added the full momentum dependence at the one-loop level, (iii) we calculated all dominant (i.e. excluding electroweak) two-loop corrections in the effective potential approach. It has been shown that the momentum effects can be sizeable and change the Higgs mass prediction by a few GeV. The effect from the threshold corrections turns out to be often more important. The importance of the two-loop corrections strongly depends on the considered parameter point. They are often a bit smaller than the two-loop corrections known from the MSSM, but we also identified regions where they can be even larger. In these regions, the  additional two-loop corrections can change the Higgs mass prediction by up to 10 GeV. We checked the impact of the presence of vectorlike states on the fine-tuning in GMSB. For this purpose, we extended the model by additional vectorlike quarks and leptons to have complete multiplets of $SU(5)$. We found that the fine-tuning can be reduced significantly compared to minimal GMSB with only the MSSM particle content.  Often, those regions with the best fine-tuning which are in agreement with the Higgs mass measurement are ruled out by gluino searches. Interestingly, we find that for heavy gluino masses the fine-tuning for heavier Higgs masses can be even better. In particular, for $m_{\tilde g} \simeq 1400$~GeV, the best fine-tuning is found for a Higgs mass of roughly 125~GeV.

\section*{Acknowledgements}
We thank Mark D. Goodsell for a fruitful collaboration to automatize the two-loop calculations with \SARAH and \SPheno and many interesting discussions in this context. This has been crucial to facilitate this project. 
\begin{appendix}
\allowdisplaybreaks
\section{Vertices}
\subsection{Vector boson Vertices with vectorlike (s)tops}
\label{app:VBvertices}
\begin{align}
\Gamma_{\tilde{u}_{{i \alpha}}\tilde{d}^*_{{j \beta}}W^-_{{\mu}}}  = & \, 
-i \frac{1}{\sqrt{2}} g_2 \delta_{\alpha \beta} \sum_{a=1}^{3}Z^{U,*}_{i a} Z_{{j a}}^{D}   
\\ 
\Gamma_{\tilde{u}_{{i \alpha}}\tilde{u}^*_{{j \beta}}Z_{{\mu}}}  = & \, 
-\frac{i}{6} \delta_{\alpha \beta} \Big(\Big(3 g_2 \cos\Theta_W   - g_1 \sin\Theta_W  \Big)\sum_{a=1}^{3}Z^{U,*}_{i a} Z_{{j a}}^{U}  \nonumber \\ 
 &-4 g_1 \sin\Theta_W  \Big(Z^{U,*}_{i 7} Z_{{j 7}}^{U}  + Z^{U,*}_{i 8} Z_{{j 8}}^{U}  + \sum_{a=1}^{3}Z^{U,*}_{i 3 + a} Z_{{j 3 + a}}^{U} \Big)\Big) 
\\
\Gamma^L_{\bar{d_{{i \alpha}}}u_{{j \beta}}W^-_{{\mu}}}  =  & \,
-i \frac{1}{\sqrt{2}} g_2 \delta_{\alpha \beta} \sum_{a=1}^{3}U^{u,*}_{L,{j a}} U_{L,{i a}}^{d}  \\ 
 \Gamma^R_{\bar{d_{{i \alpha}}}u_{{j \beta}}W^-_{{\mu}}}  =  & \,0 
\\ 
\Gamma^L_{\bar{u_{{i \alpha}}}u_{{j \beta}}Z_{{\mu}}}  =  & \,
-\frac{i}{6} \delta_{\alpha \beta} \Big(\Big(3 g_2 \cos\Theta_W   - g_1 \sin\Theta_W  \Big)\sum_{a=1}^{3}U^{u,*}_{L,{j a}} U_{L,{i a}}^{u}   -4 g_1 U^{u,*}_{L,{j 4}} \sin\Theta_W  U_{L,{i 4}}^{u} \Big)\\ 
 \Gamma^R_{\bar{u_{{i \alpha}}}u_{{j \beta}}Z_{{\mu}}}  =  & \,\frac{2 i}{3} g_1 \delta_{\alpha \beta} \sin\Theta_W  \Big(U^{u,*}_{R,{i 4}} U_{R,{j 4}}^{u}  + \sum_{a=1}^{3}U^{u,*}_{R,{i a}} U_{R,{j a}}^{u} \Big) 
\\ 
\end{align}

\subsection{Quark vertices involving vectorlike (s)tops}
\label{app:QuarkVertices}
\begin{align}
\Gamma^L_{\bar{u_{{i \alpha}}}d_{{j \beta}}W^+_{{\mu}}}  =  & \,
-i \frac{1}{\sqrt{2}} g_2 \delta_{\alpha \beta} \sum_{a=1}^{3}U^{d,*}_{L,{j a}} U_{L,{i a}}^{u}  \\ 
 \Gamma^R_{\bar{u_{{i \alpha}}}d_{{j \beta}}W^+_{{\mu}}}  =  & \,0 
\\
\Gamma^L_{\bar{u_{{i \alpha}}}u_{{j \beta}}g_{{\gamma \mu}}}  =  & \,
-\frac{i}{2} g_3 \lambda^{\gamma}_{\alpha,\beta} \Big(U^{u,*}_{L,{j 4}} U_{L,{i 4}}^{u}  + \sum_{a=1}^{3}U^{u,*}_{L,{j a}} U_{L,{i a}}^{u} \Big)\\ 
 \Gamma^R_{\bar{u_{{i \alpha}}}u_{{j \beta}}g_{{\gamma \mu}}}  =  & \,-\frac{i}{2} g_3 \lambda^{\gamma}_{\alpha,\beta} \Big(U^{u,*}_{R,{i 4}} U_{R,{j 4}}^{u}  + \sum_{a=1}^{3}U^{u,*}_{R,{i a}} U_{R,{j a}}^{u} \Big) 
\\ 
\Gamma^L_{\bar{u_{{i \alpha}}}u_{{j \beta}}\gamma_{{\mu}}}  =  & \,
-\frac{i}{6} \delta_{\alpha \beta} \Big(\Big(3 g_2 \sin\Theta_W   + g_1 \cos\Theta_W  \Big)\sum_{a=1}^{3}U^{u,*}_{L,{j a}} U_{L,{i a}}^{u}   + 4 g_1 U^{u,*}_{L,{j 4}} \cos\Theta_W  U_{L,{i 4}}^{u} \Big)\\ 
 \Gamma^R_{\bar{u_{{i \alpha}}}u_{{j \beta}}\gamma_{{\mu}}}  =  & \,-\frac{2 i}{3} g_1 \cos\Theta_W  \delta_{\alpha \beta} \Big(U^{u,*}_{R,{i 4}} U_{R,{j 4}}^{u}  + \sum_{a=1}^{3}U^{u,*}_{R,{i a}} U_{R,{j a}}^{u} \Big) 
\\ 
\Gamma^L_{\bar{u_{{i \alpha}}}u_{{j \beta}}Z_{{\mu}}}  =  & \,
-\frac{i}{6} \delta_{\alpha \beta} \Big(\Big(3 g_2 \cos\Theta_W   - g_1 \sin\Theta_W  \Big)\sum_{a=1}^{3}U^{u,*}_{L,{j a}} U_{L,{i a}}^{u}   -4 g_1 U^{u,*}_{L,{j 4}} \sin\Theta_W  U_{L,{i 4}}^{u} \Big)\\ 
 \Gamma^R_{\bar{u_{{i \alpha}}}u_{{j \beta}}Z_{{\mu}}}  =  & \,\frac{2 i}{3} g_1 \delta_{\alpha \beta} \sin\Theta_W  \Big(U^{u,*}_{R,{i 4}} U_{R,{j 4}}^{u}  + \sum_{a=1}^{3}U^{u,*}_{R,{i a}} U_{R,{j a}}^{u} \Big) 
\\ 
\Gamma^L_{\bar{u_{{i \alpha}}}u_{{j \beta}}A^0_{{k}}}  =  & \,
\frac{1}{\sqrt{2}} \delta_{\alpha \beta} \Big(U^{u,*}_{R,{i 4}} \sum_{a=1}^{3}U^{u,*}_{L,{j a}} Y_{t',{a}}   + \sum_{b=1}^{3}U^{u,*}_{L,{j b}} \sum_{a=1}^{3}U^{u,*}_{R,{i a}} Y_{u,{a b}}  \Big)Z_{{k 2}}^{A} \\ 
 \Gamma^R_{\bar{u_{{i \alpha}}}u_{{j \beta}}A^0_{{k}}}  =  & \,- \frac{1}{\sqrt{2}} \delta_{\alpha \beta} Z_{{k 2}}^{A} \Big(\sum_{a=1}^{3}Y^*_{{t'},{a}} U_{L,{i a}}^{u}  U_{R,{j 4}}^{u}  + \sum_{b=1}^{3}\sum_{a=1}^{3}Y^*_{u,{a b}} U_{R,{j a}}^{u}  U_{L,{i b}}^{u} \Big) 
\\ 
\Gamma^L_{\bar{d_{{i \alpha}}}\tilde{\chi}^-_{{j}}\tilde{u}_{{k \gamma}}}  =  & \,
i U^*_{j 2} \delta_{\alpha \gamma} \sum_{b=1}^{3}Z^{U,*}_{k b} \sum_{a=1}^{3}U^{d,*}_{R,{i a}} Y_{d,{a b}}   \\ 
 \Gamma^R_{\bar{d_{{i \alpha}}}\tilde{\chi}^-_{{j}}\tilde{u}_{{k \gamma}}}  =  & \,-i \delta_{\alpha \gamma} \Big(g_2 \sum_{a=1}^{3}Z^{U,*}_{k a} U_{L,{i a}}^{d}  V_{{j 1}}  - \Big(Z^{U,*}_{k 7} \sum_{a=1}^{3}Y^*_{{t'},{a}} U_{L,{i a}}^{d}   + \sum_{b=1}^{3}\sum_{a=1}^{3}Y^*_{u,{a b}} Z^{U,*}_{k 3 + a}  U_{L,{i b}}^{d} \Big)V_{{j 2}} \Big) 
\\
\Gamma^L_{\tilde{\chi}^0_{{i}}u_{{j \beta}}\tilde{u}^*_{{k \gamma}}}  =  & \,
-\frac{i}{6} \delta_{\beta \gamma} \Big(3 \sqrt{2} g_2 N^*_{i 2} \sum_{a=1}^{3}U^{u,*}_{L,{j a}} Z_{{k a}}^{U}  +6 N^*_{i 4} \Big(\sum_{a=1}^{3}U^{u,*}_{L,{j a}} Y_{t',{a}}  Z_{{k 7}}^{U}  + \sum_{b=1}^{3}U^{u,*}_{L,{j b}} \sum_{a=1}^{3}Y_{u,{a b}} Z_{{k 3 + a}}^{U}  \Big)\nonumber \\ 
 &+\sqrt{2} g_1 N^*_{i 1} \Big(4 U^{u,*}_{L,{j 4}} Z_{{k 8}}^{U}  + \sum_{a=1}^{3}U^{u,*}_{L,{j a}} Z_{{k a}}^{U} \Big)\Big)\\ 
 \Gamma^R_{\tilde{\chi}^0_{{i}}u_{{j \beta}}\tilde{u}^*_{{k \gamma}}}  =  & \,\frac{i}{3} \delta_{\beta \gamma} \Big(2 \sqrt{2} g_1 \sum_{a=1}^{3}Z_{{k 3 + a}}^{U} U_{R,{j a}}^{u}  N_{{i 1}} -3 \sum_{b=1}^{3}\sum_{a=1}^{3}Y^*_{u,{a b}} U_{R,{j a}}^{u}  Z_{{k b}}^{U}  N_{{i 4}} \nonumber \\ 
 &+\Big(2 \sqrt{2} g_1 N_{{i 1}} Z_{{k 7}}^{U}  -3 \sum_{a=1}^{3}Y^*_{{t'},{a}} Z_{{k a}}^{U}  N_{{i 4}} \Big)U_{R,{j 4}}^{u} \Big) 
\\
\Gamma^L_{\bar{u_{{i \alpha}}}d_{{j \beta}}H^+_{{k}}}  =  & \,
i \delta_{\alpha \beta} \Big(U^{u,*}_{R,{i 4}} \sum_{a=1}^{3}U^{d,*}_{L,{j a}} Y_{t',{a}}   + \sum_{b=1}^{3}U^{d,*}_{L,{j b}} \sum_{a=1}^{3}U^{u,*}_{R,{i a}} Y_{u,{a b}}  \Big)Z_{{k 2}}^{+} \\ 
 \Gamma^R_{\bar{u_{{i \alpha}}}d_{{j \beta}}H^+_{{k}}}  =  & \,i \delta_{\alpha \beta} \sum_{b=1}^{3}\sum_{a=1}^{3}Y^*_{d,{a b}} U_{R,{j a}}^{d}  U_{L,{i b}}^{u}  Z_{{k 1}}^{+}  
\\ 
\Gamma^L_{\tilde{g}_{{\alpha}}u_{{j \beta}}\tilde{u}^*_{{k \gamma}}}  =  & \,
-i \frac{1}{\sqrt{2}} g_3 \phi_{\tilde{g}} \lambda^{\alpha}_{\gamma,\beta} \Big(U^{u,*}_{L,{j 4}} Z_{{k 8}}^{U}  + \sum_{a=1}^{3}U^{u,*}_{L,{j a}} Z_{{k a}}^{U} \Big)\\ 
 \Gamma^R_{\tilde{g}_{{\alpha}}u_{{j \beta}}\tilde{u}^*_{{k \gamma}}}  =  & \,i \frac{1}{\sqrt{2}} g_3 \phi_{\tilde{g}}^* \lambda^{\alpha}_{\gamma,\beta} \Big(Z_{{k 7}}^{U} U_{R,{j 4}}^{u}  + \sum_{a=1}^{3}Z_{{k 3 + a}}^{U} U_{R,{j a}}^{u} \Big) 
\\ 
\Gamma^L_{\bar{u_{{i \alpha}}}u_{{j \beta}}h_{{k}}}  =  & \,
-i \frac{1}{\sqrt{2}} \delta_{\alpha \beta} \Big(U^{u,*}_{R,{i 4}} \sum_{a=1}^{3}U^{u,*}_{L,{j a}} Y_{t',{a}}   + \sum_{b=1}^{3}U^{u,*}_{L,{j b}} \sum_{a=1}^{3}U^{u,*}_{R,{i a}} Y_{u,{a b}}  \Big)Z_{{k 2}}^{H} \\ 
 \Gamma^R_{\bar{u_{{i \alpha}}}u_{{j \beta}}h_{{k}}}  =  & \,-i \frac{1}{\sqrt{2}} \delta_{\alpha \beta} Z_{{k 2}}^{H} \Big(\sum_{a=1}^{3}Y^*_{{t'},{a}} U_{L,{i a}}^{u}  U_{R,{j 4}}^{u}  + \sum_{b=1}^{3}\sum_{a=1}^{3}Y^*_{u,{a b}} U_{R,{j a}}^{u}  U_{L,{i b}}^{u} \Big) 
\end{align}

\subsection{Higgs Vertices with vectorlike (s)tops}
\label{app:HiggsVertices}
\begin{align}
\Gamma^L_{\bar{u_{{i \alpha}}}u_{{j \beta}}h_{{k}}}  =  & \,
-i \frac{1}{\sqrt{2}} \delta_{\alpha \beta} \Big(U^{u,*}_{R,{i 4}} \sum_{a=1}^{3}U^{u,*}_{L,{j a}} Y_{t',{a}}   + \sum_{b=1}^{3}U^{u,*}_{L,{j b}} \sum_{a=1}^{3}U^{u,*}_{R,{i a}} Y_{u,{a b}}  \Big)Z_{{k 2}}^{H} \\ 
 \Gamma^R_{\bar{u_{{i \alpha}}}u_{{j \beta}}h_{{k}}}  =  & \,-i \frac{1}{\sqrt{2}} \delta_{\alpha \beta} Z_{{k 2}}^{H} \Big(\sum_{a=1}^{3}Y^*_{{t'},{a}} U_{L,{i a}}^{u}  U_{R,{j 4}}^{u}  + \sum_{b=1}^{3}\sum_{a=1}^{3}Y^*_{u,{a b}} U_{R,{j a}}^{u}  U_{L,{i b}}^{u} \Big) 
\\ 
\Gamma_{h_{{i}}\tilde{u}_{{j \beta}}\tilde{u}^*_{{k \gamma}}}  = & \, 
\frac{i}{12} \delta_{\beta \gamma} \Big(\Big(-3 g_{2}^{2}  + g_{1}^{2}\Big)\sum_{a=1}^{3}Z^{U,*}_{j a} Z_{{k a}}^{U}  \Big(v_d Z_{{i 1}}^{H}  - v_u Z_{{i 2}}^{H} \Big)\nonumber \\ 
 &+2 Z^{U,*}_{j 7} \Big(3 \sqrt{2} \mu \sum_{a=1}^{3}Y^*_{{t'},{a}} Z_{{k a}}^{U}  Z_{{i 1}}^{H} -3 \sqrt{2} \sum_{a=1}^{3}T^*_{{t'},{a}} Z_{{k a}}^{U}  Z_{{i 2}}^{H} -6 v_u \sum_{b=1}^{3}\sum_{a=1}^{3}Y^*_{{t'},{a}} Y_{u,{b a}}  Z_{{k 3 + b}}^{U}  Z_{{i 2}}^{H} \nonumber \\ 
 &-2 g_{1}^{2} v_d Z_{{i 1}}^{H} Z_{{k 7}}^{U} +2 g_{1}^{2} v_u Z_{{i 2}}^{H} Z_{{k 7}}^{U} -6 v_u \sum_{a=1}^{3}|Y_{t',{a}}|^2 Z_{{i 2}}^{H} Z_{{k 7}}^{U} \Big)\nonumber \\ 
 &-2 \Big(-3 \sqrt{2} \mu \sum_{b=1}^{3}\sum_{a=1}^{3}Y^*_{u,{a b}} Z^{U,*}_{j 3 + a}  Z_{{k b}}^{U}  Z_{{i 1}}^{H} +3 \sqrt{2} M_{T'} Z^{U,*}_{j 8} \sum_{a=1}^{3}Y^*_{{t'},{a}} Z_{{k a}}^{U}  Z_{{i 2}}^{H} \nonumber \\ 
 &+3 \sqrt{2} \sum_{b=1}^{3}Z^{U,*}_{j b} \sum_{a=1}^{3}Z_{{k 3 + a}}^{U} T_{u,{a b}}   Z_{{i 2}}^{H} +6 v_u \sum_{a=1}^{3}Z^{U,*}_{j a} Y_{t',{a}}  \sum_{b=1}^{3}Y^*_{{t'},{b}} Z_{{k b}}^{U}  Z_{{i 2}}^{H} \nonumber \\ 
 &+3 \sqrt{2} \sum_{b=1}^{3}\sum_{a=1}^{3}Z^{U,*}_{j 3 + a} T^*_{u,{a b}}  Z_{{k b}}^{U}  Z_{{i 2}}^{H} +3 \sqrt{2} Z^{U,*}_{j 8} \sum_{b=1}^{3}\sum_{a=1}^{3}Y^*_{u,{a b}} m_{t',{a}}  Z_{{k b}}^{U}  Z_{{i 2}}^{H} \nonumber \\ 
 &+6 v_u \sum_{c=1}^{3}Z^{U,*}_{j 3 + c} \sum_{b=1}^{3}\sum_{a=1}^{3}Y^*_{u,{c a}} Y_{u,{b a}}  Z_{{k 3 + b}}^{U}   Z_{{i 2}}^{H} +6 v_u \sum_{c=1}^{3}\sum_{b=1}^{3}Z^{U,*}_{j b} \sum_{a=1}^{3}Y^*_{u,{a c}} Y_{u,{a b}}   Z_{{k c}}^{U}  Z_{{i 2}}^{H} \nonumber \\ 
 &+2 g_{1}^{2} \sum_{a=1}^{3}Z^{U,*}_{j 3 + a} Z_{{k 3 + a}}^{U}  \Big(v_d Z_{{i 1}}^{H}  - v_u Z_{{i 2}}^{H} \Big)+3 \sqrt{2} \sum_{a=1}^{3}Z^{U,*}_{j a} T_{t',{a}}  Z_{{i 2}}^{H} Z_{{k 7}}^{U} \nonumber \\ 
 &+6 v_u \sum_{b=1}^{3}Z^{U,*}_{j 3 + b} \sum_{a=1}^{3}Y^*_{u,{b a}} Y_{t',{a}}   Z_{{i 2}}^{H} Z_{{k 7}}^{U} \nonumber \\ 
 &-3 \sqrt{2} \mu^* Z_{{i 1}}^{H} \Big(\sum_{a=1}^{3}Z^{U,*}_{j a} Y_{t',{a}}  Z_{{k 7}}^{U}  + \sum_{b=1}^{3}Z^{U,*}_{j b} \sum_{a=1}^{3}Y_{u,{a b}} Z_{{k 3 + a}}^{U}  \Big)-2 g_{1}^{2} v_d Z^{U,*}_{j 8} Z_{{i 1}}^{H} Z_{{k 8}}^{U} \nonumber \\ 
 &+2 g_{1}^{2} v_u Z^{U,*}_{j 8} Z_{{i 2}}^{H} Z_{{k 8}}^{U} +3 \sqrt{2} M_{T'}^* \sum_{a=1}^{3}Z^{U,*}_{j a} Y_{t',{a}}  Z_{{i 2}}^{H} Z_{{k 8}}^{U} \nonumber \\ 
 &+3 \sqrt{2} \sum_{b=1}^{3}Z^{U,*}_{j b} \sum_{a=1}^{3}m^*_{{t'},{a}} Y_{u,{a b}}   Z_{{i 2}}^{H} Z_{{k 8}}^{U} \Big)\Big) 
\\ 
 \Gamma_{h_{{i}}h_{{j}}\tilde{u}_{{k \gamma}}\tilde{u}^*_{{l \delta}}}  = & \, 
\frac{i}{12} \delta_{\gamma \delta} \Big(\Big(-3 g_{2}^{2}  + g_{1}^{2}\Big)\sum_{a=1}^{3}Z^{U,*}_{k a} Z_{{l a}}^{U}  \Big(Z_{{i 1}}^{H} Z_{{j 1}}^{H}  - Z_{{i 2}}^{H} Z_{{j 2}}^{H} \Big)\nonumber \\ 
 &-4 \Big(3 \sum_{a=1}^{3}Z^{U,*}_{k a} Y_{t',{a}}  \sum_{b=1}^{3}Y^*_{{t'},{b}} Z_{{l b}}^{U}  Z_{{i 2}}^{H} Z_{{j 2}}^{H} +3 Z^{U,*}_{k 7} \sum_{b=1}^{3}\sum_{a=1}^{3}Y^*_{{t'},{a}} Y_{u,{b a}}  Z_{{l 3 + b}}^{U}  Z_{{i 2}}^{H} Z_{{j 2}}^{H} \nonumber \\ 
 &+3 \sum_{c=1}^{3}Z^{U,*}_{k 3 + c} \sum_{b=1}^{3}\sum_{a=1}^{3}Y^*_{u,{c a}} Y_{u,{b a}}  Z_{{l 3 + b}}^{U}   Z_{{i 2}}^{H} Z_{{j 2}}^{H} \nonumber \\ 
 &+3 \sum_{c=1}^{3}\sum_{b=1}^{3}Z^{U,*}_{k b} \sum_{a=1}^{3}Y^*_{u,{a c}} Y_{u,{a b}}   Z_{{l c}}^{U}  Z_{{i 2}}^{H} Z_{{j 2}}^{H} +g_{1}^{2} \sum_{a=1}^{3}Z^{U,*}_{k 3 + a} Z_{{l 3 + a}}^{U}  \Big(Z_{{i 1}}^{H} Z_{{j 1}}^{H}  - Z_{{i 2}}^{H} Z_{{j 2}}^{H} \Big)\nonumber \\ 
 &+g_{1}^{2} Z^{U,*}_{k 7} Z_{{i 1}}^{H} Z_{{j 1}}^{H} Z_{{l 7}}^{U} - g_{1}^{2} Z^{U,*}_{k 7} Z_{{i 2}}^{H} Z_{{j 2}}^{H} Z_{{l 7}}^{U} +3 Z^{U,*}_{k 7} \sum_{a=1}^{3}|Y_{t',{a}}|^2 Z_{{i 2}}^{H} Z_{{j 2}}^{H} Z_{{l 7}}^{U} \nonumber \\ 
 &+3 \sum_{b=1}^{3}Z^{U,*}_{k 3 + b} \sum_{a=1}^{3}Y^*_{u,{b a}} Y_{t',{a}}   Z_{{i 2}}^{H} Z_{{j 2}}^{H} Z_{{l 7}}^{U} - g_{1}^{2} Z^{U,*}_{k 8} Z_{{i 1}}^{H} Z_{{j 1}}^{H} Z_{{l 8}}^{U} \nonumber \\ 
 &+g_{1}^{2} Z^{U,*}_{k 8} Z_{{i 2}}^{H} Z_{{j 2}}^{H} Z_{{l 8}}^{U} \Big)\Big) 
\end{align}

\section{Renormalization Group Equations}
We give in the following the two-loop RGEs for the considered model. In general, the RGEs for a parameter $X$ are
defined by
\begin{equation}
\frac{d}{dt} X = \frac{1}{16 \pi^2} \beta_X^{(1)} +  \frac{1}{(16 \pi^2)} \beta_X^{(2)} 
\end{equation}
Here, $t = \log \left(Q/M\right)$, with $Q$ the renormalization scale
and $M$ a reference scale. For
a parameter $x$ present in the MSSM we show only the difference with
respect to the MSSM RGEs 
\begin{equation}
\Delta \beta_x^{(n)} \equiv \beta_x^{(n)} - \beta_x^{(n),MSSM}   
\end{equation}
Where $\beta_x^{(n)}$ refers to the minimal model with vectorlike top quarks discussed here. The additional difference to the UV complete version of the model is given as
\begin{equation}
\Delta^{UV} \beta_x^{(n)} \equiv \beta_x^{(n),UV} - \beta_x^{(n)}   
\end{equation}

The calculation of the RGEs in \SARAH is based on generic expressions given in Refs.~\cite{Martin:1993zk,Yamada:1994id,Jack:1999zs,Jack:2000nm,Fonseca:2011vn,Goodsell:2012fm}

\label{app:RGEs}
\subsection{Gauge Couplings}
{\allowdisplaybreaks  
\begin{align} 
\Delta \beta_{g_1}^{(1)} & =  \frac{8}{5} g_{1}^{3} \\ 
\Delta^{UV} \beta_{g_1}^{(1)} & =  \frac{7}{5} g_{1}^{3}\\
\Delta \beta_{g_1}^{(2)} & =  \frac{2}{75} g_{1}^{3} \Big(-195 \Big({Y_{t'}  Y_{t'}^*}\Big)  + 64 \Big(5 g_{3}^{2}  + g_{1}^{2}\Big)\Big)\\ 
\Delta^{UV}\beta_{g_1}^{(2)} & =  \frac{1}{75} g_{1}^{3} \Big(217 g_{1}^{2}  + 45 g_{2}^{2}  + 80 g_{3}^{2} \Big)\\ 
\Delta \beta_{g_2}^{(1)} & =  0\\ 
\Delta^{UV}\beta_{g_2}^{(1)} & =  3 g_{2}^{3} \\ 
\Delta  \beta_{g_2}^{(2)} & = -6 g_{2}^{3} \Big({Y_{t'}  Y_{t'}^*}\Big) \\ 
\Delta^{UV}\beta_{g_2}^{(2)} & =  \frac{1}{5} g_{2}^{3} \Big(105 g_{2}^{2}  + 80 g_{3}^{2}  + g_{1}^{2}\Big)\\ 
\Delta  \beta_{g_3}^{(1)} & = g_{3}^{3}\\ 
\Delta^{UV}\beta_{g_3}^{(1)} & =  2 g_{3}^{3}\\
\Delta  \beta_{g_3}^{(2)} & = \frac{2}{15} g_{3}^{3} \Big(-30 \Big({Y_{t'}  Y_{t'}^*}\Big)  + 85 g_{3}^{2}  + 8 g_{1}^{2} \Big) \\
\Delta^{UV}\beta_{g_3}^{(2)} & =  \frac{2}{15} g_{3}^{3} \Big(170 g_{3}^{2}  + 45 g_{2}^{2}  + g_{1}^{2}\Big)
\end{align}
} 

\subsection{Gaugino Mass Parameters}
{\allowdisplaybreaks  
\begin{align} 
\Delta  \beta_{M_1}^{(1)} & =  \frac{16}{5} g_{1}^{2} M_1 \\ 
\Delta^{UV} \beta_{M_1}^{(1)} & =  \frac{14}{5} g_{1}^{2} M_1 \\ 
\Delta  \beta_{M_1}^{(2)} & =  \frac{4}{75} g_{1}^{2} \Big(-195 M_1 \Big({Y_{t'}  Y_{t'}^*}\Big)  + 195 \Big({Y_{t'}^*  T_{t'}}\Big)  + 64 \Big(2 g_{1}^{2} M_1  + 5 g_{3}^{2} \Big(M_1 + M_3\Big)\Big)\Big)\\ 
\Delta^{UV} \beta_{M_1}^{(2)} & =  \frac{2}{75} g_{1}^{2} \Big(434 g_{1}^{2} M_1  + 5 \Big(16 g_{3}^{2} \Big(M_1 + M_3\Big) + 9 g_{2}^{2} \Big(M_1 + M_2\Big)\Big)\Big)\\ 
\Delta  \beta_{M_2}^{(1)} & =  0\\ 
\Delta^{UV} \beta_{M_2}^{(1)} & =  6 g_{2}^{2} M_2 \\ 
\Delta  \beta_{M_2}^{(2)} & =  12 g_{2}^{2} \Big(- M_2 \Big({Y_{t'}  Y_{t'}^*}\Big)  + \Big({Y_{t'}^*  T_{t'}}\Big)\Big)\\ 
\Delta^{UV} \beta_{M_2}^{(2)} & =  \frac{2}{5} g_{2}^{2} \Big(10 \Big(21 g_{2}^{2} M_2  + 8 g_{3}^{2} \Big(M_3 + M_2\Big)\Big) + g_{1}^{2} \Big(M_1 + M_2\Big)\Big)\\ 
\Delta  \beta_{M_3}^{(1)} & =  2 g_{3}^{2} M_3 \\ 
\Delta^{UV} \beta_{M_3}^{(1)} & =  4 g_{3}^{2} M_3 \\ 
\Delta  \beta_{M_3}^{(2)} & =  \frac{8}{15} g_{3}^{2} \Big(-15 M_3 \Big({Y_{t'}  Y_{t'}^*}\Big)  + 15 \Big({Y_{t'}^*  T_{t'}}\Big)  + 4 g_{1}^{2} M_1  + 4 g_{1}^{2} M_3  + 85 g_{3}^{2} M_3 \Big) \\
\Delta^{UV} \beta_{M_3}^{(2)} & =  \frac{4}{15} g_{3}^{2} \Big(5 \Big(68 g_{3}^{2} M_3  + 9 g_{2}^{2} \Big(M_3 + M_2\Big)\Big) + g_{1}^{2} \Big(M_1 + M_3\Big)\Big)
\end{align}
}

\subsection{Trilinear Superpotential Parameters}
{\allowdisplaybreaks  
\begin{align} 
\Delta  \beta_{Y_d}^{(1)} & =  Y_{t',{j}} \Big({Y_d  Y_{t'}^*}\Big)_{i} \\ 
\Delta  \beta_{Y_d}^{(2)} & =  -3 {Y_d  Y_{u}^{\dagger}  Y_u} \Big({Y_{t'}  Y_{t'}^*}\Big) +Y_d \Big(-3 \Big({Y_{t'}  {Y_{d}^{\dagger}  Y_d  Y_{t'}^*}}\Big)  + \frac{8}{75} \Big(50 g_{3}^{4}  + 7 g_{1}^{4} \Big)\Big)-2 \Big({Y_d  Y_{t'}^*}\Big)_{i} \Big({Y_{d}^{T}  Y_d^*  Y_{t'}}\Big)_{j} \nonumber \\ 
 &-2 \Big({Y_d  Y_{t'}^*}\Big)_{i} \Big({Y_{u}^{T}  Y_u^*  Y_{t'}}\Big)_{j} +\frac{1}{5} Y_{t',{j}} \Big(-10 \Big({Y_d  Y_{u}^{\dagger}  Y_u  Y_{t'}^*}\Big)_{i}  + \Big(-15 \mbox{Tr}\Big({Y_u  Y_{u}^{\dagger}}\Big)  \nonumber \\
 & -25 \Big({Y_{t'}  Y_{t'}^*}\Big)  + 4 g_{1}^{2} \Big)\Big({Y_d  Y_{t'}^*}\Big)_{i} \Big)\\ 
 \Delta^{UV} \beta_{Y_d}^{(2)} & = \frac{1}{75} \Big(49 g_{1}^{4}  + 675 g_{2}^{4}  + 800 g_{3}^{4} \Big)Y_d \\ 
\Delta  \beta_{Y_e}^{(1)} & =  0\\ 
\Delta  \beta_{Y_e}^{(2)} & =  \frac{3}{25} Y_e \Big(24 g_{1}^{4}  -25 \Big({Y_{t'}  {Y_{d}^{\dagger}  Y_d  Y_{t'}^*}}\Big) \Big)\\ 
\Delta^{UV} \beta_{Y_e}^{(2)} & = \frac{9}{25} \Big(25 g_{2}^{4}  + 7 g_{1}^{4} \Big)Y_e \\ 
\Delta  \beta_{Y_u}^{(1)} & = 3 \Big(Y_{t',{j}} \Big({Y_u  Y_{t'}^*}\Big)_{i}  + Y_u \Big({Y_{t'}  Y_{t'}^*}\Big) \Big)\\ 
\Delta  \beta_{Y_u}^{(2)} & = -9 {Y_u  Y_{u}^{\dagger}  Y_u} \Big({Y_{t'}  Y_{t'}^*}\Big) \nonumber \\ 
 &+Y_u \Big(-18 \Big({Y_{t'}  {Y_{u}^{\dagger}  Y_u  Y_{t'}^*}}\Big)  -3 \Big({Y_{t'}  {Y_{d}^{\dagger}  Y_d  Y_{t'}^*}}\Big)  -9 \Big(\Big({Y_{t'}  Y_{t'}^*}\Big)\Big)^{2}  + \frac{104}{75} g_{1}^{4}  + \frac{16}{3} g_{3}^{4}  \nonumber \\
 &+ \frac{4}{5} \Big(20 g_{3}^{2}  + g_{1}^{2}\Big)\Big({Y_{t'}  Y_{t'}^*}\Big) \Big)-4 \Big({Y_u  Y_{t'}^*}\Big)_{i} \Big({Y_{u}^{T}  Y_u^*  Y_{t'}}\Big)_{j}  +\frac{1}{5} Y_{t',{j}} \Big(-10 \Big(2 \Big({Y_u  Y_{u}^{\dagger}  Y_u  Y_{t'}^*}\Big)_{i}  \nonumber \\
 &+ \Big({Y_u  Y_{d}^{\dagger}  Y_d  Y_{t'}^*}\Big)_{i}\Big) + \Big(2 g_{1}^{2}  + 30 g_{2}^{2}  -45 \mbox{Tr}\Big({Y_u  Y_{u}^{\dagger}}\Big)  -65 \Big({Y_{t'}  Y_{t'}^*}\Big) \Big)\Big({Y_u  Y_{t'}^*}\Big)_{i} \Big) \\
 \Delta^{UV}  \beta_{Y_u}^{(2)} & =  \frac{1}{75} \Big(675 g_{2}^{4}  + 800 g_{3}^{4}  + 91 g_{1}^{4} \Big)Y_u \\
\beta_{Y_{t',{i}}}^{(1)} & =  \Big(-3 g_{2}^{2}  + 3 \mbox{Tr}\Big({Y_u  Y_{u}^{\dagger}}\Big)  + 6 \Big({Y_{t'}  Y_{t'}^*}\Big)  -\frac{13}{15} g_{1}^{2}  -\frac{16}{3} g_{3}^{2} \Big)Y_{t',{i}}  + 3 \Big({Y_{u}^{T}  Y_u^*  Y_{t'}}\Big)_{i}  + \Big({Y_{d}^{T}  Y_d^*  Y_{t'}}\Big)_{i}\\ 
\beta_{Y_{t',{i}}}^{(2)} & =  +\Big(\frac{3367}{450} g_{1}^{4} +g_{1}^{2} g_{2}^{2} +\frac{15}{2} g_{2}^{4} +\frac{136}{45} g_{1}^{2} g_{3}^{2} +8 g_{2}^{2} g_{3}^{2} +\frac{32}{9} g_{3}^{4} -22 \Big(\Big({Y_{t'}  Y_{t'}^*}\Big)\Big)^{2} -5 \Big({Y_{t'}  {Y_{d}^{\dagger}  Y_d  Y_{t'}^*}}\Big) \nonumber \\ 
 &-22 \Big({Y_{t'}  {Y_{u}^{\dagger}  Y_u  Y_{t'}^*}}\Big) +\Big({Y_{t'}  Y_{t'}^*}\Big) \Big(16 g_{3}^{2}  + 6 g_{2}^{2}  -9 \mbox{Tr}\Big({Y_u  Y_{u}^{\dagger}}\Big)  + \frac{6}{5} g_{1}^{2} \Big)+\frac{4}{5} g_{1}^{2} \mbox{Tr}\Big({Y_u  Y_{u}^{\dagger}}\Big) \nonumber \\ 
 &+16 g_{3}^{2} \mbox{Tr}\Big({Y_u  Y_{u}^{\dagger}}\Big) -3 \mbox{Tr}\Big({Y_d  Y_{u}^{\dagger}  Y_u  Y_{d}^{\dagger}}\Big) -9 \mbox{Tr}\Big({Y_u  Y_{u}^{\dagger}  Y_u  Y_{u}^{\dagger}}\Big) \Big)Y_{t',{i}} \nonumber \\ 
 &+\Big(-3 \mbox{Tr}\Big({Y_d  Y_{d}^{\dagger}}\Big)  + \frac{2}{5} g_{1}^{2}  - \mbox{Tr}\Big({Y_e  Y_{e}^{\dagger}}\Big) \Big)\Big({Y_{d}^{T}  Y_d^*  Y_{t'}}\Big)_{i} +\frac{2}{5} g_{1}^{2} \Big({Y_{u}^{T}  Y_u^*  Y_{t'}}\Big)_{i} +6 g_{2}^{2} \Big({Y_{u}^{T}  Y_u^*  Y_{t'}}\Big)_{i} \nonumber \\
 &-13 \Big({Y_{t'}  Y_{t'}^*}\Big) \Big({Y_{u}^{T}  Y_u^*  Y_{t'}}\Big)_{i} -9 \mbox{Tr}\Big({Y_u  Y_{u}^{\dagger}}\Big) \Big({Y_{u}^{T}  Y_u^*  Y_{t'}}\Big)_{i} -2 \Big({Y_{d}^{T}  Y_d^*  Y_{d}^{T}  Y_d^*  Y_{t'}}\Big)_{i}\nonumber \\ 
 & -2 \Big({Y_{u}^{T}  Y_u^*  Y_{d}^{T}  Y_d^*  Y_{t'}}\Big)_{i} -4 \Big({Y_{u}^{T}  Y_u^*  Y_{u}^{T}  Y_u^*  Y_{t'}}\Big)_{i} \\
\Delta^{UV}  \beta_{Y_{t',{{i}}}}^{(2)} & =  \frac{1}{75} \Big(675 g_{2}^{4}  + 800 g_{3}^{4}  + 91 g_{1}^{4} \Big)Y_{t',{{i}}}  
\end{align}} 

\subsection{Bilinear Superpotential Parameters}
{\allowdisplaybreaks  \begin{align} 
\Delta  \beta_{\mu}^{(1)} & =  3 \mu \Big({Y_{t'}  Y_{t'}^*}\Big) \\ 
\Delta  \beta_{\mu}^{(2)} & =  +\frac{4}{5} \Big(20 g_{3}^{2}  + g_{1}^{2}\Big)\mu \Big({Y_{t'}  Y_{t'}^*}\Big) -9 \mu \Big(\Big({Y_{t'}  Y_{t'}^*}\Big)\Big)^{2} \nonumber \\ 
 &+\frac{6}{25} \mu \Big(-25 \Big({Y_{t'}  {Y_{d}^{\dagger}  Y_d  Y_{t'}^*}}\Big)  + 4 g_{1}^{4}  -75 \Big({Y_{t'}  {Y_{u}^{\dagger}  Y_u  Y_{t'}^*}}\Big) \Big)\\ 
\Delta^{UV} \beta_{\mu}^{(2)} & =  9 g_{2}^{4} \mu  + \frac{21}{25} g_{1}^{4} \mu \\  
\beta_{M_{T'}}^{(1)} & =  
\frac{2}{15} \Big(15 M_{T'} \Big({Y_{t'}  Y_{t'}^*}\Big)  + 15 \Big({Y_{t'}  {Y_{u}^{\dagger}  m_{t'}}}\Big)  -8 \Big(5 g_{3}^{2}  + g_{1}^{2}\Big)M_{T'} \Big)\\ 
\beta_{M_{T'}}^{(2)} & =  
+\frac{2096}{225} g_{1}^{4} M_{T'} +\frac{256}{45} g_{1}^{2} g_{3}^{2} M_{T'} +\frac{32}{9} g_{3}^{4} M_{T'} -8 M_{T'} \Big(\Big({Y_{t'}  Y_{t'}^*}\Big)\Big)^{2} -2 M_{T'} \Big({Y_{t'}  {Y_{d}^{\dagger}  Y_d  Y_{t'}^*}}\Big) \nonumber \\ 
 &-2 M_{T'} \Big({Y_{t'}  {Y_{u}^{\dagger}  Y_u  Y_{t'}^*}}\Big) -2 \Big({Y_{t'}  {Y_{d}^{\dagger}  Y_d  Y_{u}^{\dagger}  m_{t'}}}\Big) -2 \Big({Y_{t'}  {Y_{u}^{\dagger}  Y_u  Y_{u}^{\dagger}  m_{t'}}}\Big) \nonumber \\ 
 &+\Big({Y_{t'}  {Y_{u}^{\dagger}  m_{t'}}}\Big) \Big(6 g_{2}^{2}  -6 \mbox{Tr}\Big({Y_u  Y_{u}^{\dagger}}\Big)  -\frac{2}{5} g_{1}^{2} \Big)\nonumber \\ 
 &+\Big({Y_{t'}  Y_{t'}^*}\Big) \Big(6 g_{2}^{2} M_{T'}  -6 M_{T'} \mbox{Tr}\Big({Y_u  Y_{u}^{\dagger}}\Big)  -8 \Big({Y_{t'}  {Y_{u}^{\dagger}  m_{t'}}}\Big)  -\frac{2}{5} g_{1}^{2} M_{T'} \Big)\\ 
\Delta^{UV} \beta_{M_{T'}}^{(2)} & =  \frac{16}{75} \Big(50 g_{3}^{4}  + 7 g_{1}^{4} \Big)M_{T'} \\  
\beta_{m_{t',{i}}}^{(1)} & =  
2 \Big(M_{T'} \Big({Y_u  Y_{t'}^*}\Big)_{i}  + \Big({Y_u  Y_{u}^{\dagger}  m_{t'}}\Big)_{i}\Big) -\frac{16}{15} \Big(5 g_{3}^{2}  + g_{1}^{2}\Big)m_{t',{i}} \\ 
\beta_{m_{t',{i}}}^{(2)} & =  
+\frac{16}{225} \Big(131 g_{1}^{4}  + 50 g_{3}^{4}  + 80 g_{1}^{2} g_{3}^{2} \Big)m_{t',{i}} \nonumber \\ 
 &-\frac{2}{5} \Big(\Big(20 M_{T'} \Big({Y_{t'}  Y_{t'}^*}\Big)  + 5 \Big({Y_{t'}  {Y_{u}^{\dagger}  m_{t'}}}\Big)  + M_{T'} \Big(-15 g_{2}^{2}  + 15 \mbox{Tr}\Big({Y_u  Y_{u}^{\dagger}}\Big)  + g_{1}^{2}\Big)\Big)\Big({Y_u  Y_{t'}^*}\Big)_{i} \nonumber \\ 
 &+\Big(-15 g_{2}^{2}  + 15 \mbox{Tr}\Big({Y_u  Y_{u}^{\dagger}}\Big)  + 15 \Big({Y_{t'}  Y_{t'}^*}\Big)  + g_{1}^{2}\Big)\Big({Y_u  Y_{u}^{\dagger}  m_{t'}}\Big)_{i} \nonumber \\ 
 &+5 \Big(M_{T'} \Big({Y_u  Y_{d}^{\dagger}  Y_d  Y_{t'}^*}\Big)_{i}  + M_{T'} \Big({Y_u  Y_{u}^{\dagger}  Y_u  Y_{t'}^*}\Big)_{i}  + \Big({Y_u  Y_{d}^{\dagger}  Y_d  Y_{u}^{\dagger}  m_{t'}}\Big)_{i} + \Big({Y_u  Y_{u}^{\dagger}  Y_u  Y_{u}^{\dagger}  m_{t'}}\Big)_{i}\Big)\Big)\\
\Delta^{UV} \beta_{m_{t',{{i_1}}}}^{(2)} & =  \frac{16}{75} \Big(50 g_{3}^{4}  + 7 g_{1}^{4} \Big)m_{t',{{i_1}}}  \\
\beta_{M_{E'}}^{(1)} & =  -\frac{12}{5} g_{1}^{2} M_{E'} \\ 
\beta_{M_{E'}}^{(2)} & =  \frac{648}{25} g_{1}^{4} M_{E'} \\ 
\beta_{M_{Q'}}^{(1)} & =  
-\frac{1}{15} \Big(45 g_{2}^{2}  + 80 g_{3}^{2}  + g_{1}^{2}\Big)M_{Q'} \\ 
\beta_{M_{Q'}}^{(2)} & =  
\frac{1}{450} \Big(10 g_{1}^{2} \Big(16 g_{3}^{2}  + 9 g_{2}^{2} \Big) + 25 \Big(256 g_{3}^{4}  + 288 g_{2}^{2} g_{3}^{2}  + 297 g_{2}^{4} \Big) + 289 g_{1}^{4} \Big)M_{Q'} 
\end{align}}

\subsection{Trilinear Soft-Breaking Parameters}
{\allowdisplaybreaks  \begin{align} 
\Delta  \beta_{T_d}^{(1)} & =  2 \Big({Y_d  Y_{t'}^*}\Big)_{i} T_{t',{j}}  + Y_{t',{j}} \Big({T_d  Y_{t'}^*}\Big)_{i} \\ 
\Delta  \beta_{T_d}^{(2)} & =  -6 {Y_d  Y_{u}^{\dagger}  T_u} \Big({Y_{t'}  Y_{t'}^*}\Big) -3 {T_d  Y_{u}^{\dagger}  Y_u} \Big({Y_{t'}  Y_{t'}^*}\Big) -6 {Y_d  Y_{u}^{\dagger}  Y_u} \Big({Y_{t'}^*  T_{t'}}\Big) \nonumber \\ 
 &+Y_d \Big(-6 \Big({{Y_{d}^{\dagger}  Y_d  Y_{t'}^*}  T_{t'}}\Big)  -6 \Big({Y_{t'}^*  {T_{d}^{T}  Y_d^*  Y_{t'}}}\Big)  -\frac{32}{75} \Big(50 g_{3}^{4} M_3  + 7 g_{1}^{4} M_1 \Big)\Big)+\frac{56}{75} g_{1}^{4} T_d +\frac{16}{3} g_{3}^{4} T_d \nonumber \\ 
 &-3 \Big({Y_{t'}  {Y_{d}^{\dagger}  Y_d  Y_{t'}^*}}\Big) T_d -4 \Big({T_d  Y_{t'}^*}\Big)_{i} \Big({Y_{d}^{T}  Y_d^*  Y_{t'}}\Big)_{j} -4 \Big({Y_d  Y_{t'}^*}\Big)_{i} \Big({Y_{d}^{T}  Y_d^*  T_{t'}}\Big)_{j} -2 \Big({T_d  Y_{t'}^*}\Big)_{i} \Big({Y_{u}^{T}  Y_u^*  Y_{t'}}\Big)_{j} \nonumber \\ 
 &-4 \Big({Y_d  Y_{t'}^*}\Big)_{i} \Big({Y_{u}^{T}  Y_u^*  T_{t'}}\Big)_{j} -2 \Big({Y_d  Y_{t'}^*}\Big)_{i} \Big({T_{d}^{T}  Y_d^*  Y_{t'}}\Big)_{j} -4 \Big({Y_d  Y_{t'}^*}\Big)_{i} \Big({T_{u}^{T}  Y_u^*  Y_{t'}}\Big)_{j} \nonumber \\ 
 &-\frac{1}{5} Y_{t',{j}} \Big(\Big(30 \mbox{Tr}\Big({Y_{u}^{\dagger}  T_u}\Big)  + 50 \Big({Y_{t'}^*  T_{t'}}\Big)  + 8 g_{1}^{2} M_1 \Big)\Big({Y_d  Y_{t'}^*}\Big)_{i} +\Big(15 \mbox{Tr}\Big({Y_u  Y_{u}^{\dagger}}\Big)  + 25 \Big({Y_{t'}  Y_{t'}^*}\Big) \nonumber \\ 
 & -4 g_{1}^{2} \Big)\Big({T_d  Y_{t'}^*}\Big)_{i} +10 \Big(2 \Big({Y_d  Y_{u}^{\dagger}  T_u  Y_{t'}^*}\Big)_{i}  + \Big({T_d  Y_{u}^{\dagger}  Y_u  Y_{t'}^*}\Big)_{i}\Big)\Big)+\frac{8}{5} g_{1}^{2} \Big({Y_d  Y_{t'}^*}\Big)_{i} T_{t',{j}} \nonumber \\ 
 & -10 \Big({Y_{t'}  Y_{t'}^*}\Big) \Big({Y_d  Y_{t'}^*}\Big)_{i} T_{t',{j}} -6 \mbox{Tr}\Big({Y_u  Y_{u}^{\dagger}}\Big) \Big({Y_d  Y_{t'}^*}\Big)_{i} T_{t',{j}} -4 \Big({Y_d  Y_{u}^{\dagger}  Y_u  Y_{t'}^*}\Big)_{i} T_{t',{j}} \\ \Delta^{UV} \beta_{T_d}^{(2)} & =  
\frac{1}{75} \Big(-4 \Big(49 g_{1}^{4} M_1  + 675 g_{2}^{4} M_2  + 800 g_{3}^{4} M_3 \Big)Y_d  + \Big(49 g_{1}^{4}  + 675 g_{2}^{4}  + 800 g_{3}^{4} \Big)T_d \Big)\\ 
\Delta^{UV} \beta_{T_d}^{(2)} & = \frac{1}{75} \Big(-4 \Big(49 g_{1}^{4} M_1  + 675 g_{2}^{4} M_2  + 800 g_{3}^{4} M_3 \Big)Y_d  + \Big(49 g_{1}^{4}  + 675 g_{2}^{4}  + 800 g_{3}^{4} \Big)T_d \Big)\\ 
\Delta  \beta_{T_e}^{(2)} & =  -\frac{3}{25} \Big(\Big(-24 g_{1}^{4}  + 25 \Big({Y_{t'}  {Y_{d}^{\dagger}  Y_d  Y_{t'}^*}}\Big) \Big)T_e  + Y_e \Big(50 \Big({{Y_{d}^{\dagger}  Y_d  Y_{t'}^*}  T_{t'}}\Big)  \nonumber \\ 
 &+ 50 \Big({Y_{t'}^*  {T_{d}^{T}  Y_d^*  Y_{t'}}}\Big)  + 96 g_{1}^{4} M_1 \Big)\Big)\\ 
\Delta^{UV} \beta_{T_e}^{(2)} & =  -\frac{9}{25} \Big(- \Big(25 g_{2}^{4}  + 7 g_{1}^{4} \Big)T_e  + 4 \Big(25 g_{2}^{4} M_2  + 7 g_{1}^{4} M_1 \Big)Y_e \Big)\\ 
\Delta  \beta_{T_u}^{(1)} & =  
3 \Big({Y_{t'}  Y_{t'}^*}\Big) T_u  + 4 \Big({Y_u  Y_{t'}^*}\Big)_{i} T_{t',{j}}  + 5 Y_{t',{j}} \Big({T_u  Y_{t'}^*}\Big)_{i}  + 6 Y_u \Big({Y_{t'}^*  T_{t'}}\Big) \\ 
\Delta  \beta_{T_u}^{(2)} & =  
-12 {Y_u  Y_{u}^{\dagger}  T_u} \Big({Y_{t'}  Y_{t'}^*}\Big) -15 {T_u  Y_{u}^{\dagger}  Y_u} \Big({Y_{t'}  Y_{t'}^*}\Big) -18 {Y_u  Y_{u}^{\dagger}  Y_u} \Big({Y_{t'}^*  T_{t'}}\Big) \nonumber \\ 
 &-\frac{2}{75} Y_u \Big(208 g_{1}^{4} M_1 +800 g_{3}^{4} M_3 +225 \Big({Y_{t'}^*  {T_{d}^{T}  Y_d^*  Y_{t'}}}\Big) +1350 \Big({Y_{t'}^*  {T_{u}^{T}  Y_u^*  Y_{t'}}}\Big) -60 g_{1}^{2} \Big({Y_{t'}^*  T_{t'}}\Big) \nonumber \\ 
 &-1200 g_{3}^{2} \Big({Y_{t'}^*  T_{t'}}\Big) +30 \Big({Y_{t'}  Y_{t'}^*}\Big) \Big(2 g_{1}^{2} M_1  + 40 g_{3}^{2} M_3  + 45 \Big({Y_{t'}^*  T_{t'}}\Big) \Big)+225 \Big({{Y_{d}^{\dagger}  Y_d  Y_{t'}^*}  T_{t'}}\Big) \nonumber \\ 
 &+1350 \Big({{Y_{u}^{\dagger}  Y_u  Y_{t'}^*}  T_{t'}}\Big) \Big)+\frac{104}{75} g_{1}^{4} T_u +\frac{16}{3} g_{3}^{4} T_u +\frac{4}{5} g_{1}^{2} \Big({Y_{t'}  Y_{t'}^*}\Big) T_u +16 g_{3}^{2} \Big({Y_{t'}  Y_{t'}^*}\Big) T_u \nonumber \\ 
 &-9 \Big(\Big({Y_{t'}  Y_{t'}^*}\Big)\Big)^{2} T_u -3 \Big({Y_{t'}  {Y_{d}^{\dagger}  Y_d  Y_{t'}^*}}\Big) T_u -18 \Big({Y_{t'}  {Y_{u}^{\dagger}  Y_u  Y_{t'}^*}}\Big) T_u -6 \Big({T_u  Y_{t'}^*}\Big)_{i} \Big({Y_{u}^{T}  Y_u^*  Y_{t'}}\Big)_{j} \nonumber \\ 
 &-8 \Big({Y_u  Y_{t'}^*}\Big)_{i} \Big({Y_{u}^{T}  Y_u^*  T_{t'}}\Big)_{j}-6 \Big({Y_u  Y_{t'}^*}\Big)_{i} \Big({T_{u}^{T}  Y_u^*  Y_{t'}}\Big)_{j} \nonumber \\ 
 &-\frac{1}{5} Y_{t',{j}} \Big(2 \Big(2 g_{1}^{2} M_1  + 30 g_{2}^{2} M_2  + 45 \mbox{Tr}\Big({Y_{u}^{\dagger}  T_u}\Big)  + 65 \Big({Y_{t'}^*  T_{t'}}\Big) \Big)\Big({Y_u  Y_{t'}^*}\Big)_{i} \nonumber \\ 
 &+5 \Big(-3 \Big(4 g_{2}^{2}  -5 \mbox{Tr}\Big({Y_u  Y_{u}^{\dagger}}\Big)  -7 \Big({Y_{t'}  Y_{t'}^*}\Big) \Big)\Big({T_u  Y_{t'}^*}\Big)_{i}  + 4 \Big({T_u  Y_{d}^{\dagger}  Y_d  Y_{t'}^*}\Big)_{i}  + 4 \Big({Y_u  Y_{d}^{\dagger}  T_d  Y_{t'}^*}\Big)_{i}  \nonumber \\ 
 &+ 6 \Big({T_u  Y_{u}^{\dagger}  Y_u  Y_{t'}^*}\Big)_{i}  + 8 \Big({Y_u  Y_{u}^{\dagger}  T_u  Y_{t'}^*}\Big)_{i} \Big)\Big)+\frac{6}{5} g_{1}^{2} \Big({Y_u  Y_{t'}^*}\Big)_{i} T_{t',{j}} +6 g_{2}^{2} \Big({Y_u  Y_{t'}^*}\Big)_{i} T_{t',{j}} \nonumber \\ 
 &-18 \Big({Y_{t'}  Y_{t'}^*}\Big) \Big({Y_u  Y_{t'}^*}\Big)_{i} T_{t',{j}} -12 \mbox{Tr}\Big({Y_u  Y_{u}^{\dagger}}\Big) \Big({Y_u  Y_{t'}^*}\Big)_{i} T_{t',{j}} \nonumber \\ 
 &-2 \Big({Y_u  Y_{d}^{\dagger}  Y_d  Y_{t'}^*}\Big)_{i} T_{t',{j}} -6 \Big({Y_u  Y_{u}^{\dagger}  Y_u  Y_{t'}^*}\Big)_{i} T_{t',{j}} \\
\Delta^{UV}  \beta_{T_u}^{(2)} & =  
\frac{1}{75} \Big(-4 \Big(675 g_{2}^{4} M_2  + 800 g_{3}^{4} M_3  + 91 g_{1}^{4} M_1 \Big)Y_u  + \Big(675 g_{2}^{4}  + 800 g_{3}^{4}  + 91 g_{1}^{4} \Big)T_u \Big) \\
\beta_{T_{t',{i}}}^{(1)} & =  
+\Big(11 \Big({Y_{t'}^*  T_{t'}}\Big)  + 6 g_{2}^{2} M_2  + 6 \mbox{Tr}\Big({Y_{u}^{\dagger}  T_u}\Big)  + \frac{26}{15} g_{1}^{2} M_1  + \frac{32}{3} g_{3}^{2} M_3 \Big)Y_{t',{i}} +\Big({Y_{d}^{T}  Y_d^*  T_{t'}}\Big)_{i} \nonumber \\ 
 & +5 \Big({Y_{u}^{T}  Y_u^*  T_{t'}}\Big)_{i} +2 \Big({T_{d}^{T}  Y_d^*  Y_{t'}}\Big)_{i} +4 \Big({T_{u}^{T}  Y_u^*  Y_{t'}}\Big)_{i} -\frac{13}{15} g_{1}^{2} T_{t',{i}} -3 g_{2}^{2} T_{t',{i}} -\frac{16}{3} g_{3}^{2} T_{t',{i}} \nonumber \\ 
 &+7 \Big({Y_{t'}  Y_{t'}^*}\Big) T_{t',{i}} +3 \mbox{Tr}\Big({Y_u  Y_{u}^{\dagger}}\Big) T_{t',{i}} \\ 
\beta_{T_{t',{i}}}^{(2)} & =  
+\Big(-\frac{6734}{225} g_{1}^{4} M_1 -2 g_{1}^{2} g_{2}^{2} M_1 -\frac{272}{45} g_{1}^{2} g_{3}^{2} M_1 -\frac{272}{45} g_{1}^{2} g_{3}^{2} M_3 -16 g_{2}^{2} g_{3}^{2} M_3 -\frac{128}{9} g_{3}^{4} M_3 \nonumber \\ 
 &-2 g_{1}^{2} g_{2}^{2} M_2 -30 g_{2}^{4} M_2 -16 g_{2}^{2} g_{3}^{2} M_2 -10 \Big({Y_{t'}^*  {T_{d}^{T}  Y_d^*  Y_{t'}}}\Big) -44 \Big({Y_{t'}^*  {T_{u}^{T}  Y_u^*  Y_{t'}}}\Big) \nonumber \\ 
 &+\frac{8}{5} g_{1}^{2} \Big({Y_{t'}^*  T_{t'}}\Big) +12 g_{2}^{2} \Big({Y_{t'}^*  T_{t'}}\Big) +32 g_{3}^{2} \Big({Y_{t'}^*  T_{t'}}\Big) -10 \Big({{Y_{d}^{\dagger}  Y_d  Y_{t'}^*}  T_{t'}}\Big) -42 \Big({{Y_{u}^{\dagger}  Y_u  Y_{t'}^*}  T_{t'}}\Big) \nonumber \\ 
 &-\frac{8}{5} g_{1}^{2} M_1 \mbox{Tr}\Big({Y_u  Y_{u}^{\dagger}}\Big) -32 g_{3}^{2} M_3 \mbox{Tr}\Big({Y_u  Y_{u}^{\dagger}}\Big) -15 \Big({Y_{t'}^*  T_{t'}}\Big) \mbox{Tr}\Big({Y_u  Y_{u}^{\dagger}}\Big) +\frac{8}{5} g_{1}^{2} \mbox{Tr}\Big({Y_{u}^{\dagger}  T_u}\Big) \nonumber \\ 
 &+32 g_{3}^{2} \mbox{Tr}\Big({Y_{u}^{\dagger}  T_u}\Big) -\frac{1}{5} \Big({Y_{t'}  Y_{t'}^*}\Big) \Big(12 g_{1}^{2} M_1  + 160 g_{3}^{2} M_3  + 415 \Big({Y_{t'}^*  T_{t'}}\Big)  + 60 g_{2}^{2} M_2  + 90 \mbox{Tr}\Big({Y_{u}^{\dagger}  T_u}\Big) \Big)\nonumber \\ 
 &-6 \mbox{Tr}\Big({Y_d  Y_{u}^{\dagger}  T_u  Y_{d}^{\dagger}}\Big) -6 \mbox{Tr}\Big({Y_u  Y_{d}^{\dagger}  T_d  Y_{u}^{\dagger}}\Big) -36 \mbox{Tr}\Big({Y_u  Y_{u}^{\dagger}  T_u  Y_{u}^{\dagger}}\Big) \Big)Y_{t',{i}} \nonumber \\ 
 &-\frac{2}{5} \Big(15 \mbox{Tr}\Big({Y_{d}^{\dagger}  T_d}\Big)  + 2 g_{1}^{2} M_1  + 5 \mbox{Tr}\Big({Y_{e}^{\dagger}  T_e}\Big) \Big)\Big({Y_{d}^{T}  Y_d^*  Y_{t'}}\Big)_{i} +\frac{2}{5} g_{1}^{2} \Big({Y_{d}^{T}  Y_d^*  T_{t'}}\Big)_{i} \nonumber \\ 
 & -3 \mbox{Tr}\Big({Y_d  Y_{d}^{\dagger}}\Big) \Big({Y_{d}^{T}  Y_d^*  T_{t'}}\Big)_{i} - \mbox{Tr}\Big({Y_e  Y_{e}^{\dagger}}\Big) \Big({Y_{d}^{T}  Y_d^*  T_{t'}}\Big)_{i} -\frac{4}{5} g_{1}^{2} M_1 \Big({Y_{u}^{T}  Y_u^*  Y_{t'}}\Big)_{i} \nonumber \\ 
 &-12 g_{2}^{2} M_2 \Big({Y_{u}^{T}  Y_u^*  Y_{t'}}\Big)_{i} -24 \Big({Y_{t'}^*  T_{t'}}\Big) \Big({Y_{u}^{T}  Y_u^*  Y_{t'}}\Big)_{i} -18 \mbox{Tr}\Big({Y_{u}^{\dagger}  T_u}\Big) \Big({Y_{u}^{T}  Y_u^*  Y_{t'}}\Big)_{i} +12 g_{2}^{2} \Big({Y_{u}^{T}  Y_u^*  T_{t'}}\Big)_{i} \nonumber \\ 
 &-23 \Big({Y_{t'}  Y_{t'}^*}\Big) \Big({Y_{u}^{T}  Y_u^*  T_{t'}}\Big)_{i} -15 \mbox{Tr}\Big({Y_u  Y_{u}^{\dagger}}\Big) \Big({Y_{u}^{T}  Y_u^*  T_{t'}}\Big)_{i} +\frac{4}{5} g_{1}^{2} \Big({T_{d}^{T}  Y_d^*  Y_{t'}}\Big)_{i} \nonumber \\ 
 &-6 \mbox{Tr}\Big({Y_d  Y_{d}^{\dagger}}\Big) \Big({T_{d}^{T}  Y_d^*  Y_{t'}}\Big)_{i} -2 \mbox{Tr}\Big({Y_e  Y_{e}^{\dagger}}\Big) \Big({T_{d}^{T}  Y_d^*  Y_{t'}}\Big)_{i} +\frac{6}{5} g_{1}^{2} \Big({T_{u}^{T}  Y_u^*  Y_{t'}}\Big)_{i} +6 g_{2}^{2} \Big({T_{u}^{T}  Y_u^*  Y_{t'}}\Big)_{i}
 \nonumber \\ 
 & -18 \Big({Y_{t'}  Y_{t'}^*}\Big) \Big({T_{u}^{T}  Y_u^*  Y_{t'}}\Big)_{i}-12 \mbox{Tr}\Big({Y_u  Y_{u}^{\dagger}}\Big) \Big({T_{u}^{T}  Y_u^*  Y_{t'}}\Big)_{i} -2 \Big({Y_{d}^{T}  Y_d^*  Y_{d}^{T}  Y_d^*  T_{t'}}\Big)_{i}  \nonumber \\ 
 &-4 \Big({Y_{d}^{T}  Y_d^*  T_{d}^{T}  Y_d^*  Y_{t'}}\Big)_{i} -4 \Big({Y_{u}^{T}  Y_u^*  Y_{d}^{T}  Y_d^*  T_{t'}}\Big)_{i} -6 \Big({Y_{u}^{T}  Y_u^*  Y_{u}^{T}  Y_u^*  T_{t'}}\Big)_{i} -4 \Big({Y_{u}^{T}  Y_u^*  T_{d}^{T}  Y_d^*  Y_{t'}}\Big)_{i}  \nonumber \\ 
 &-8 \Big({Y_{u}^{T}  Y_u^*  T_{u}^{T}  Y_u^*  Y_{t'}}\Big)_{i} -4 \Big({T_{d}^{T}  Y_d^*  Y_{d}^{T}  Y_d^*  Y_{t'}}\Big)_{i} -2 \Big({T_{u}^{T}  Y_u^*  Y_{d}^{T}  Y_d^*  Y_{t'}}\Big)_{i} -6 \Big({T_{u}^{T}  Y_u^*  Y_{u}^{T}  Y_u^*  Y_{t'}}\Big)_{i} \nonumber \\ 
 &+\frac{3367}{450} g_{1}^{4} T_{t',{i}} +g_{1}^{2} g_{2}^{2} T_{t',{i}} +\frac{15}{2} g_{2}^{4} T_{t',{i}} +\frac{136}{45} g_{1}^{2} g_{3}^{2} T_{t',{i}} +8 g_{2}^{2} g_{3}^{2} T_{t',{i}} \nonumber \\ 
 &+\frac{32}{9} g_{3}^{4} T_{t',{i}} +2 g_{1}^{2} \Big({Y_{t'}  Y_{t'}^*}\Big) T_{t',{i}} +6 g_{2}^{2} \Big({Y_{t'}  Y_{t'}^*}\Big) T_{t',{i}} +16 g_{3}^{2} \Big({Y_{t'}  Y_{t'}^*}\Big) T_{t',{i}} -27 \Big(\Big({Y_{t'}  Y_{t'}^*}\Big)\Big)^{2} T_{t',{i}} \nonumber \\ 
 &-5 \Big({Y_{t'}  {Y_{d}^{\dagger}  Y_d  Y_{t'}^*}}\Big) T_{t',{i}} -24 \Big({Y_{t'}  {Y_{u}^{\dagger}  Y_u  Y_{t'}^*}}\Big) T_{t',{i}} +\frac{4}{5} g_{1}^{2} \mbox{Tr}\Big({Y_u  Y_{u}^{\dagger}}\Big) T_{t',{i}} \nonumber \\ 
 &+16 g_{3}^{2} \mbox{Tr}\Big({Y_u  Y_{u}^{\dagger}}\Big) T_{t',{i}} -12 \Big({Y_{t'}  Y_{t'}^*}\Big) \mbox{Tr}\Big({Y_u  Y_{u}^{\dagger}}\Big) T_{t',{i}} -3 \mbox{Tr}\Big({Y_d  Y_{u}^{\dagger}  Y_u  Y_{d}^{\dagger}}\Big) T_{t',{i}} \nonumber \\ 
 &-9 \mbox{Tr}\Big({Y_u  Y_{u}^{\dagger}  Y_u  Y_{u}^{\dagger}}\Big) T_{t',{i}} \\
 \Delta^{UV}  \beta_{T_{t',{{i_1}}}}^{(2)} & =  
\frac{1}{75} \Big(-4 \Big(675 g_{2}^{4} M_2  + 800 g_{3}^{4} M_3  + 91 g_{1}^{4} M_1 \Big)Y_{t',{{i_1}}}  + \Big(675 g_{2}^{4}  + 800 g_{3}^{4}  + 91 g_{1}^{4} \Big)T_{t',{{i_1}}} \Big)
\end{align}}

\subsection{Bilinear Soft-Breaking Parameters}
{\allowdisplaybreaks  \begin{align} 
\Delta \beta_{B_\mu}^{(1)} & =  
3 B_{\mu} \Big({Y_{t'}  Y_{t'}^*}\Big)  + 6 \mu \Big({Y_{t'}^*  T_{t'}}\Big) \\ 
\Delta \beta_{B_\mu}^{(2)} & =  
+B_{\mu} \Big(-18 \Big({Y_{t'}  {Y_{u}^{\dagger}  Y_u  Y_{t'}^*}}\Big)  -6 \Big({Y_{t'}  {Y_{d}^{\dagger}  Y_d  Y_{t'}^*}}\Big)  -9 \Big(\Big({Y_{t'}  Y_{t'}^*}\Big)\Big)^{2}  + \frac{24}{25} g_{1}^{4}  + \frac{4}{5} \Big(20 g_{3}^{2}  + g_{1}^{2}\Big)\Big({Y_{t'}  Y_{t'}^*}\Big) \Big)\nonumber \\ 
 &-\frac{4}{25} \mu \Big(24 g_{1}^{4} M_1 +75 \Big({Y_{t'}^*  {T_{d}^{T}  Y_d^*  Y_{t'}}}\Big) +225 \Big({Y_{t'}^*  {T_{u}^{T}  Y_u^*  Y_{t'}}}\Big) -10 g_{1}^{2} \Big({Y_{t'}^*  T_{t'}}\Big) -200 g_{3}^{2} \Big({Y_{t'}^*  T_{t'}}\Big) \nonumber \\ 
 &+5 \Big({Y_{t'}  Y_{t'}^*}\Big) \Big(2 g_{1}^{2} M_1  + 40 g_{3}^{2} M_3  + 45 \Big({Y_{t'}^*  T_{t'}}\Big) \Big)+75 \Big({{Y_{d}^{\dagger}  Y_d  Y_{t'}^*}  T_{t'}}\Big) +225 \Big({{Y_{u}^{\dagger}  Y_u  Y_{t'}^*}  T_{t'}}\Big) \Big)\\ 
\Delta^{UV} \beta_{B_{\mu}}^{(2)} & =  
\frac{3}{25} \Big(-4 \Big(75 g_{2}^{4} M_2  + 7 g_{1}^{4} M_1 \Big)\mu  + \Big(75 g_{2}^{4}  + 7 g_{1}^{4} \Big)B_{\mu} \Big)\\ 
\beta_{B_{T'}}^{(1)} & =  
\frac{2}{15} \Big(16 g_{1}^{2} M_1 M_{T'} +80 g_{3}^{2} M_3 M_{T'} +B_{T'} \Big(15 \Big({Y_{t'}  Y_{t'}^*}\Big)  -8 \Big(5 g_{3}^{2}  + g_{1}^{2}\Big)\Big)+15 \Big({Y_{t'}  {Y_{u}^{\dagger}  B_{t'}}}\Big)\nonumber \\ 
 & +30 M_{T'} \Big({Y_{t'}^*  T_{t'}}\Big) +30 \Big({{Y_{u}^{\dagger}  m_{t'}}  T_{t'}}\Big) \Big)\\ 
\beta_{B_{T'}}^{(2)} & =  
-\frac{8384}{225} g_{1}^{4} M_1 M_{T'} -\frac{512}{45} g_{1}^{2} g_{3}^{2} M_1 M_{T'} -\frac{512}{45} g_{1}^{2} g_{3}^{2} M_3 M_{T'} -\frac{128}{9} g_{3}^{4} M_3 M_{T'} +\frac{4}{5} g_{1}^{2} M_1 \Big({Y_{t'}  {Y_{u}^{\dagger}  m_{t'}}}\Big) \nonumber \\ 
 &-12 g_{2}^{2} M_2 \Big({Y_{t'}  {Y_{u}^{\dagger}  m_{t'}}}\Big) -\frac{2}{5} g_{1}^{2} \Big({Y_{t'}  {Y_{u}^{\dagger}  B_{t'}}}\Big) +6 g_{2}^{2} \Big({Y_{t'}  {Y_{u}^{\dagger}  B_{t'}}}\Big) \nonumber \\ 
 &-2 \Big({Y_{t'}  {Y_{d}^{\dagger}  Y_d  Y_{u}^{\dagger}  B_{t'}}}\Big) -2 \Big({Y_{t'}  {Y_{u}^{\dagger}  Y_u  Y_{u}^{\dagger}  B_{t'}}}\Big) -4 M_{T'} \Big({Y_{t'}^*  {T_{d}^{T}  Y_d^*  Y_{t'}}}\Big) \nonumber \\ 
 &-4 M_{T'} \Big({Y_{t'}^*  {T_{u}^{T}  Y_u^*  Y_{t'}}}\Big) -\frac{4}{5} g_{1}^{2} M_{T'} \Big({Y_{t'}^*  T_{t'}}\Big) +12 g_{2}^{2} M_{T'} \Big({Y_{t'}^*  T_{t'}}\Big) -16 \Big({Y_{t'}  {Y_{u}^{\dagger}  m_{t'}}}\Big) \Big({Y_{t'}^*  T_{t'}}\Big) \nonumber \\ 
 &-4 \Big({{Y_{u}^{\dagger}  m_{t'}}  {T_{d}^{T}  Y_d^*  Y_{t'}}}\Big) -4 \Big({{Y_{u}^{\dagger}  m_{t'}}  {T_{u}^{T}  Y_u^*  Y_{t'}}}\Big) -\frac{4}{5} g_{1}^{2} \Big({{Y_{u}^{\dagger}  m_{t'}}  T_{t'}}\Big) \nonumber \\ 
 &+12 g_{2}^{2} \Big({{Y_{u}^{\dagger}  m_{t'}}  T_{t'}}\Big) -4 M_{T'} \Big({{Y_{d}^{\dagger}  Y_d  Y_{t'}^*}  T_{t'}}\Big) -4 M_{T'} \Big({{Y_{u}^{\dagger}  Y_u  Y_{t'}^*}  T_{t'}}\Big) \nonumber \\ 
 &-4 \Big({{Y_{d}^{\dagger}  Y_d  Y_{u}^{\dagger}  m_{t'}}  T_{t'}}\Big) -4 \Big({{Y_{u}^{\dagger}  Y_u  Y_{u}^{\dagger}  m_{t'}}  T_{t'}}\Big) -6 \Big({Y_{t'}  {Y_{u}^{\dagger}  B_{t'}}}\Big) \mbox{Tr}\Big({Y_u  Y_{u}^{\dagger}}\Big) \nonumber \\ 
 &-12 M_{T'} \Big({Y_{t'}^*  T_{t'}}\Big) \mbox{Tr}\Big({Y_u  Y_{u}^{\dagger}}\Big) -12 \Big({{Y_{u}^{\dagger}  m_{t'}}  T_{t'}}\Big) \mbox{Tr}\Big({Y_u  Y_{u}^{\dagger}}\Big) \nonumber \\ 
 &+\frac{2}{225} B_{T'} \Big(1048 g_{1}^{4} +640 g_{1}^{2} g_{3}^{2} +400 g_{3}^{4} -900 \Big(\Big({Y_{t'}  Y_{t'}^*}\Big)\Big)^{2} -225 \Big({Y_{t'}  {Y_{d}^{\dagger}  Y_d  Y_{t'}^*}}\Big) \nonumber \\ 
 &-225 \Big({Y_{t'}  {Y_{u}^{\dagger}  Y_u  Y_{t'}^*}}\Big) -45 \Big({Y_{t'}  Y_{t'}^*}\Big) \Big(-15 g_{2}^{2}  + 15 \mbox{Tr}\Big({Y_u  Y_{u}^{\dagger}}\Big)  + g_{1}^{2}\Big)\Big)\nonumber \\ 
 &-12 \Big({Y_{t'}  {Y_{u}^{\dagger}  m_{t'}}}\Big) \mbox{Tr}\Big({Y_{u}^{\dagger}  T_u}\Big) \nonumber \\ 
 &+\Big({Y_{t'}  Y_{t'}^*}\Big) \Big(\frac{4}{5} g_{1}^{2} M_1 M_{T'} -12 g_{2}^{2} M_2 M_{T'} -8 \Big({Y_{t'}  {Y_{u}^{\dagger}  B_{t'}}}\Big) -32 M_{T'} \Big({Y_{t'}^*  T_{t'}}\Big) -16 \Big({{Y_{u}^{\dagger}  m_{t'}}  T_{t'}}\Big) \nonumber \\ 
 &-12 M_{T'} \mbox{Tr}\Big({Y_{u}^{\dagger}  T_u}\Big) \Big)\\ 
\Delta^{UV}  \beta_{B_{T'}}^{(2)} & =  
-\frac{16}{75} \Big(4 \Big(50 g_{3}^{4} M_3  + 7 g_{1}^{4} M_1 \Big)M_{T'}  - \Big(50 g_{3}^{4}  + 7 g_{1}^{4} \Big)B_{T'} \Big)\\ 
\beta_{B_{t',{i}}}^{(1)} & =  
\frac{2}{15} \Big(15 \Big(2 M_{T'} \Big({T_u  Y_{t'}^*}\Big)_{i}  + 2 \Big({T_u  Y_{u}^{\dagger}  m_{t'}}\Big)_{i}  + B_{T'} \Big({Y_u  Y_{t'}^*}\Big)_{i}  + \Big({Y_u  Y_{u}^{\dagger}  B_{t'}}\Big)_{i}\Big) \nonumber \\ 
 &+ 16 \Big(5 g_{3}^{2} M_3  + g_{1}^{2} M_1 \Big)m_{t',{i}}  -8 \Big(5 g_{3}^{2}  + g_{1}^{2}\Big)B_{t',{i}} \Big)\\ 
\beta_{B_{t',{i}}}^{(2)} & =  
-\frac{64}{225} \Big(131 g_{1}^{4} M_1  + 40 g_{1}^{2} g_{3}^{2} \Big(M_1 + M_3\Big) + 50 g_{3}^{4} M_3 \Big)m_{t',{i}} +\frac{16}{225} \Big(131 g_{1}^{4}  + 50 g_{3}^{4}  + 80 g_{1}^{2} g_{3}^{2} \Big)B_{t',{i}} \nonumber \\ 
 &+\frac{4}{5} g_{1}^{2} M_1 M_{T'} \Big({Y_u  Y_{t'}^*}\Big)_{i} -12 g_{2}^{2} M_2 M_{T'} \Big({Y_u  Y_{t'}^*}\Big)_{i} -\frac{2}{5} g_{1}^{2} B_{T'} \Big({Y_u  Y_{t'}^*}\Big)_{i} +6 g_{2}^{2} B_{T'} \Big({Y_u  Y_{t'}^*}\Big)_{i} \nonumber \\ 
 &-8 B_{T'} \Big({Y_{t'}  Y_{t'}^*}\Big) \Big({Y_u  Y_{t'}^*}\Big)_{i} -2 \Big({Y_{t'}  {Y_{u}^{\dagger}  B_{t'}}}\Big) \Big({Y_u  Y_{t'}^*}\Big)_{i} -16 M_{T'} \Big({Y_{t'}^*  T_{t'}}\Big) \Big({Y_u  Y_{t'}^*}\Big)_{i} \nonumber \\ 
 &-4 \Big({{Y_{u}^{\dagger}  m_{t'}}  T_{t'}}\Big) \Big({Y_u  Y_{t'}^*}\Big)_{i} -6 B_{T'} \mbox{Tr}\Big({Y_u  Y_{u}^{\dagger}}\Big) \Big({Y_u  Y_{t'}^*}\Big)_{i} -12 M_{T'} \mbox{Tr}\Big({Y_{u}^{\dagger}  T_u}\Big) \Big({Y_u  Y_{t'}^*}\Big)_{i} \nonumber \\ 
 &-\frac{4}{5} g_{1}^{2} M_{T'} \Big({T_u  Y_{t'}^*}\Big)_{i} +12 g_{2}^{2} M_{T'} \Big({T_u  Y_{t'}^*}\Big)_{i} -16 M_{T'} \Big({Y_{t'}  Y_{t'}^*}\Big) \Big({T_u  Y_{t'}^*}\Big)_{i} -4 \Big({Y_{t'}  {Y_{u}^{\dagger}  m_{t'}}}\Big) \Big({T_u  Y_{t'}^*}\Big)_{i} \nonumber \\ 
 &-12 M_{T'} \mbox{Tr}\Big({Y_u  Y_{u}^{\dagger}}\Big) \Big({T_u  Y_{t'}^*}\Big)_{i} +\frac{4}{5} g_{1}^{2} M_1 \Big({Y_u  Y_{u}^{\dagger}  m_{t'}}\Big)_{i} -12 g_{2}^{2} M_2 \Big({Y_u  Y_{u}^{\dagger}  m_{t'}}\Big)_{i} \nonumber \\ 
 &-12 \Big({Y_{t'}^*  T_{t'}}\Big) \Big({Y_u  Y_{u}^{\dagger}  m_{t'}}\Big)_{i} -12 \mbox{Tr}\Big({Y_{u}^{\dagger}  T_u}\Big) \Big({Y_u  Y_{u}^{\dagger}  m_{t'}}\Big)_{i} -\frac{2}{5} g_{1}^{2} \Big({Y_u  Y_{u}^{\dagger}  B_{t'}}\Big)_{i} +6 g_{2}^{2} \Big({Y_u  Y_{u}^{\dagger}  B_{t'}}\Big)_{i} \nonumber \\ 
 &-6 \Big({Y_{t'}  Y_{t'}^*}\Big) \Big({Y_u  Y_{u}^{\dagger}  B_{t'}}\Big)_{i} -6 \mbox{Tr}\Big({Y_u  Y_{u}^{\dagger}}\Big) \Big({Y_u  Y_{u}^{\dagger}  B_{t'}}\Big)_{i} -\frac{4}{5} g_{1}^{2} \Big({T_u  Y_{u}^{\dagger}  m_{t'}}\Big)_{i} +12 g_{2}^{2} \Big({T_u  Y_{u}^{\dagger}  m_{t'}}\Big)_{i} \nonumber \\ 
 & -12 \Big({Y_{t'}  Y_{t'}^*}\Big) \Big({T_u  Y_{u}^{\dagger}  m_{t'}}\Big)_{i} -12 \mbox{Tr}\Big({Y_u  Y_{u}^{\dagger}}\Big) \Big({T_u  Y_{u}^{\dagger}  m_{t'}}\Big)_{i} -2 B_{T'} \Big({Y_u  Y_{d}^{\dagger}  Y_d  Y_{t'}^*}\Big)_{i} \nonumber \\ 
 &-4 M_{T'} \Big({Y_u  Y_{d}^{\dagger}  T_d  Y_{t'}^*}\Big)_{i} -2 B_{T'} \Big({Y_u  Y_{u}^{\dagger}  Y_u  Y_{t'}^*}\Big)_{i} -4 M_{T'} \Big({Y_u  Y_{u}^{\dagger}  T_u  Y_{t'}^*}\Big)_{i} -4 M_{T'} \Big({T_u  Y_{d}^{\dagger}  Y_d  Y_{t'}^*}\Big)_{i} \nonumber \\ 
 &-4 M_{T'} \Big({T_u  Y_{u}^{\dagger}  Y_u  Y_{t'}^*}\Big)_{i} -2 \Big({Y_u  Y_{d}^{\dagger}  Y_d  Y_{u}^{\dagger}  B_{t'}}\Big)_{i} -4 \Big({Y_u  Y_{d}^{\dagger}  T_d  Y_{u}^{\dagger}  m_{t'}}\Big)_{i} -2 \Big({Y_u  Y_{u}^{\dagger}  Y_u  Y_{u}^{\dagger}  B_{t'}}\Big)_{i} \nonumber \\ 
 &-4 \Big({Y_u  Y_{u}^{\dagger}  T_u  Y_{u}^{\dagger}  m_{t'}}\Big)_{i} -4 \Big({T_u  Y_{d}^{\dagger}  Y_d  Y_{u}^{\dagger}  m_{t'}}\Big)_{i} -4 \Big({T_u  Y_{u}^{\dagger}  Y_u  Y_{u}^{\dagger}  m_{t'}}\Big)_{i} \\
\Delta^{UV} \beta_{B_{t',{{i_1}}}}^{(2)} & =  
-\frac{16}{75} \Big(4 \Big(50 g_{3}^{4} M_3  + 7 g_{1}^{4} M_1 \Big)m_{t',{{i_1}}}  - \Big(50 g_{3}^{4}  + 7 g_{1}^{4} \Big)B_{t',{{i_1}}} \Big)\\
\beta_{B_{Q'}}^{(1)} & =  
\frac{1}{15} \Big(2 \Big(45 g_{2}^{2} M_2  + 80 g_{3}^{2} M_3  + g_{1}^{2} M_1 \Big)M_{Q'}  - \Big(45 g_{2}^{2}  + 80 g_{3}^{2}  + g_{1}^{2}\Big)B_{Q'} \Big)\\ 
\beta_{B_{Q'}}^{(2)} & =  
\frac{1}{450} \Big(-4 \Big(25 \Big(144 g_{2}^{2} g_{3}^{2} \Big(M_3 + M_2\Big) + 256 g_{3}^{4} M_3  + 297 g_{2}^{4} M_2 \Big) + 289 g_{1}^{4} M_1  \nonumber \\ 
 &+ 5 g_{1}^{2} \Big(16 g_{3}^{2} \Big(M_1 + M_3\Big) + 9 g_{2}^{2} \Big(M_1 + M_2\Big)\Big)\Big)M_{Q'}+\Big(10 g_{1}^{2} \Big(16 g_{3}^{2}  + 9 g_{2}^{2} \Big) \nonumber \\ 
 &+ 25 \Big(256 g_{3}^{4}  + 288 g_{2}^{2} g_{3}^{2}   + 297 g_{2}^{4} \Big) + 289 g_{1}^{4} \Big)B_{Q'} \Big)\\ 
\beta_{B_{E'}}^{(1)} & =  
\frac{12}{5} g_{1}^{2} \Big(2 M_1 M_{E'}  - B_{E'} \Big)\\ 
\beta_{B_{E'}}^{(2)} & =  
-\frac{648}{25} g_{1}^{4} \Big(4 M_1 M_{E'}  - B_{E'} \Big)
\end{align}}

\subsection{Soft-Breaking Scalar Masses}
Traces:
\begin{align} 
\sigma_{1,1} & = \sqrt{\frac{3}{5}} g_1 \Big(-2 \mbox{Tr}\Big({m_u^2}\Big)  + 2 m_{\tilde{\bar{t}}'}^2  -2 m_{\tilde{t}'}^2  - \mbox{Tr}\Big({m_l^2}\Big)  - m_{H_d}^2  + m_{H_u}^2 + \nonumber \\ 
 & \mbox{Tr}\Big({m_d^2}\Big) + \mbox{Tr}\Big({m_e^2}\Big) + \mbox{Tr}\Big({m_q^2}\Big)\Big)\\ 
\Delta^{UV}\sigma_{1,1} & = \sqrt{\frac{3}{5}} g_1 \Big(- m_{\tilde{\bar{e}}'}^2  - m_{\tilde{\bar{q}}'}^2  + m_{\tilde{q}'}^2 + m_{\tilde{e}'}^2\Big)\\ 
\sigma_{2,11} & = \frac{1}{10} g_{1}^{2} \Big(2 \mbox{Tr}\Big({m_d^2}\Big)  + 3 \mbox{Tr}\Big({m_l^2}\Big)  + 3 m_{H_d}^2  + 3 m_{H_u}^2  + 6 \mbox{Tr}\Big({m_e^2}\Big)  + 8 \mbox{Tr}\Big({m_u^2}\Big)  \nonumber \\ 
 &+ 8 m_{\tilde{\bar{t}}'}^2  + 8 m_{\tilde{t}'}^2  + \mbox{Tr}\Big({m_q^2}\Big)\Big)\\ 
\Delta^{UV}\sigma_{2,11} & = \frac{1}{10} g_{1}^{2} \Big(6 \Big(m_{\tilde{e}'}^2 + m_{\tilde{\bar{e}}'}^2\Big) + m_{\tilde{q}'}^2 + m_{\tilde{\bar{q}}'}^2\Big)\\ 
\sigma_{3,1} & = \frac{1}{20} \frac{1}{\sqrt{15}} g_1 \Big(-9 g_{1}^{2} m_{H_d}^2 -45 g_{2}^{2} m_{H_d}^2 +9 g_{1}^{2} m_{H_u}^2 +45 g_{2}^{2} m_{H_u}^2 -32 g_{1}^{2} m_{\tilde{t}'}^2 -160 g_{3}^{2} m_{\tilde{t}'}^2 \nonumber \\ 
 &+32 g_{1}^{2} m_{\tilde{\bar{t}}'}^2 +160 g_{3}^{2} m_{\tilde{\bar{t}}'}^2+120 {m_{\tilde u \tilde t'}^2  Y_u  Y_{t'}^*} -30 {Y_{t'}  m_q^{2 *}  Y_{t'}^*} -90 m_{H_u}^2 \Big({Y_{t'}  Y_{t'}^*}\Big)  \nonumber \\ 
 &+120 m_{\tilde{t}'}^2 \Big({Y_{t'}  Y_{t'}^*}\Big) +4 g_{1}^{2} \mbox{Tr}\Big({m_d^2}\Big) +80 g_{3}^{2} \mbox{Tr}\Big({m_d^2}\Big) \nonumber \\ 
 &+36 g_{1}^{2} \mbox{Tr}\Big({m_e^2}\Big) -9 g_{1}^{2} \mbox{Tr}\Big({m_l^2}\Big) -45 g_{2}^{2} \mbox{Tr}\Big({m_l^2}\Big) +g_{1}^{2} \mbox{Tr}\Big({m_q^2}\Big) +45 g_{2}^{2} \mbox{Tr}\Big({m_q^2}\Big) +80 g_{3}^{2} \mbox{Tr}\Big({m_q^2}\Big) \nonumber \\ 
 &-32 g_{1}^{2} \mbox{Tr}\Big({m_u^2}\Big) -160 g_{3}^{2} \mbox{Tr}\Big({m_u^2}\Big) +90 m_{H_d}^2 \mbox{Tr}\Big({Y_d  Y_{d}^{\dagger}}\Big) +30 m_{H_d}^2 \mbox{Tr}\Big({Y_e  Y_{e}^{\dagger}}\Big) -90 m_{H_u}^2 \mbox{Tr}\Big({Y_u  Y_{u}^{\dagger}}\Big) \nonumber \\ 
 &-60 \mbox{Tr}\Big({Y_d  Y_{d}^{\dagger}  m_d^{2 *}}\Big) -30 \mbox{Tr}\Big({Y_d  m_q^{2 *}  Y_{d}^{\dagger}}\Big) -60 \mbox{Tr}\Big({Y_e  Y_{e}^{\dagger}  m_e^{2 *}}\Big) +30 \mbox{Tr}\Big({Y_e  m_l^{2 *}  Y_{e}^{\dagger}}\Big) \nonumber \\ 
 &+120 \mbox{Tr}\Big({Y_u  Y_{u}^{\dagger}  m_u^{2 *}}\Big) -30 \mbox{Tr}\Big({Y_u  m_q^{2 *}  Y_{u}^{\dagger}}\Big) \Big)\\ 
\Delta^{UV}\sigma_{3,1} & = \frac{1}{20} \frac{1}{\sqrt{15}} g_1 \Big(5 \Big(16 g_{3}^{2}  + 9 g_{2}^{2} \Big)\Big(- m_{\tilde{\bar{q}}'}^2  + m_{\tilde{q}'}^2\Big) + g_{1}^{2} \Big(-36 m_{\tilde{\bar{e}}'}^2  + 36 m_{\tilde{e}'}^2  - m_{\tilde{\bar{q}}'}^2  + m_{\tilde{q}'}^2\Big)\Big)\\ 
 \sigma_{2,2} & = \frac{1}{2} \Big(3 \mbox{Tr}\Big({m_q^2}\Big)  + m_{H_d}^2 + m_{H_u}^2 + \mbox{Tr}\Big({m_l^2}\Big)\Big)\\ 
\Delta^{UV}\sigma_{2,2} & = \frac{3}{2} \Big(m_{\tilde{q}'}^2 + m_{\tilde{\bar{q}}'}^2\Big)\\ 
 \sigma_{2,3} & = \frac{1}{2} \Big(2 \mbox{Tr}\Big({m_q^2}\Big)  + m_{\tilde{t}'}^2 + m_{\tilde{\bar{t}}'}^2 + \mbox{Tr}\Big({m_d^2}\Big) + \mbox{Tr}\Big({m_u^2}\Big)\Big) \\
\Delta^{UV}\sigma_{2,3} & = m_{\tilde{q}'}^2 + m_{\tilde{\bar{q}}'}^2
\end{align}

{\allowdisplaybreaks  \begin{align} 
\Delta  \beta_{m_q^2}^{(1)} & =  
2 T^*_{{t'},{i}} T_{t',{j}}  + Y^*_{{t'},{i}} \Big(2 \Big(m_{H_u}^2 + m_{\tilde{t}'}^2\Big)Y_{t',{j}}  + \Big({m_q^{2 *}  Y_{t'}}\Big)_{j}\Big) + Y_{t',{j}} \Big(2 \Big({Y_{u}^{\dagger}  m_{\tilde u \tilde t'}^2}\Big)_{i}  + \Big({m_q^2  Y_{t'}^*}\Big)_{i}\Big)\\ 
\Delta  \beta_{m_q^2}^{(2)} & =  
+32 g_{3}^{4} {\bf 1} |M_3|^2 -12 m_{H_u}^2 {Y_{u}^{\dagger}  Y_u} \Big({Y_{t'}  Y_{t'}^*}\Big) -6 m_{\tilde{t}'}^2 {Y_{u}^{\dagger}  Y_u} \Big({Y_{t'}  Y_{t'}^*}\Big) -6 {T_{u}^{\dagger}  T_u} \Big({Y_{t'}  Y_{t'}^*}\Big) \nonumber \\ 
 &-3 {m_q^2  Y_{u}^{\dagger}  Y_u} \Big({Y_{t'}  Y_{t'}^*}\Big) -6 {Y_{u}^{\dagger}  m_u^2  Y_u} \Big({Y_{t'}  Y_{t'}^*}\Big) -3 {Y_{u}^{\dagger}  Y_u  m_q^2} \Big({Y_{t'}  Y_{t'}^*}\Big) \nonumber \\ 
 &-6 {Y_{u}^{\dagger}  T_u} \Big({Y_{t'}  T_{t'}^*}\Big) -6 {Y_{u}^{\dagger}  Y_u} \Big({Y_{t'}  {m_q^2  Y_{t'}^*}}\Big) -6 {Y_{u}^{\dagger}  Y_u} \Big({Y_{t'}  {Y_{u}^{\dagger}  m_{\tilde u \tilde t'}^2}}\Big) \nonumber \\ 
 &-6 {T_{u}^{\dagger}  Y_u} \Big({Y_{t'}^*  T_{t'}}\Big) -6 {Y_{u}^{\dagger}  Y_u} \Big({T_{t'}  T_{t'}^*}\Big) +\frac{8}{5} g_{1}^{2} m_{H_u}^2 Y^*_{{t'},{i}} Y_{t',{j}} +\frac{8}{5} g_{1}^{2} m_{\tilde{t}'}^2 Y^*_{{t'},{i}} Y_{t',{j}} \nonumber \\ 
 &-\frac{8}{5} g_{1}^{2} M_1 T^*_{{t'},{i}} Y_{t',{j}} -20 m_{H_u}^2 Y^*_{{t'},{i}} \Big({Y_{t'}  Y_{t'}^*}\Big) Y_{t',{j}} -20 m_{\tilde{t}'}^2 Y^*_{{t'},{i}} \Big({Y_{t'}  Y_{t'}^*}\Big) Y_{t',{j}} -10 Y^*_{{t'},{i}} \Big({Y_{t'}  {m_q^2  Y_{t'}^*}}\Big) Y_{t',{j}} \nonumber \\ 
 &-10 Y^*_{{t'},{i}} \Big({Y_{t'}  {Y_{u}^{\dagger}  m_{\tilde u \tilde t'}^2}}\Big) Y_{t',{j}} -10 T^*_{{t'},{i}} \Big({Y_{t'}^*  T_{t'}}\Big) Y_{t',{j}} -10 Y^*_{{t'},{i}} \Big({T_{t'}  T_{t'}^*}\Big) Y_{t',{j}} \nonumber \\ 
 &-12 m_{H_u}^2 Y^*_{{t'},{i}} \mbox{Tr}\Big({Y_u  Y_{u}^{\dagger}}\Big) Y_{t',{j}} -6 m_{\tilde{t}'}^2 Y^*_{{t'},{i}} \mbox{Tr}\Big({Y_u  Y_{u}^{\dagger}}\Big) Y_{t',{j}} -6 T^*_{{t'},{i}} \mbox{Tr}\Big({Y_{u}^{\dagger}  T_u}\Big) Y_{t',{j}} \nonumber \\ 
 &-6 Y^*_{{t'},{i}} \mbox{Tr}\Big({T_u^*  T_{u}^{T}}\Big) Y_{t',{j}} -6 Y^*_{{t'},{i}} \mbox{Tr}\Big({m_q^2  Y_{u}^{\dagger}  Y_u}\Big) Y_{t',{j}} -6 Y^*_{{t'},{i}} \mbox{Tr}\Big({m_u^2  Y_u  Y_{u}^{\dagger}}\Big) Y_{t',{j}} +\frac{4}{5} g_{1}^{2} Y_{t',{j}} \Big({m_q^2  Y_{t'}^*}\Big)_{i} \nonumber \\ 
 &-5 \Big({Y_{t'}  Y_{t'}^*}\Big) Y_{t',{j}} \Big({m_q^2  Y_{t'}^*}\Big)_{i} -3 \mbox{Tr}\Big({Y_u  Y_{u}^{\dagger}}\Big) Y_{t',{j}} \Big({m_q^2  Y_{t'}^*}\Big)_{i} +\frac{8}{5} g_{1}^{2} Y_{t',{j}} \Big({Y_{u}^{\dagger}  m_{\tilde u \tilde t'}^2}\Big)_{i}  \nonumber \\ 
 &-10 \Big({Y_{t'}  Y_{t'}^*}\Big) Y_{t',{j}} \Big({Y_{u}^{\dagger}  m_{\tilde u \tilde t'}^2}\Big)_{i} -6 \mbox{Tr}\Big({Y_u  Y_{u}^{\dagger}}\Big) Y_{t',{j}} \Big({Y_{u}^{\dagger}  m_{\tilde u \tilde t'}^2}\Big)_{i} +\frac{4}{5} g_{1}^{2} Y^*_{{t'},{i}} \Big({m_q^{2 *}  Y_{t'}}\Big)_{j} \nonumber \\ 
 &-5 Y^*_{{t'},{i}} \Big({Y_{t'}  Y_{t'}^*}\Big) \Big({m_q^{2 *}  Y_{t'}}\Big)_{j} -3 Y^*_{{t'},{i}} \mbox{Tr}\Big({Y_u  Y_{u}^{\dagger}}\Big) \Big({m_q^{2 *}  Y_{t'}}\Big)_{j} -8 m_{H_u}^2 Y_{t',{j}} \Big({Y_{u}^{\dagger}  Y_u  Y_{t'}^*}\Big)_{i}  \nonumber \\ 
 &-4 m_{\tilde{t}'}^2 Y_{t',{j}} \Big({Y_{u}^{\dagger}  Y_u  Y_{t'}^*}\Big)_{i}-2 \Big({m_q^{2 *}  Y_{t'}}\Big)_{j} \Big({Y_{u}^{\dagger}  Y_u  Y_{t'}^*}\Big)_{i} -4 Y_{t',{j}} \Big({Y_{u}^{\dagger}  T_u  T_{t'}^*}\Big)_{i} -4 Y_{t',{j}} \Big({T_{u}^{\dagger}  T_u  Y_{t'}^*}\Big)_{i} \nonumber \\ 
 &-8 m_{H_u}^2 Y^*_{{t'},{i}} \Big({Y_{u}^{T}  Y_u^*  Y_{t'}}\Big)_{j} -4 m_{\tilde{t}'}^2 Y^*_{{t'},{i}} \Big({Y_{u}^{T}  Y_u^*  Y_{t'}}\Big)_{j} -2 \Big({m_q^2  Y_{t'}^*}\Big)_{i} \Big({Y_{u}^{T}  Y_u^*  Y_{t'}}\Big)_{j}  \nonumber \\ 
 & -4 \Big({Y_{u}^{\dagger}  m_{\tilde u \tilde t'}^2}\Big)_{i} \Big({Y_{u}^{T}  Y_u^*  Y_{t'}}\Big)_{j} -4 T^*_{{t'},{i}} \Big({Y_{u}^{T}  Y_u^*  T_{t'}}\Big)_{j}-4 Y^*_{{t'},{i}} \Big({Y_{u}^{T}  T_u^*  T_{t'}}\Big)_{j} -4 T^*_{{t'},{i}} \Big({T_{u}^{T}  Y_u^*  Y_{t'}}\Big)_{j}  \nonumber \\ 
 &-4 Y^*_{{t'},{i}} \Big({T_{u}^{T}  T_u^*  Y_{t'}}\Big)_{j} -2 Y_{t',{j}} \Big({m_q^2  Y_{u}^{\dagger}  Y_u  Y_{t'}^*}\Big)_{i} -4 Y_{t',{j}} \Big({Y_{u}^{\dagger}  m_u^2  Y_u  Y_{t'}^*}\Big)_{i} -4 Y_{t',{j}} \Big({Y_{u}^{\dagger}  Y_u  m_q^2  Y_{t'}^*}\Big)_{i} \nonumber \\ 
 &-4 Y_{t',{j}} \Big({Y_{u}^{\dagger}  Y_u  Y_{u}^{\dagger}  m_{\tilde u \tilde t'}^2}\Big)_{i} -2 Y^*_{{t'},{i}} \Big({m_q^{2 *}  Y_{u}^{T}  Y_u^*  Y_{t'}}\Big)_{j} -4 Y^*_{{t'},{i}} \Big({Y_{u}^{T}  m_u^{2 *}  Y_u^*  Y_{t'}}\Big)_{j} -4 Y^*_{{t'},{i}} \Big({Y_{u}^{T}  Y_u^*  m_q^{2 *}  Y_{t'}}\Big)_{j} \nonumber \\ 
 &+\frac{8}{25} g_{1}^{2} M_1^* \Big(2 g_{1}^{2} M_1 {\bf 1}  + 5 Y^*_{{t'},{i}} \Big(2 M_1 Y_{t',{j}}  - T_{t',{j}} \Big)\Big)+\frac{8}{5} g_{1}^{2} T^*_{{t'},{i}} T_{t',{j}} -10 T^*_{{t'},{i}} \Big({Y_{t'}  Y_{t'}^*}\Big) T_{t',{j}} \nonumber \\ 
 &-10 Y^*_{{t'},{i}} \Big({Y_{t'}  T_{t'}^*}\Big) T_{t',{j}} -6 T^*_{{t'},{i}} \mbox{Tr}\Big({Y_u  Y_{u}^{\dagger}}\Big) T_{t',{j}} -6 Y^*_{{t'},{i}} \mbox{Tr}\Big({T_u^*  Y_{u}^{T}}\Big) T_{t',{j}} -4 \Big({Y_{u}^{\dagger}  Y_u  T_{t'}^*}\Big)_{i} T_{t',{j}}  \nonumber \\ 
 &-4 \Big({T_{u}^{\dagger}  Y_u  Y_{t'}^*}\Big)_{i} T_{t',{j}} \\ 
\Delta^{UV} \beta_{m_q^2}^{(2)} & =  \frac{2}{25} {\bf 1} \Big(675 g_{2}^{4} |M_2|^2  + 7 g_{1}^{4} |M_1|^2  + 800 g_{3}^{4} |M_3|^2 \Big)\\ 
\Delta \beta_{m_l^2}^{(2)} & =  \frac{144}{25} g_{1}^{4} {\bf 1} |M_1|^2 \\ 
\Delta^{UV} \beta_{m_l^2}^{(2)} & =  \frac{18}{25} {\bf 1} \Big(75 g_{2}^{4} |M_2|^2  + 7 g_{1}^{4} |M_1|^2 \Big)\\ 
\Delta \beta_{m_{H_d}^2}^{(2)} & =  
+\frac{144}{25} g_{1}^{4} |M_1|^2 \nonumber \\ 
 &-6 \Big(\Big(m_{H_d}^2 + m_{H_u}^2 + m_{\tilde{t}'}^2\Big)\Big({Y_{t'}  {Y_{d}^{\dagger}  Y_d  Y_{t'}^*}}\Big) +\Big({Y_{t'}  {T_{d}^{\dagger}  T_d  Y_{t'}^*}}\Big)+\Big({Y_{t'}  {m_q^2  Y_{d}^{\dagger}  Y_d  Y_{t'}^*}}\Big)\nonumber \\ 
 &+\Big({Y_{t'}  {Y_{d}^{\dagger}  m_d^2  Y_d  Y_{t'}^*}}\Big)+\Big({Y_{t'}  {Y_{d}^{\dagger}  Y_d  m_q^2  Y_{t'}^*}}\Big)+\Big({Y_{t'}  {Y_{d}^{\dagger}  Y_d  Y_{u}^{\dagger}  m_{\tilde u \tilde t'}^2}}\Big)+\Big({{Y_{d}^{\dagger}  Y_d  T_{t'}^*}  T_{t'}}\Big)\nonumber \\ 
 &+\Big({{T_{d}^{\dagger}  Y_d  Y_{t'}^*}  T_{t'}}\Big)+\Big({{T_{d}^{T}  Y_d^*  Y_{t'}}  T_{t'}^*}\Big)\Big)\\ 
\Delta^{UV} \beta_{m_{H_d}^2}^{(2)} & =  
54 g_{2}^{4} |M_2|^2  + \frac{126}{25} g_{1}^{4} |M_1|^2 \\ 
\Delta \beta_{m_{H_u}^2}^{(1)} & =  
6 \Big(\Big(m_{H_u}^2 + m_{\tilde{t}'}^2\Big)\Big({Y_{t'}  Y_{t'}^*}\Big)  + \Big({Y_{t'}  {m_q^2  Y_{t'}^*}}\Big) + \Big({Y_{t'}  {Y_{u}^{\dagger}  m_{\tilde u \tilde t'}^2}}\Big) + \Big({T_{t'}  T_{t'}^*}\Big)\Big)\\ 
\Delta \beta_{m_{H_u}^2}^{(2)} & =  
-36 \Big(m_{H_u}^2 + m_{\tilde{t}'}^2\Big)\Big(\Big({Y_{t'}  Y_{t'}^*}\Big)\Big)^{2} +\frac{8}{25} g_{1}^{2} M_1^* \Big(10 M_1 \Big({Y_{t'}  Y_{t'}^*}\Big)  + 18 g_{1}^{2} M_1  -5 \Big({Y_{t'}^*  T_{t'}}\Big) \Big)\nonumber \\ 
 &+\frac{4}{5} \Big({Y_{t'}  Y_{t'}^*}\Big) \Big(2 g_{1}^{2} m_{H_u}^2 +40 g_{3}^{2} m_{H_u}^2 +2 g_{1}^{2} m_{\tilde{t}'}^2 +40 g_{3}^{2} m_{\tilde{t}'}^2 +80 g_{3}^{2} |M_3|^2 -45 \Big({Y_{t'}  {m_q^2  Y_{t'}^*}}\Big) \nonumber \\ 
 &-45 \Big({Y_{t'}  {Y_{u}^{\dagger}  m_{\tilde u \tilde t'}^2}}\Big) -45 \Big({T_{t'}  T_{t'}^*}\Big) \Big)\nonumber \\ 
 &-\frac{2}{5} \Big(-4 \Big(20 g_{3}^{2}  + g_{1}^{2}\Big)\Big({Y_{t'}  {m_q^2  Y_{t'}^*}}\Big) -4 g_{1}^{2} \Big({Y_{t'}  {Y_{u}^{\dagger}  m_{\tilde u \tilde t'}^2}}\Big) -80 g_{3}^{2} \Big({Y_{t'}  {Y_{u}^{\dagger}  m_{\tilde u \tilde t'}^2}}\Big) \nonumber \\ 
 &+15 m_{H_d}^2 \Big({Y_{t'}  {Y_{d}^{\dagger}  Y_d  Y_{t'}^*}}\Big) +15 m_{H_u}^2 \Big({Y_{t'}  {Y_{d}^{\dagger}  Y_d  Y_{t'}^*}}\Big) +15 m_{\tilde{t}'}^2 \Big({Y_{t'}  {Y_{d}^{\dagger}  Y_d  Y_{t'}^*}}\Big) \nonumber \\ 
 &+180 m_{H_u}^2 \Big({Y_{t'}  {Y_{u}^{\dagger}  Y_u  Y_{t'}^*}}\Big) +90 m_{\tilde{t}'}^2 \Big({Y_{t'}  {Y_{u}^{\dagger}  Y_u  Y_{t'}^*}}\Big) +15 \Big({Y_{t'}  {T_{d}^{\dagger}  T_d  Y_{t'}^*}}\Big) \nonumber \\ 
 &+90 \Big({Y_{t'}  {T_{u}^{\dagger}  T_u  Y_{t'}^*}}\Big) +15 \Big({Y_{t'}  {m_q^2  Y_{d}^{\dagger}  Y_d  Y_{t'}^*}}\Big) +90 \Big({Y_{t'}  {m_q^2  Y_{u}^{\dagger}  Y_u  Y_{t'}^*}}\Big) \nonumber \\ 
 &+15 \Big({Y_{t'}  {Y_{d}^{\dagger}  m_d^2  Y_d  Y_{t'}^*}}\Big) +15 \Big({Y_{t'}  {Y_{d}^{\dagger}  Y_d  m_q^2  Y_{t'}^*}}\Big) +15 \Big({Y_{t'}  {Y_{d}^{\dagger}  Y_d  Y_{u}^{\dagger}  m_{\tilde u \tilde t'}^2}}\Big) \nonumber \\ 
 &+90 \Big({Y_{t'}  {Y_{u}^{\dagger}  m_u^2  Y_u  Y_{t'}^*}}\Big) +90 \Big({Y_{t'}  {Y_{u}^{\dagger}  Y_u  m_q^2  Y_{t'}^*}}\Big) +90 \Big({Y_{t'}  {Y_{u}^{\dagger}  Y_u  Y_{u}^{\dagger}  m_{\tilde u \tilde t'}^2}}\Big) +80 g_{3}^{2} M_3^* \Big({Y_{t'}^*  T_{t'}}\Big) \nonumber \\ 
 &+\Big({Y_{t'}  T_{t'}^*}\Big) \Big(4 g_{1}^{2} M_1  + 80 g_{3}^{2} M_3  + 90 \Big({Y_{t'}^*  T_{t'}}\Big) \Big)+15 \Big({{Y_{d}^{\dagger}  Y_d  T_{t'}^*}  T_{t'}}\Big) +90 \Big({{Y_{u}^{\dagger}  Y_u  T_{t'}^*}  T_{t'}}\Big) \nonumber \\ 
 &+15 \Big({{T_{d}^{\dagger}  Y_d  Y_{t'}^*}  T_{t'}}\Big) +90 \Big({{T_{u}^{\dagger}  Y_u  Y_{t'}^*}  T_{t'}}\Big) +15 \Big({{T_{d}^{T}  Y_d^*  Y_{t'}}  T_{t'}^*}\Big) +90 \Big({{T_{u}^{T}  Y_u^*  Y_{t'}}  T_{t'}^*}\Big) \nonumber \\ 
 &-4 g_{1}^{2} \Big({T_{t'}  T_{t'}^*}\Big) -80 g_{3}^{2} \Big({T_{t'}  T_{t'}^*}\Big) \Big)\\ 
\Delta^{UV} \beta_{m_{H_u}^2}^{(2)} & =  
54 g_{2}^{4} |M_2|^2  + \frac{126}{25} g_{1}^{4} |M_1|^2 \\ 
\Delta \beta_{m_d^2}^{(2)} & =  
+\frac{64}{25} g_{1}^{4} {\bf 1} |M_1|^2 +32 g_{3}^{4} {\bf 1} |M_3|^2-2 \Big(2 \Big({Y_d  T_{t'}^*}\Big)_{i} \Big({Y_d^*  T_{t'}}\Big)_{j} +2 \Big({T_d^*  Y_{t'}}\Big)_{j} \Big({T_d  Y_{t'}^*}\Big)_{i}  \nonumber \\ 
 &+2 \Big({Y_d^*  Y_{t'}}\Big)_{j} \Big({T_d  T_{t'}^*}\Big)_{i} +\Big({Y_d^*  Y_{t'}}\Big)_{j} \Big({m_d^2  Y_d  Y_{t'}^*}\Big)_{i} +2 \Big({Y_d^*  Y_{t'}}\Big)_{j} \Big({Y_d  m_q^2  Y_{t'}^*}\Big)_{i} +2 \Big({Y_d^*  Y_{t'}}\Big)_{j} \Big({Y_d  Y_{u}^{\dagger}  m_{\tilde u \tilde t'}^2}\Big)_{i} \nonumber \\ 
 &+\Big({Y_d  Y_{t'}^*}\Big)_{i} \Big(2 \Big(m_{H_d}^2 + m_{H_u}^2 + m_{\tilde{t}'}^2\Big)\Big({Y_d^*  Y_{t'}}\Big)_{j}  + 2 \Big({T_d^*  T_{t'}}\Big)_{j}  + 2 \Big({Y_d^*  m_q^{2 *}  Y_{t'}}\Big)_{j}  + \Big({m_d^{2 *}  Y_d^*  Y_{t'}}\Big)_{j}\Big)\Big)\\ 
\Delta^{UV} \beta_{m_d^2}^{(2)} & =  
\frac{8}{25} {\bf 1} \Big(200 g_{3}^{4} |M_3|^2  + 7 g_{1}^{4} |M_1|^2 \Big)\\ 
\Delta \beta_{m_u^2}^{(1)} & =  
2 m_{\tilde u \tilde t',{i}}^{2} \Big({Y_u^*  Y_{t'}}\Big)_{j} \\ 
\Delta \beta_{m_u^2}^{(2)} & =  
+\frac{256}{25} g_{1}^{4} {\bf 1} |M_1|^2 +32 g_{3}^{4} {\bf 1} |M_3|^2 \nonumber \\ 
 &-\frac{2}{5} \Big(30 {T_u  Y_{u}^{\dagger}} \Big({Y_{t'}  T_{t'}^*}\Big) +30 {Y_u  Y_{u}^{\dagger}} \Big({Y_{t'}  {m_q^2  Y_{t'}^*}}\Big) +30 {Y_u  Y_{u}^{\dagger}} \Big({Y_{t'}  {Y_{u}^{\dagger}  m_{\tilde u \tilde t'}^2}}\Big) \nonumber \\ 
 &+30 {Y_u  T_{u}^{\dagger}} \Big({Y_{t'}^*  T_{t'}}\Big) +30 {Y_u  Y_{u}^{\dagger}} \Big({T_{t'}  T_{t'}^*}\Big) +g_{1}^{2} m_{\tilde u \tilde t',{i}}^{2} \Big({Y_u^*  Y_{t'}}\Big)_{j} -15 g_{2}^{2} m_{\tilde u \tilde t',{i}}^{2} \Big({Y_u^*  Y_{t'}}\Big)_{j} \nonumber \\ 
 &+15 m_{\tilde u \tilde t',{i}}^{2} \mbox{Tr}\Big({Y_u  Y_{u}^{\dagger}}\Big) \Big({Y_u^*  Y_{t'}}\Big)_{j} +20 m_{H_u}^2 \Big({Y_u  Y_{t'}^*}\Big)_{i} \Big({Y_u^*  Y_{t'}}\Big)_{j} +10 m_{\tilde{t}'}^2 \Big({Y_u  Y_{t'}^*}\Big)_{i} \Big({Y_u^*  Y_{t'}}\Big)_{j} \nonumber \\ 
 &+5 \Big({Y_{t'}  Y_{t'}^*}\Big) \Big(6 \Big(2 m_{H_u}^2  + m_{\tilde{t}'}^2\Big){Y_u  Y_{u}^{\dagger}} +3 \Big(2 {T_u  T_{u}^{\dagger}}  + 2 {Y_u  m_q^2  Y_{u}^{\dagger}}  + {m_u^2  Y_u  Y_{u}^{\dagger}} + {Y_u  Y_{u}^{\dagger}  m_u^2}\Big)\nonumber \\ 
 &+4 m_{\tilde u \tilde t',{i}}^{2} \Big({Y_u^*  Y_{t'}}\Big)_{j} \Big)+10 \Big({Y_u  T_{t'}^*}\Big)_{i} \Big({Y_u^*  T_{t'}}\Big)_{j} +10 \Big({Y_u  Y_{t'}^*}\Big)_{i} \Big({T_u^*  T_{t'}}\Big)_{j} +10 \Big({T_u^*  Y_{t'}}\Big)_{j} \Big({T_u  Y_{t'}^*}\Big)_{i} \nonumber \\ 
 &+10 \Big({Y_u^*  Y_{t'}}\Big)_{j} \Big({T_u  T_{t'}^*}\Big)_{i} +5 \Big({Y_u^*  Y_{t'}}\Big)_{j} \Big({m_u^2  Y_u  Y_{t'}^*}\Big)_{i} \nonumber \\ 
 &+10 \Big({Y_u^*  Y_{t'}}\Big)_{j} \Big({Y_u  m_q^2  Y_{t'}^*}\Big)_{i} +10 \Big({Y_u^*  Y_{t'}}\Big)_{j} \Big({Y_u  Y_{u}^{\dagger}  m_{\tilde u \tilde t'}^2}\Big)_{i} +5 \Big({Y_u  Y_{t'}^*}\Big)_{i} \Big({m_u^{2 *}  Y_u^*  Y_{t'}}\Big)_{j}\nonumber \\ 
 & +10 \Big({Y_u  Y_{t'}^*}\Big)_{i} \Big({Y_u^*  m_q^{2 *}  Y_{t'}}\Big)_{j} +5 m_{\tilde u \tilde t',{i}}^{2} \Big({Y_u^*  Y_{d}^{T}  Y_d^*  Y_{t'}}\Big)_{j} +5 m_{\tilde u \tilde t',{i}}^{2} \Big({Y_u^*  Y_{u}^{T}  Y_u^*  Y_{t'}}\Big)_{j} \Big)\\ 
\Delta^{UV} \beta_{m_u^2}^{(2)} & =  
\frac{32}{25} {\bf 1} \Big(50 g_{3}^{4} |M_3|^2  + 7 g_{1}^{4} |M_1|^2 \Big)\\ 
\Delta \beta_{m_e^2}^{(2)} & =  \frac{576}{25} g_{1}^{4} {\bf 1} |M_1|^2 \\ 
\Delta^{UV} \beta_{m_e^2}^{(2)} & =  
\frac{504}{25} g_{1}^{4} {\bf 1} |M_1|^2 \\ 
\beta_{m_{\tilde{t}'}^2}^{(1)} & =  
-\frac{32}{15} g_{1}^{2} |M_1|^2 -\frac{32}{3} g_{3}^{2} |M_3|^2 +4 m_{H_u}^2 \Big({Y_{t'}  Y_{t'}^*}\Big) +4 m_{\tilde{t}'}^2 \Big({Y_{t'}  Y_{t'}^*}\Big) +4 \Big({Y_{t'}  {m_q^2  Y_{t'}^*}}\Big) \nonumber \\ 
 &+2 \Big({Y_{t'}  {Y_{u}^{\dagger}  m_{\tilde u \tilde t'}^2}}\Big) +4 \Big({T_{t'}  T_{t'}^*}\Big) -4 \frac{1}{\sqrt{15}} g_1 \sigma_{1,1} \\ 
 \beta_{m_{\tilde{t}'}^2}^{(2)} & =  
\frac{2}{225} \Big(2 g_{1}^{2} M_1^* \Big(45 \Big({Y_{t'}^*  T_{t'}}\Big)  + 8 \Big(393 g_{1}^{2} M_1  + 40 g_{3}^{2} M_3  + 80 g_{3}^{2} M_1 \Big) -90 M_1 \Big({Y_{t'}  Y_{t'}^*}\Big) \Big)\nonumber \\ 
 &+5 \Big(16 g_{3}^{2} \Big(-15 g_{3}^{2} M_3  + 8 g_{1}^{2} \Big(2 M_3  + M_1\Big)\Big)M_3^* \nonumber \\ 
 &-3 \Big(240 \Big(m_{H_u}^2 + m_{\tilde{t}'}^2\Big)\Big(\Big({Y_{t'}  Y_{t'}^*}\Big)\Big)^{2} +6 g_{1}^{2} \Big({Y_{t'}  {m_q^2  Y_{t'}^*}}\Big) -90 g_{2}^{2} \Big({Y_{t'}  {m_q^2  Y_{t'}^*}}\Big) +3 g_{1}^{2} \Big({Y_{t'}  {Y_{u}^{\dagger}  m_{\tilde u \tilde t'}^2}}\Big) \nonumber \\ 
 &-45 g_{2}^{2} \Big({Y_{t'}  {Y_{u}^{\dagger}  m_{\tilde u \tilde t'}^2}}\Big) +30 m_{H_d}^2 \Big({Y_{t'}  {Y_{d}^{\dagger}  Y_d  Y_{t'}^*}}\Big) +30 m_{H_u}^2 \Big({Y_{t'}  {Y_{d}^{\dagger}  Y_d  Y_{t'}^*}}\Big) \nonumber \\ 
 &+30 m_{\tilde{t}'}^2 \Big({Y_{t'}  {Y_{d}^{\dagger}  Y_d  Y_{t'}^*}}\Big) +60 m_{H_u}^2 \Big({Y_{t'}  {Y_{u}^{\dagger}  Y_u  Y_{t'}^*}}\Big) +30 m_{\tilde{t}'}^2 \Big({Y_{t'}  {Y_{u}^{\dagger}  Y_u  Y_{t'}^*}}\Big) \nonumber \\ 
 &+30 \Big({Y_{t'}  {T_{d}^{\dagger}  T_d  Y_{t'}^*}}\Big) +30 \Big({Y_{t'}  {T_{u}^{\dagger}  T_u  Y_{t'}^*}}\Big) +30 \Big({Y_{t'}  {m_q^2  Y_{d}^{\dagger}  Y_d  Y_{t'}^*}}\Big) +30 \Big({Y_{t'}  {m_q^2  Y_{u}^{\dagger}  Y_u  Y_{t'}^*}}\Big) \nonumber \\ 
 &+30 \Big({Y_{t'}  {Y_{d}^{\dagger}  m_d^2  Y_d  Y_{t'}^*}}\Big) +30 \Big({Y_{t'}  {Y_{d}^{\dagger}  Y_d  m_q^2  Y_{t'}^*}}\Big) +15 \Big({Y_{t'}  {Y_{d}^{\dagger}  Y_d  Y_{u}^{\dagger}  m_{\tilde u \tilde t'}^2}}\Big) \nonumber \\ 
 &+30 \Big({Y_{t'}  {Y_{u}^{\dagger}  m_u^2  Y_u  Y_{t'}^*}}\Big) +30 \Big({Y_{t'}  {Y_{u}^{\dagger}  Y_u  m_q^2  Y_{t'}^*}}\Big) +15 \Big({Y_{t'}  {Y_{u}^{\dagger}  Y_u  Y_{u}^{\dagger}  m_{\tilde u \tilde t'}^2}}\Big) +90 g_{2}^{2} M_2^* \Big({Y_{t'}^*  T_{t'}}\Big) \nonumber \\ 
 &+30 \Big({{Y_{d}^{\dagger}  Y_d  T_{t'}^*}  T_{t'}}\Big) +30 \Big({{Y_{u}^{\dagger}  Y_u  T_{t'}^*}  T_{t'}}\Big) +30 \Big({{T_{d}^{\dagger}  Y_d  Y_{t'}^*}  T_{t'}}\Big) +30 \Big({{T_{u}^{\dagger}  Y_u  Y_{t'}^*}  T_{t'}}\Big) \nonumber \\ 
 &+30 \Big({{T_{d}^{T}  Y_d^*  Y_{t'}}  T_{t'}^*}\Big) +30 \Big({{T_{u}^{T}  Y_u^*  Y_{t'}}  T_{t'}^*}\Big) +6 g_{1}^{2} \Big({T_{t'}  T_{t'}^*}\Big) -90 g_{2}^{2} \Big({T_{t'}  T_{t'}^*}\Big) -80 g_{3}^{4} \sigma_{2,3} -16 g_{1}^{2} \sigma_{2,11} \nonumber \\ 
 &+8 \sqrt{15} g_1 \sigma_{3,1} +90 \Big({Y_{t'}  {m_q^2  Y_{t'}^*}}\Big) \mbox{Tr}\Big({Y_u  Y_{u}^{\dagger}}\Big) +45 \Big({Y_{t'}  {Y_{u}^{\dagger}  m_{\tilde u \tilde t'}^2}}\Big) \mbox{Tr}\Big({Y_u  Y_{u}^{\dagger}}\Big) \nonumber \\ 
 &+90 \Big({T_{t'}  T_{t'}^*}\Big) \mbox{Tr}\Big({Y_u  Y_{u}^{\dagger}}\Big) +\Big({Y_{t'}  T_{t'}^*}\Big) \Big(240 \Big({Y_{t'}^*  T_{t'}}\Big)  -6 g_{1}^{2} M_1  + 90 g_{2}^{2} M_2  + 90 \mbox{Tr}\Big({Y_{u}^{\dagger}  T_u}\Big) \Big)\nonumber \\ 
 &+90 \Big({Y_{t'}^*  T_{t'}}\Big) \mbox{Tr}\Big({T_u^*  Y_{u}^{T}}\Big) +6 \Big({Y_{t'}  Y_{t'}^*}\Big) \Big(g_{1}^{2} m_{H_u}^2 -15 g_{2}^{2} m_{H_u}^2 +g_{1}^{2} m_{\tilde{t}'}^2 -15 g_{2}^{2} m_{\tilde{t}'}^2 -30 g_{2}^{2} |M_2|^2\nonumber \\ 
 & +40 \Big({Y_{t'}  {m_q^2  Y_{t'}^*}}\Big) +30 \Big({Y_{t'}  {Y_{u}^{\dagger}  m_{\tilde u \tilde t'}^2}}\Big) \nonumber \\ 
 &+40 \Big({T_{t'}  T_{t'}^*}\Big) +30 m_{H_u}^2 \mbox{Tr}\Big({Y_u  Y_{u}^{\dagger}}\Big) +15 m_{\tilde{t}'}^2 \mbox{Tr}\Big({Y_u  Y_{u}^{\dagger}}\Big) +15 \mbox{Tr}\Big({T_u^*  T_{u}^{T}}\Big) +15 \mbox{Tr}\Big({m_q^2  Y_{u}^{\dagger}  Y_u}\Big) \nonumber \\ 
 &+15 \mbox{Tr}\Big({m_u^2  Y_u  Y_{u}^{\dagger}}\Big) \Big)\Big)\Big)\Big)\\ 
\Delta^{UV} \beta_{m_{\tilde{t}'}^2}^{(2)} & =  
64 g_{3}^{4} |M_3|^2  + \frac{224}{25} g_{1}^{4} |M_1|^2 \\ 
\beta_{m_{\tilde{\bar{t}}'}^2}^{(1)} & =  
-\frac{4}{15} \Big(40 g_{3}^{2} |M_3|^2  + 8 g_{1}^{2} |M_1|^2  - \sqrt{15} g_1 \sigma_{1,1} \Big)\\ 
\beta_{m_{\tilde{\bar{t}}'}^2}^{(2)} & =  
\frac{16}{225} \Big(2 g_{1}^{2} \Big(393 g_{1}^{2} M_1  + 40 g_{3}^{2} \Big(2 M_1  + M_3\Big)\Big)M_1^* \nonumber \\ 
 &+5 \Big(2 g_{3}^{2} \Big(-15 g_{3}^{2} M_3  + 8 g_{1}^{2} \Big(2 M_3  + M_1\Big)\Big)M_3^*  + 30 g_{3}^{4} \sigma_{2,3}  + 3 \sqrt{15} g_1 \sigma_{3,1}  + 6 g_{1}^{2} \sigma_{2,11} \Big)\Big)\\ 
\beta_{m_{\tilde u \tilde t',{i}}^{2}}^{(1)} & =  
2 \Big(\Big(2 m_{H_u}^2  + m_{\tilde{t}'}^2\Big)\Big({Y_u^*  Y_{t'}}\Big)_{i}  + 2 \Big({T_u^*  T_{t'}}\Big)_{i}  + 2 \Big({Y_u^*  m_q^{2 *}  Y_{t'}}\Big)_{i}  + \Big({m_u^{2 *}  Y_u^*  Y_{t'}}\Big)_{i}\Big)\\ 
\beta_{m_{\tilde u \tilde t',{i}}^{2}}^{(2)} & =  
-\frac{2}{5} \Big(\Big(2 g_{1}^{2} m_{H_u}^2 -30 g_{2}^{2} m_{H_u}^2 +g_{1}^{2} m_{\tilde{t}'}^2 -15 g_{2}^{2} m_{\tilde{t}'}^2 +4 g_{1}^{2} |M_1|^2 -60 g_{2}^{2} |M_2|^2 +80 m_{H_u}^2 \Big({Y_{t'}  Y_{t'}^*}\Big) \nonumber \\ 
 &+60 m_{\tilde{t}'}^2 \Big({Y_{t'}  Y_{t'}^*}\Big) +40 \Big({Y_{t'}  {m_q^2  Y_{t'}^*}}\Big) +40 \Big({Y_{t'}  {Y_{u}^{\dagger}  m_{\tilde u \tilde t'}^2}}\Big) +40 \Big({T_{t'}  T_{t'}^*}\Big) +60 m_{H_u}^2 \mbox{Tr}\Big({Y_u  Y_{u}^{\dagger}}\Big) \nonumber \\ 
 &+15 m_{\tilde{t}'}^2 \mbox{Tr}\Big({Y_u  Y_{u}^{\dagger}}\Big) +30 \mbox{Tr}\Big({T_u^*  T_{u}^{T}}\Big) +30 \mbox{Tr}\Big({m_q^2  Y_{u}^{\dagger}  Y_u}\Big) +30 \mbox{Tr}\Big({m_u^2  Y_u  Y_{u}^{\dagger}}\Big) \Big)\Big({Y_u^*  Y_{t'}}\Big)_{i} \nonumber \\ 
 &-2 g_{1}^{2} M_1^* \Big({Y_u^*  T_{t'}}\Big)_{i} +30 g_{2}^{2} M_2^* \Big({Y_u^*  T_{t'}}\Big)_{i} +40 \Big({Y_{t'}  T_{t'}^*}\Big) \Big({Y_u^*  T_{t'}}\Big)_{i} +30 \mbox{Tr}\Big({T_u^*  Y_{u}^{T}}\Big) \Big({Y_u^*  T_{t'}}\Big)_{i} \nonumber \\ 
 & -2 g_{1}^{2} M_1 \Big({T_u^*  Y_{t'}}\Big)_{i} +30 g_{2}^{2} M_2 \Big({T_u^*  Y_{t'}}\Big)_{i} +40 \Big({Y_{t'}^*  T_{t'}}\Big) \Big({T_u^*  Y_{t'}}\Big)_{i} +30 \mbox{Tr}\Big({Y_{u}^{\dagger}  T_u}\Big) \Big({T_u^*  Y_{t'}}\Big)_{i} \nonumber \\ 
 &+2 g_{1}^{2} \Big({T_u^*  T_{t'}}\Big)_{i} -30 g_{2}^{2} \Big({T_u^*  T_{t'}}\Big)_{i}+40 \Big({Y_{t'}  Y_{t'}^*}\Big) \Big({T_u^*  T_{t'}}\Big)_{i} +30 \mbox{Tr}\Big({Y_u  Y_{u}^{\dagger}}\Big) \Big({T_u^*  T_{t'}}\Big)_{i}  \nonumber \\ 
 &+g_{1}^{2} \Big({m_u^{2 *}  Y_u^*  Y_{t'}}\Big)_{i} -15 g_{2}^{2} \Big({m_u^{2 *}  Y_u^*  Y_{t'}}\Big)_{i} +20 \Big({Y_{t'}  Y_{t'}^*}\Big) \Big({m_u^{2 *}  Y_u^*  Y_{t'}}\Big)_{i} \nonumber \\ 
 &+15 \mbox{Tr}\Big({Y_u  Y_{u}^{\dagger}}\Big) \Big({m_u^{2 *}  Y_u^*  Y_{t'}}\Big)_{i} +2 g_{1}^{2} \Big({Y_u^*  m_q^{2 *}  Y_{t'}}\Big)_{i} -30 g_{2}^{2} \Big({Y_u^*  m_q^{2 *}  Y_{t'}}\Big)_{i}  \nonumber \\ 
 &+40 \Big({Y_{t'}  Y_{t'}^*}\Big) \Big({Y_u^*  m_q^{2 *}  Y_{t'}}\Big)_{i} +30 \mbox{Tr}\Big({Y_u  Y_{u}^{\dagger}}\Big) \Big({Y_u^*  m_q^{2 *}  Y_{t'}}\Big)_{i} +10 m_{H_d}^2 \Big({Y_u^*  Y_{d}^{T}  Y_d^*  Y_{t'}}\Big)_{i} \nonumber \\ 
 &+10 m_{H_u}^2 \Big({Y_u^*  Y_{d}^{T}  Y_d^*  Y_{t'}}\Big)_{i} +5 m_{\tilde{t}'}^2 \Big({Y_u^*  Y_{d}^{T}  Y_d^*  Y_{t'}}\Big)_{i} +10 \Big({Y_u^*  Y_{d}^{T}  T_d^*  T_{t'}}\Big)_{i} \nonumber \\ 
 &+20 m_{H_u}^2 \Big({Y_u^*  Y_{u}^{T}  Y_u^*  Y_{t'}}\Big)_{i} +5 m_{\tilde{t}'}^2 \Big({Y_u^*  Y_{u}^{T}  Y_u^*  Y_{t'}}\Big)_{i} +10 \Big({Y_u^*  Y_{u}^{T}  T_u^*  T_{t'}}\Big)_{i} \nonumber \\ 
 &+10 \Big({Y_u^*  T_{d}^{T}  T_d^*  Y_{t'}}\Big)_{i} +10 \Big({Y_u^*  T_{u}^{T}  T_u^*  Y_{t'}}\Big)_{i} +10 \Big({T_u^*  Y_{d}^{T}  Y_d^*  T_{t'}}\Big)_{i} +10 \Big({T_u^*  Y_{u}^{T}  Y_u^*  T_{t'}}\Big)_{i} \nonumber \\ 
 & +10 \Big({T_u^*  T_{d}^{T}  Y_d^*  Y_{t'}}\Big)_{i} +10 \Big({T_u^*  T_{u}^{T}  Y_u^*  Y_{t'}}\Big)_{i} +5 \Big({m_u^{2 *}  Y_u^*  Y_{d}^{T}  Y_d^*  Y_{t'}}\Big)_{i} +5 \Big({m_u^{2 *}  Y_u^*  Y_{u}^{T}  Y_u^*  Y_{t'}}\Big)_{i} \nonumber \\ 
 &+10 \Big({Y_u^*  m_q^{2 *}  Y_{d}^{T}  Y_d^*  Y_{t'}}\Big)_{i} +10 \Big({Y_u^*  m_q^{2 *}  Y_{u}^{T}  Y_u^*  Y_{t'}}\Big)_{i} +10 \Big({Y_u^*  Y_{d}^{T}  m_d^{2 *}  Y_d^*  Y_{t'}}\Big)_{i} +10 \Big({Y_u^*  Y_{d}^{T}  Y_d^*  m_q^{2 *}  Y_{t'}}\Big)_{i} \nonumber \\ 
 &+10 \Big({Y_u^*  Y_{u}^{T}  m_u^{2 *}  Y_u^*  Y_{t'}}\Big)_{i} +10 \Big({Y_u^*  Y_{u}^{T}  Y_u^*  m_q^{2 *}  Y_{t'}}\Big)_{i} \Big)
 \Delta^{UV} \beta_{m_{\tilde{\bar{t}}'}^2}^{(2)} & =  
64 g_{3}^{4} |M_3|^2  + \frac{224}{25} g_{1}^{4} |M_1|^2 \\ 
\beta_{m_{\tilde{q}'}^2}^{(1)} & =  
-6 g_{2}^{2} |M_2|^2  + \frac{1}{\sqrt{15}} g_1 \sigma_{1,1}  -\frac{2}{15} g_{1}^{2} |M_1|^2  -\frac{32}{3} g_{3}^{2} |M_3|^2 \\ 
\beta_{m_{\tilde{q}'}^2}^{(2)} & =  
+\frac{2}{5} g_{1}^{2} g_{2}^{2} |M_2|^2 +87 g_{2}^{4} |M_2|^2 +32 g_{2}^{2} g_{3}^{2} |M_2|^2 \nonumber \\ 
 &+\frac{1}{225} g_{1}^{2} \Big(5 \Big(16 g_{3}^{2} \Big(2 M_1  + M_3\Big) + 9 g_{2}^{2} \Big(2 M_1  + M_2\Big)\Big) + 867 g_{1}^{2} M_1 \Big)M_1^* \nonumber \\ 
 &+\frac{16}{45} g_{3}^{2} \Big(15 \Big(10 g_{3}^{2} M_3  + 3 g_{2}^{2} \Big(2 M_3  + M_2\Big)\Big) + g_{1}^{2} \Big(2 M_3  + M_1\Big)\Big)M_3^* +\frac{1}{5} g_{1}^{2} g_{2}^{2} M_1 M_2^*\nonumber \\ 
 & +16 g_{2}^{2} g_{3}^{2} M_3 M_2^* +6 g_{2}^{4} \sigma_{2,2} +\frac{32}{3} g_{3}^{4} \sigma_{2,3} +\frac{2}{15} g_{1}^{2} \sigma_{2,11} +4 \frac{1}{\sqrt{15}} g_1 \sigma_{3,1} \\ 
\beta_{m_{\tilde{\bar{q}}'}^2}^{(1)} & =  
-6 g_{2}^{2} |M_2|^2  - \frac{1}{\sqrt{15}} g_1 \sigma_{1,1}  -\frac{2}{15} g_{1}^{2} |M_1|^2  -\frac{32}{3} g_{3}^{2} |M_3|^2 \\ 
\beta_{m_{\tilde{\bar{q}}'}^2}^{(2)} & =  
+\frac{2}{5} g_{1}^{2} g_{2}^{2} |M_2|^2 +87 g_{2}^{4} |M_2|^2 +32 g_{2}^{2} g_{3}^{2} |M_2|^2 \nonumber \\ 
 &+\frac{1}{225} g_{1}^{2} \Big(5 \Big(16 g_{3}^{2} \Big(2 M_1  + M_3\Big) + 9 g_{2}^{2} \Big(2 M_1  + M_2\Big)\Big) + 867 g_{1}^{2} M_1 \Big)M_1^* \nonumber \\ 
 &+\frac{16}{45} g_{3}^{2} \Big(15 \Big(10 g_{3}^{2} M_3  + 3 g_{2}^{2} \Big(2 M_3  + M_2\Big)\Big) + g_{1}^{2} \Big(2 M_3  + M_1\Big)\Big)M_3^* +\frac{1}{5} g_{1}^{2} g_{2}^{2} M_1 M_2^* \nonumber \\ 
 &+16 g_{2}^{2} g_{3}^{2} M_3 M_2^* +6 g_{2}^{4} \sigma_{2,2} +\frac{32}{3} g_{3}^{4} \sigma_{2,3} +\frac{2}{15} g_{1}^{2} \sigma_{2,11} -4 \frac{1}{\sqrt{15}} g_1 \sigma_{3,1} \\ 
\beta_{m_{\tilde{e}'}^2}^{(1)} & =  
\frac{2}{5} g_1 \Big(-12 g_1 |M_1|^2  + \sqrt{15} \sigma_{1,1} \Big)\\ 
\beta_{m_{\tilde{e}'}^2}^{(2)} & =  
\frac{8}{25} g_1 \Big(15 g_1 \sigma_{2,11}  + 486 g_{1}^{3} |M_1|^2  + 5 \sqrt{15} \sigma_{3,1} \Big)\\ 
\beta_{m_{\tilde{\bar{e}}'}^2}^{(1)} & =  
-\frac{2}{5} g_1 \Big(12 g_1 |M_1|^2  + \sqrt{15} \sigma_{1,1} \Big)\\ 
\beta_{m_{\tilde{\bar{e}}'}^2}^{(2)} & =  
\frac{8}{25} g_1 \Big(15 g_1 \sigma_{2,11}  + 486 g_{1}^{3} |M_1|^2  -5 \sqrt{15} \sigma_{3,1} \Big)
\end{align}}

\subsection{Vacuum expectation values}
{\allowdisplaybreaks  \begin{align} 
\Delta \beta_{v_d}^{(2)} & =  
3 v_d \Big({Y_{t'}  {Y_{d}^{\dagger}  Y_d  Y_{t'}^*}}\Big)  -\frac{6}{25} g_{1}^{4} v_d \\ 
\Delta^{UV} \beta_{v_d}^{(2)} & =  
-\frac{3}{100} \Big(75 g_{2}^{4}  + 7 g_{1}^{4} \Big)v_d \\ 
\Delta \beta_{v_u}^{(1)} & =  
-3 v_u \Big({Y_{t'}  Y_{t'}^*}\Big) \\ 
\Delta \beta_{v_u}^{(2)} & =  
-\frac{1}{50} v_u \Big(5 \Big(5 \Big(32 g_{3}^{2}  + 9 g_{2}^{2} \text{Xi} \Big) + g_{1}^{2} \Big(9 \text{Xi}  + 8\Big)\Big)\Big({Y_{t'}  Y_{t'}^*}\Big) -450 \Big(\Big({Y_{t'}  Y_{t'}^*}\Big)\Big)^{2} \nonumber \\ 
 &+6 \Big(-25 \Big({Y_{t'}  {Y_{d}^{\dagger}  Y_d  Y_{t'}^*}}\Big)  + 2 \Big(-75 \Big({Y_{t'}  {Y_{u}^{\dagger}  Y_u  Y_{t'}^*}}\Big)  + g_{1}^{4}\Big)\Big)\Big)\\
\Delta^{UV} \beta_{v_u}^{(2)} & =  
-\frac{3}{100} \Big(75 g_{2}^{4}  + 7 g_{1}^{4} \Big)v_u  
 \end{align}}

\end{appendix}

\bibliographystyle{JHEP-2}
\bibliography{lit}

\end{document}